\documentclass[prd,aps,10pt,a4paper,twocolumn,showpacs,superscriptaddress,nofootinbib,nobibnotes,floatfix,showkeys]{revtex4-1}

\usepackage{graphicx}
\usepackage{bm}

\newcommand{\OBS}{{\rm obs}}
\newcommand{\GRAV}{{\rm grav}}
\newcommand{\MAX}{{\rm max}}
\newcommand{\MIN}{{\rm min}}
\newcommand{\KIN}{{\rm kin}}
\newcommand{\TOT}{{\rm tot}}
\newcommand{\CIRC}{{\rm circ}}
\newcommand{\STAT}{{\rm stat}}

\newcommand{\PER}{{\rm per}}
\newcommand{\INF}{{\rm inf}}
\newcommand{\CRIT}{{\rm crit}}

\newcommand{\ddd}{{\rm d}}

\newcommand{\SST}{${}^{\rm st}$}
\newcommand{\RRD}{${}^{\rm rd}$}
\newcommand{\NND}{${}^{\rm nd}$}
\newcommand{\TTH}{${}^{\rm th}$}

\graphicspath{{FIGS/}}

\begin{document}

\title{Seeing relativity -- I. Ray tracing in a Schwarzschild metric to
  explore the maximal analytic extension of the metric and making a
  proper rendering of the stars}

\author{Alain Riazuelo} 

\email{riazuelo@iap.fr}

\affiliation{Sorbonne Universit\'e, CNRS, UMR~7095, Institut
  d’Astrophysique de Paris, 98 bis boulevard Arago, 75014~Paris,
  France}

\begin{abstract}
  We present an implementation of a ray tracing code in the
  Schwarzschild metric. We aim at building a numerical code with a
  correct implementation of both special (aberration, amplification,
  Doppler) and general (deflection of light, lensing, gravitational
  redshift) relativistic effects so as to simulate what an observer
  with arbitrary velocity would see near, or possibly within, the
  black hole. We also pay some specific attention to perform a
  satisfactory rendering of stars. Using this code, we then show
  several unexplored features of the maximal analytical extension of
  the metric. In particular, we study the aspect of the second
  asymptotic region of the metric as seen by an observer crossing the
  horizon. We also address several aspects related to the white hole
  region (i.e., past singularity) seen both from outside the black
  hole, inside the future horizon and inside the past horizon, which
  gives rise to the most counter-intuitive effects.
\end{abstract}




\maketitle

\section{Introduction}

The visual aspect of black holes is a frequent public outreach
question but for decades, did not generate much interest among
physicists or astronomers. A somewhat caricatural example of this is
given by S.~Chandrasekhar famous book on black
holes~\cite{chandrasekhar83}, where the visual aspect of a
Schwarzschild black hole is summarized in a tiny picture on page~130,
the caption of which being rather pedantic and obscure for a non
expert reader since the author does not describe the picture as a
sketch of the angular size of a Schwarzschild black hole as a function
of the distance of a static observer, but rather talks about the
``cone of avoidance'' of null geodesics (which technically means the
same thing).

Still, the problem of black hole visualization had already drawn some
sparse attention at the time of Chandrasekhar book in the more
difficult context of the Kerr metric, the earliest work being that of
Bardeen in 1972~\cite{bardeen72} and later
Luminet~\cite{luminet79}. An increasing number of work were published
afterward, a very incomplete subset of which lies in
Refs.~\cite{fukue88,viergutz93,marck95,fanton97,falcke00,hamilton04,beckwith05}.
In the recent years, a much larger amount of work has been performed on
black hole visualization.  The main reason for this growing interest
came from obvious astronomical constraints: the largest black hole (in
term of angular size) seen from Earth, Sgr~A*, has an angular diameter
of order of 60~$\mu$as~\cite{gillessen09}, which is unobservable by
conventional astronomical devices, but which should be at reach within
less than a decade with the advent of long baseline interferometry in
the millimetric domain, thus making the actual aspect of a black hole
silhouette become a problem of astronomical relevance. Consequently,
most of those recent works are motivated by actual astrophysical
observational projects of our Galactic center such as
GRAVITY~\cite{gravity} or the Event Horizon Telescope~\cite{eht}, see,
e.g., \cite{vincent11,broderick11,chan13,broderick14}, but some others
were focused on what could be seen if an observer stood close to a
black hole~\cite{muller10,thorne15} thus being less relevant from an
observational perspective, but more focused on the diversity of
physical effects that can arise in the vicinity of a black hole. Our
work fits within this second category.

The case of special relativity is much simpler and has deserved an
earlier attention as early as 70~years ago with the pioneering
sketches of Gamow~\cite{gamow}. (Although it is often said that
special relativity was also a source of artistic inspiration for
S.~Dali in his famous painting ``La persist\`encia de la Memoria''
(The Persistence of Memory, 1931), it actually does not seem to be the
case~\cite{dali}.) The increasingly easier access to large computing
facilities has progressively allowed the completion of excellent works
by, for example Ruder and Nollert~\cite{ruder-nollert} or Searle {\it
  et al.}~\cite{searle}.

The aim of this paper is to present here a numerical code that
implements most of the relativistic effects that arise when simulating
what an observer would see in a black hole metric, focusing here on
the Schwarzschild one. The Schwarzschild metric is the simplest black
hole metric that exists. It is both astrophysically relevant
(contrarily to the Reissner-Nordstrom one) and simple to study thanks
to its spherical symmetry (contrarily to the Kerr on Kerr-Newmann
metrics). In particular, as we shall see, it is possible to perform a
rather nice and efficient rendering of pointlike light sources (i.e.,
stars) thanks to spherical symmetry, an issue which was not addressed
satisfactorily till now. Although initially made for teaching purpose,
these simulations allowed an exhaustive study of the metric and was
very easy to adapt to the whole Schwarzschild metric, i.e. its maximal
analytic extension, or Kruskal-Szekeres coordinates, where we found a
series of unexpected visual effect which where rather counter-intuitive
even for an experienced relativity scientist.

The paper is structured as follows. The camera (i.e., the way to
project a part of the celestial sphere on a two-dimensional screen) is
described in Sec.~\ref{sec_cam}. We then address the way to simulate
the way a background sky is distorted both by special relativistic
effects and the presence of a gravitational field
(Sec.~\ref{sec_cel1}). This first, naive, method can be significantly
improved in term of computational time by explicitly using the fact
that the metric is spherically symmetric as explained in
Sec.~\ref{sec_cel2}.  The results of this section are then used to
make a very rapid and satisfactory rendering of the stars (or any
pointlike light source), as explained in Sec.~\ref{sec_stars}.  In
Sec.~\ref{sec_data}, we list all the data that we use in order to
produce examples of realistic images. We computed many pictures using
this method which are useful in getting a better representation of
special and general relativistic effects. Since many of these
correspond more a new way to show old results rather than actual new
results, we have put then in Appendix~\ref{app_img}. Still keeping in
mind that this work can have some obvious popular science application,
we explain in Sec.\ref{sec_plane} how to rather easily adapt them for
planetarium projection using the standards of this field. This being
done, we present in Sec.~\ref{sec_hor} some results we obtained when
exploring the parameter space of the simulated images, and that were
not, or not significantly emphasized in existing literature. For
example, we compute the angular size of the horizon both at horizon
crossing and close to the singularity. Then, we explore in
Sec.~\ref{sec_Kruskal} some new features which happen when considering
the maximal analytic extension of the metric, both inside and outside
the horizon.

In what follows, we shall use the $(+---)$ convention for the metric
signature. We place ourselves in a coordinates systems such that
$c= G = 1$, and keep the second as time unit, so that a distance of
$1$ corresponds to one light second (i.e.
approximately~$3 \times 10^5\;{\rm km}$) and a mass of $1$
corresponds to around $2\times10^5$ Solar masses. As long as our in
interested in making single images, the mass of the black hole does
not matter, however, since only the ratio $r / M$, $r$ being the
radial coordinate, matters. Unless otherwise specified we shall place
ourselves in a spherical version of the so-called Schwarzschild
coordinates, where the line element is written
\begin{equation}
\label{ds2}
\ddd s^2 = 
   \left(1 - \frac{2 M}{r} \right) \ddd t^2
 - \frac{\ddd r^2}{1 - \frac{2 M}{r}}
 - r^2 \left(\ddd \theta^2 + \sin^2 \theta \ddd \varphi^2 \right) .
\end{equation}

\section{Defining the observer and its camera}
\label{sec_cam}

We assume that the images we want to simulate are those that would be
seen by an observer is endowed with a four-velocity $u_\OBS^\mu$ that
either results from a chosen trajectory (possibly following a geodesic
or not) or from any user defined data. We do not take into account the
fact that the observer is unlikely to survive to his/her journey
within the black hole environment either because of arbitrarily large
acceleration imposed by some non geodesic motion (staying static just
above the horizon, for example) or because of arbitrarily large tidal
effects endured close or within a black hole of sufficiently small
mass. We simulate here standard images and do not address much the
interesting issue of stereoscopic vision as was done in
Ref.~\cite{hamilton10}.

The camera orientation is described by a set of three unit orthogonal
spacelike vectors, $X^\mu$, $Y^\mu$, $Z^\mu$ all of which are
orthogonal to $u_\OBS^\mu$. We define the orientation of the camera by
the following assumptions:
\begin{itemize}

\item Any pixel of the screen can be seen as pointing toward a
  spacelike direction $N^\mu$ belonging to the $X^\mu$, $Y^\mu$,
  $Z^\mu$ hyperplane, where $N_\mu N^\mu = - 1$, i.e., a photon
  hitting the screen coming from this direction possesses a wave
  vector proportional to $u_\OBS^\mu - N^\mu$, which is evidently a
  null vector;

\item The $Z^\mu$ direction points toward the center of the screen,
  i.e. photons hitting the center of the screen possesses a wavevector
  proportional to $u_\OBS^\mu - Z^\mu$;

\item The $X^\mu$ direction is associated to the central horizontal
  line of the screen in the sense that any photon hitting this part of
  the screen possesses a wavevector proportional to $u_\OBS - a Z^\mu
  - c X^\mu$, where $a^2 + c^2 = 1$, the value of $a$ and $c$ being
  determined by the pixel position and the choice of projection (see
  below);

\item The $Y^\mu$ direction is associated to the central vertical line
  of the screen in the sense that any photon hitting this part of the
  screen possesses a wavevector proportional to $u_\OBS - a ' Z^\mu -
  b Y^\mu$, where $a'^2 + b^2 = 1$ (same remark as for $a$ and $c$
  above).

\end{itemize}

Moreover, we do not focus here on simulating images seen from a large
distance, in which one can perform a flat sky approximation. Instead,
the viewing angle of the picture is supposed to be of same order of
normal viewing condition, an opening angle of $90^\circ$ being a
relevant choice for this purpose. The exact association between a
pixel of coordinates $(i, j)$ and the corresponding direction
$N^\mu (i, j)$ is somewhat arbitrary and depends on the choice of
projection one wants to use. A natural choice is spherical projection,
which reproduces the exact view on the screen provided that the person
viewing the screen is set at the proper position with respect to it,
that depends on the view opening angle. When not at the proper
position, any shape on the screen becomes distorted. For example, the
circular silhouette of a Schwarzschild black hole no longer appears
circular. In order to evade this problem, a natural choice corresponds
to stereographic projection for which the circular shape of the black
hole silhouette is always a circle on the screen. We shall make this
choice for flat projections. If one denotes $\alpha$ the half of the
opening angle of the picture along the horizontal direction, and $R$
and $C$ the number of rows and columns of the screen then the pixel
coordinates $(i, j)$ of a given direction $N^\mu$ is given by
\begin{eqnarray}
\label{pix_i}
i & = &   \frac{C + 1}{2} 
        + \frac{C}{2 \tan(\alpha / 2)} 
          \frac{- X_\mu N^\mu}{1 + (- N_\mu Z^\mu)} , \\
\label{pix_j}
j & = &   \frac{R + 1}{2} 
        + \frac{C}{2 \tan(\alpha / 2)} 
          \frac{- Y_\mu N^\mu}{1 + (- N_\mu Z^\mu)} .
\end{eqnarray}
(The minus sign at the denominator of each formula comes from the
signature convention we adopt here.)  We assume here that the pixel
coordinates ranges from $1$ to $C$ in along the horizontal direction,
number $1$ being on the left, and from $1$ to $R$ in the vertical
direction, from top to bottom, following the usual computer
convention. In the usual case where both $R$ and $C$ are even, the
center of the screen (i.e., $N^\mu = Z^\mu$) does not correspond to a
pixel but to the common corner of the four adjacent central pixels of
coordinates $(C/2, R/2)$, $(C/2 + 1, R/2)$, $(C/2, R/2 + 1)$ and
$(C/2 + 1, R/2 + 1)$. We also assume that pixels are square, i.e, that
pixel aspect ratio is exactly 1. In the case where one produces
pictures with non square pixels such as for standard video formats
(e.g., 576i~4:3, which has a pixel ratio of 12:11), then one has to
rescale the second formula~(\ref{pix_j}) along the vertical direction
according to the imposed pixel ratio.

The inverse transform that allows to compute the direction $N^\mu$ as
a function of pixel coordinates can be written in two steps by
defining the intermediate spacelike vector $W^\mu$,
\begin{eqnarray}
W^\mu & = &   Z^\mu
            + \frac{2 \tan(\alpha / 2)}{C} \times \\  \nonumber & & 
              \left[  \left(i - \frac{C + 1}{2}\right) X^\mu 
                    + \left(j - \frac{R + 1}{2}\right) Y^\mu \right] , \\
N^\mu & = & - Z^\mu + 2 \frac{W^\mu}{- W_\mu W^\mu} .
\end{eqnarray}

Now, the way one defines the vectors $X^\mu$, $Y^\mu$, $Z^\mu$ is done
as follows:
\begin{itemize}

\item We start from what we call a reference tetrad that is defined
  for each event of the space-time. This tetrad is orthonormal and is
  made of one timelike vector $T^\mu$ and three spacelike vectors
  $R^\mu$, $\Theta^\mu$, $\Phi^\mu$, the label of which are of course
  related to the coordinate system~(\ref{ds2}) (see later).

\item All these vectors are Lorentz transformed according to the unique
  Lorentz boost $\Lambda^\mu_{\;\nu}$ that transforms the tetrad
  timelike vector $T^\mu$ into the observer's velocity $u_\OBS^\mu$
  and leaves invariant any vector orthogonal to both of them. 

\item The three spacelike vector obtained after performing the Lorentz
  boost on $\Theta^\mu$, $\Phi^\mu$, $R^\mu$, i.e., $A^\mu \equiv
  \Lambda^\mu_{\;\nu} \Theta^\nu$, $B^\mu \equiv \Lambda^\mu_{\;\nu}
  \Phi^\nu$, $C^\mu \equiv \Lambda^\mu_{\;\nu} R^\nu$ are then rotated
  by a space rotation $R^\mu_{\;\nu}$ that leaves $u_\OBS^\mu$
  invariant in order to give the three camera vectors $X^\mu =
  R^\mu_{\;\nu} A^\nu$, $Y^\mu = R^\mu_{\;\nu} B^\nu$, $Z^\mu =
  R^\mu_{\;\nu} C^\nu$.

\end{itemize}

Simple algebra allows to prove that the $\Lambda^\mu_{\;\nu}$
components are (see, e.g.,~\cite{gourgoulhon10}):
\begin{equation}
\label{Lambda_cov}
\Lambda^\mu_{\;\nu}
 =   \delta^\mu_\nu
   - \frac{1}{1 + \gamma} (T^\mu + u_\OBS^\mu) (T_\nu + u_{\OBS\;\nu})
   + 2 u_\OBS^\mu T_\nu ,
\end{equation}
where we have defined $\gamma$ as the Lorentz factor
\begin{equation}
\gamma \equiv T_\mu u_\OBS^\mu ,
\end{equation}
so that one indeed has $\Lambda^\mu_{\;\nu} T^\nu = u_\OBS^\mu$ and
$\Lambda^\mu_{\;\nu} (- u_\OBS^\nu + 2 \gamma T^\nu) = T^\mu$, as
expected.

Before computing the components of this matrix one of course has to
define those of the reference tetrad vectors. A natural choice can be
to use a normalized version of the standard spherical Schwarzschild
coordinates (i.e., $T^\mu \propto \partial / \partial t$,
$R^\mu \propto \partial / \partial r$ and so on), however such a
choice is only valid outside the black hole, since the then defined
$T^\mu$ would no longer be timelike within the black hole. It is
therefore more appropriate to define the tetrad that can be associated
to a freely falling observer starting from infinity with zero velocity
and zero angular momentum. This is a rather common choice of Zero
Angular Momentum Observers (ZAMO), see, e.g.,~\cite{thorne86}.  Such
set of observers' four-velocity will correspond to the vector
$T^\mu$. Then we define $R^\mu$ as the unique unit spacelike vector
spanned by $\partial / \partial t$ and $\partial / \partial r$ that is
orthogonal to $T^\mu$ and that reduces to $\partial / \partial r$ at
infinity. We keep $\Theta^\mu$ and $\Phi^\mu$ unchanged as compared to
the first ansatz above. Anticipating on the analysis of the maximal
analytic extension of the metric, we use a subscript ${\rm I}$ to the
vectors $T^\mu$ and $R^\mu$ as we will need to perform another choice
for those vectors in some situations. Those vector components are:
\begin{eqnarray}
\label{tetrad_start}
T^\mu_{\rm I} = \left(\begin{array}{c}
  \frac{1}{1 - \frac{2 M}{r}} \\
 - \sqrt{\frac{2 M}{r}} \\ 0 \\ 0 \end{array} \right) \;&,&\quad
R^\mu_{\rm I} = \left(\begin{array}{c}
 - \frac{\sqrt{\frac{2 M}{r}} }{1 - \frac{2 M}{r} } \\ 
 1 \\ 0 \\ 0 \end{array} \right) , \\
\label{ThetaPhi}
\Theta^\mu = \left(\begin{array}{c}
 0 \\ 0 \\ \frac{1}{r} \\ 0 \end{array} \right) \;&,&\quad
\Phi^\mu = \left(\begin{array}{c}
 0 \\ 0 \\ 0 \\ \frac{1}{r \sin \theta} \end{array} \right) .
\label{tetrad_end}
\end{eqnarray}
This tetrad is defined everywhere in the astrophysically relevant part
of the Schwarzschild metric, regardless one is inside or outside the
black hole (except, of course, at $r = 0$).

Regarding the rotation $R^\mu_\nu$, we do not compute its components
explicitly. Rather, we define its three associated Euler angle. We
first rotate $A^\mu$ and $B^\mu$ by an angle $\phi$ around
$C^\mu$. Then, we rotate the newly obtained $A'^\mu$ and $C^\mu$ by
and angle $\vartheta$ around the newly obtained $B'^\mu$, and finally,
we perform a new rotation by an angle $\psi$ around the new
$C'^\mu$. In other words, we process along the following sequence:
\begin{eqnarray}
\left(\begin{array}{c} A'^\mu \\ B'^\mu \end{array} \right)
 & = & 
\left(\begin{array}{cc} \cos \phi & \sin \phi \\
 - \sin \phi & \cos \phi 
      \end{array} \right)
\left(\begin{array}{c} A^\mu \\ B^\mu \end{array} \right) , \\
\left(\begin{array}{c} Z^\mu \\ A''^\mu \end{array} \right)
 & = & 
\left(\begin{array}{cc} \cos \vartheta & \sin \vartheta \\
 - \sin \vartheta & \cos \vartheta 
      \end{array} \right)
\left(\begin{array}{c} C^\mu \\ A'^\mu \end{array} \right) , \\
\left(\begin{array}{c} X^\mu \\ Y^\mu \end{array} \right)
 & = & 
\left(\begin{array}{cc} \cos \psi & \sin \psi \\
 - \sin \psi & \cos \psi 
      \end{array} \right)
\left(\begin{array}{c} A''^\mu \\ B'^\mu \end{array} \right) . 
\end{eqnarray}

\section{Drawing the celestial sphere -- Naive version}
\label{sec_cel1}

As stated above, any photon seen originating from unit spacelike
direction $N^\mu$ orthogonal to the observer velocity $u_\OBS^\mu$ is
endowed with a wave vector proportional to $k^\mu \propto u_\OBS -
N^\mu$. Once such vector is defined (up to some unimportant
proportionality constant), knowing from which direction on the
celestial sphere it originates amounts to propagate it backward in
time (i.e. backward in its affine parameter $p$) the geodesic equation
\begin{equation}
\frac{\ddd k^\mu}{\ddd p} + \Gamma^\mu_{\nu\rho} k^\nu k^\rho = 0 ,
\end{equation}
where the $\Gamma^\mu_{\nu\rho}$ are the usual Christoffel symbols. 

\subsection{First case -- Observer outside the horizon}

If one works in the usual Schwarzschild spherical coordinates, then
this set of equations is written as (see, e.g.,\cite{carroll97}):
\begin{eqnarray}
\label{deb_ode1} 
\frac{\ddd k^t}{\ddd p} & = & - \frac{A'}{A} k^t k^r , \\
\frac{\ddd k^r}{\ddd p}
 & = & - \frac{1}{2} A A' (k^t)^2
       + \frac{1}{2} \frac{A'}{A} (k^r)^2 \\ \nonumber & &
       + A r \left((k^\theta)^2 + \sin^2 \theta (k^\varphi)^2 \right) , \\
\frac{\ddd k^\theta}{\ddd p}
 & = & - \frac{2}{r} k^r k^\varphi 
       + \sin \theta \cos \theta (k^\varphi)^2 , \\
\label{fin_ode1}
\frac{\ddd k^\varphi}{\ddd p}
 & = & - \frac{2}{r} k^r k^\varphi 
       - 2 \frac{\cos \theta}{\sin \theta} k^\theta k^\varphi ,
\end{eqnarray}
where we have set
\begin{equation}
\label{def_A}
A \equiv 1 - \frac{2 M}{r} ,
\end{equation}
and where the prime denotes a derivative with respect to the $r$
coordinate (i.e., $A' = 2 M / r^2$). These equation have of course to
be solved together with the position equation:
\begin{equation}
\label{def_ode2}
\frac{\ddd x^\mu}{\ddd p} = k^\mu .
\end{equation}
In practice, this set of equations is solved using an adaptative
4\TTH order Runge-Kutta method inspired from the Numerical
Recipes~\cite{numrec}. One step that has to be implemented with some
care is the choice of the time step. As we will see later, two
numerically tricky zones lie at $r = 2 M$ (horizon crossing) and
$r = 3 M$ (light circle crossing, where radial motion can be slow and
unstable whereas orthoradial motion must be carefully computed; see
Appendix~\ref{app_geod}). Therefore, the timestep choice is
essentially given by the following choices:
\begin{itemize}
\item If the geodesic is recessing away from the black hole (when
  propagated backward in time) with $r$ already larger than $4 M$, the
  timestep is chosen proportional to $r^2$, so that infinity (or, in
  practice, a very large value of $r / M$) is reached after a few steps
  (typically 4 or 5);

\item If the geodesic approaches the black hole, then one chooses a
  timestep proportional to $r - 4 M$;

\item If one lies within the $r = 4 M$ sphere, then a sufficiently
  small timestep is chosen so as to insure both stability of the
  integration and monitoring a possible horizon crossing (see later).
\end{itemize}

If one is outside the black hole and if we do not consider the maximal
analytic extension of the metric, then this equation has to be solved
only when the geodesics originates from infinity. By setting the
constants of motion $E$ and $L^2$ by their standard definition, i.e.,
\begin{eqnarray}
\label{def_E}
E & \equiv & \pi_t = g_{t \mu} k^\mu = A k^t , \\
\label{def_L2}
L^2 & \equiv & r^4 \left((k^\theta)^2 + \sin^2 \theta (k^\varphi)^2 \right) ,
\end{eqnarray}
then the geodesic equation needs to be computed if and only if 
\begin{enumerate}

\item $L^2 / E^2 \geq 27 M^2$ and $r > 3 M$, or

\item $L^2 / E^2 \leq 27 M^2$ and $k^r < 0$.

\end{enumerate}
(Since this is a fairly well-known result we just recall it here, but
for the sake of completeness derive it in Appendix~\ref{app_geod}.)

If none of these conditions are satisfied, this means that the
geodesic originates from the past event horizon, or, from an
astrophysically realistic point of view, from the infinitely
redshifted surface of the collapsing object which gave birth to the
black hole as it was passing through the horizon. If one does not
works within the maximal analytic extension of the metric, then such
geodesics do not carry any photon and the corresponding pixel is
black. If, on the contrary, one works in the maximal analytic
extension and want to compute from which part of the singularity a
given null geodesic originates from, then the geodesic can be
propagated back to the past singularity and imaged provided one
decides of some emission properties of the past singularity
(see~\ref{sec_Kruskal}).

For geodesics originating from infinity, one obtains at the end of
integration a wavevector $k_\infty^\mu$ whose only non negligible
components are $k_\infty^t$ and $k_\infty^r$, the two others tending
to 0 when $r$ tends to infinity because of angular momentum
conservation. Regarding the position, $r$ and $t$ both tend to minus
infinity with their difference $r - t$ being almost constant, and
$\theta$ and $\varphi$ tend to be constant. This is because as long as
the radial coordinate $r$ is much larger than the impact parameter $b
\equiv L / E$, the geodesic can be considered as (almost) purely
radial and originating from the direction defined by the above
mentioned $\theta$ and $\varphi$.

In addition, one can compute the redshift $z$ of the photons we
receive. This is done through the standard formula
\begin{equation}
1 + z = \frac{k^t_\infty}{k^\mu u_{\OBS\;\mu}}
      = \frac{E}{k^\mu u_{\OBS\;\mu}} , 
\end{equation}
where the numerator is evaluated (as the subscript indicates) at
infinity, whereas the denominator is computed at observer's position
before integration. Note that this formula include both the kinetic
and gravitational redshift.

Once the initial direction of the photon and the redshift are known,
we can draw the corresponding pixel.

\subsection{Second case -- Observer inside the horizon and/or within
  the maximal analytic extension}
\label{ssec_max}

The set~(\ref{deb_ode1}--\ref{fin_ode1}) is valid only when geodesics
do not cross the horizon. Therefore, if the observer is within the
horizon, it is not possible to simulate what he/she sees of the
celestial sphere using these equations. In order to do so, one has to
use another system of coordinates, the most natural of which being
that of Kruskal~\cite{kruskal60}. In this case, the subset of
coordinates $(t, r)$ have to be replaced by the subset $(U, V)$
defined by the following procedure. First, we define the so-called
``tortoise'' coordinate $r^*$ by
\begin{equation}
\label{def_rstar}
r^* = r + 2 M \ln \left|\frac{r}{2 M} - 1 \right|,
\end{equation}
so that 
\begin{equation}
\ddd r^* = \frac{\ddd r}{1 - \frac{2 M}{r}}.
\end{equation}
Here, $r^*$ is a growing function of $r$ outside the horizon and a
decreasing function of $r$ inside (regardless one considers the
maximal analytic extension or not).  Then, we define the null outgoing
and ingoing coordinates $u$ and $v$ such as
\begin{eqnarray}
u & \equiv & t + r^* , \\
v & \equiv & t - r^* ,
\end{eqnarray}
and finally $U$ and $V$ are defined through
\begin{eqnarray}
\label{defU}
U & \equiv & \epsilon \exp \left(+ \frac{u}{4 M} \right) , \\
\label{defV} 
V & \equiv & \eta \exp \left(- \frac{v}{4 M} \right) .
\end{eqnarray}
The constant $\epsilon, \eta = \pm 1$ are then chosen so that both $U$
and $V$ are future-oriented. Anticipating on what we will do in
\S\ref{sec_hor}, \ref{sec_Kruskal}, we will need to know the values of
$\epsilon$ and $\eta$ in all the regions of the maximal analytic
extension of the metric. Since this is rarely done in the literature
(see, e.g., ~\cite{carroll97,hawking73,wald84}), their value are
summarized in Table~\ref{table_eps_eta}, using the following labels
for the regions: we call I our asymptotic region, II the black hole
interior beyond the future event horizon, III the other asymptotic
region and and IV the region beyond the past event horizon. (In
Ref.~\cite{hawking73}, our regions III and IV are called I' and II',
respectively, and IV and III in Ref.~\cite{wald84}).
\begin{table}
\begin{tabular}{|c||p{4cm}|c|c|}
  \hline
  Region & Remark & $\epsilon$ & $\eta$ \\
  \hline
  \hline
  I & $t$ is future-oriented & $1$ & $-1$ \\
  II &  $r$ is past-oriented & $1$ & $1$ \\
  III & $t$ is past-oriented & $-1$ & $1$ \\
  IV & $r$ is future-oriented & $-1$ & $-1$ \\
  \hline
\end{tabular}
\caption{Values of the parameters $\epsilon$ and $\eta$ as defined in 
Eqns~(\ref{defU}--\ref{defV}). Their value allow the null coordinates 
$U$ and $V$ to with proper time regardless the observer position within 
the maximal extension of the Schwarzschild metric. Which of the
variable $t$ or $r$ of the standard Schwarzschild coordinates that is
bound to grow or decrease with proper time from any timelike geodesics
is also given.}
\label{table_eps_eta}
\end{table}
This choice for $\epsilon$ and $\eta$ also possesses the nice property
to ensure that the product $U V$ can be expressed as a function of $r$
only, without any reference to $\epsilon$ or $\eta$~\footnote{This is
  because the product $\epsilon \eta$ has the same sign as $2 M - r$
  so that it compensate to absolute value that arises in the
  definition of $r^*$ in Eq.~(\ref{def_rstar}).}:
\begin{equation}
\label{res_UV}
U V = - \exp\left(\frac{r}{2 M} \right) \frac{r - 2 M}{2 M} .
\end{equation}
This being set, the set of geodesic
equations~(\ref{deb_ode1}--\ref{fin_ode1}) has then to be replaced by
the following set of equations
\begin{eqnarray}
\ddot U & = &
- \frac{\partial_U F}{F} \dot U^2
- \frac{r}{F} \partial_U r \; 
  \left(\dot \theta^2 + \sin^2 \theta \dot \varphi^2 \right) , \\
\ddot V & = &
- \frac{\partial_V F}{F} \dot V^2
- \frac{r}{F} \partial_V r \; 
  \left(\dot \theta^2 + \sin^2 \theta \dot \varphi^2 \right) , \\
\ddot \theta & = & 
  \cos \theta \sin \theta \dot \varphi^2
- \frac{2 \dot \theta}{r}(\dot U \partial_U r + \dot V \partial_V r) , \\
\ddot \varphi & = & 
- 2 \frac{\cos \theta}{\sin \theta} \dot \theta \dot \varphi
- \frac{2 \dot \varphi}{r}(\dot U \partial_U r + \dot V \partial_V r) , \\
\end{eqnarray}
where we have set
\begin{equation}
F \equiv g_{UV} = - \frac{8 M^2 A}{U V} ,
\end{equation}
$A$ being defined in Eq.~(\ref{def_A}) and where we have used the
following intermediate quantities
\begin{eqnarray}
\partial_U r & = & \frac{2 M A}{U} , \\
\partial_V r & = & \frac{2 M A}{V} , \\
\partial_U F & = & \frac{2 M A F'}{U} , \\
\partial_U F & = & \frac{2 M A F'}{V} ,
\end{eqnarray}
in which $F'$ denotes the derivative of $F$ with respect to $r$, which
is quite straightforwardly deduced from Eq.~(\ref{res_UV}) and can be
written as
\begin{equation}
F' = \left(- \frac{1}{2 M} + A'(r) \right) \frac{F}{A} .
\end{equation}
Since $F'$ only appears when multiplied by $A$ the set of equation is
regular at horizon crossing. 

The set of variables $U, V$ is well suited for horizon crossing and is
in principle defined everywhere on the maximal analytic extension of
the manifold. In practice it is however not possible to use it
everywhere since the exponential dependence of both $U$ and $V$ in
term of $r$ and $t$ make it numerically impossible to use as soon as
one goes several Schwarzschild radii away from the black
hole. Therefore, we adopt the following procedure in order to choose
the coordinate system:
\begin{itemize}
\item From the knowledge of constants $E$ and $L^2$ and position, we
  determine whether the geodesic we are interested in has any chance
  to cross horizon;

\item If not, then we use the Schwarzschild coordinates $r, t$;

\item If so, then we check whether one is close to the horizon;

\item If not, then we keep the Schwarzschild coordinates;

\item If so, then we switch to Kruskal coordinates;

\item Then we keep on using Kruskal coordinates till horizon has been
  crossed, in which case we switch back to Schwarzschild coordinates.

\item In the specific case on is interested in an observer in region
  II crossing null geodesics coming from region IV, we always keep the
  Kruskal coordinates.

\end{itemize}

The only arbitrariness here lies in where exactly we decide to switch
from one coordinate system to the other. In practice, the radial
motion of photon is determined by and effective potential $V(r)$ given
in Eq.~(\ref{def_pot}). This potential show that in some case, a
geodesic may spend a large amount of time around $r = 3 M$ (the
so-called light circle or photon sphere) around which the one
dimensional radial motion is unstable. Therefore, we choose to impose
that the light circle crossing is made using Schwarzschild coordinates
$r, t$.

\subsection{From geodesic equation solution to RGB values}
\label{ssec_rgb}

Whichever method we used to solve the geodesic equation, we can
compute, for any pixel of our screen, from which the corresponding
null geodesic originates from. If it originates from the past
horizon, the corresponding pixel is black (unless the special case of
Sec.~\ref{sec_Kruskal}), or it corresponds to a direction of the
celestial sphere. Assuming that we have a map of this celestial
sphere, i.e. from a full sky survey of the sky (see \S\ref{sec_data}),
we can determine which pixel of the celestial sphere the direction we
found corresponds to. However, this is not the end of the story.

In a perfectly realistic situation, we should have at our disposal a
spectral map of the celestial sphere, i.e., spectroscopic data for
each direction of it. Then we would modify the spectrum according to
the computed redshift and then compute the eye response to that
observed spectrum and then deduce the corresponding RGB coordinates of
the pixel. But in practice, we are limited by the fact that we do not
have such spectral information. This therefore gives the correct
colors of the sky seen by the observer only when redshift is
negligible, and there is no simple way to compute the color or
intensity change of the celestial sphere due to the redshift,
therefore we shall compensate this by various visual artifacts. For
example, it is possible to define by hand the spectral information for
each direction of the celestial sphere that matches the pixel color
(black body plus emission lines, for example). However, we found that
if we deal properly with the stars (see \S\ref{sec_stars}), the
rendering of the rest of the celestial sphere was not of crucial
importance as the overall rendering was, by far, dominated by the
stars. Therefore, we chose to let the pixel hue unchanged and simply
shift the intensity of the corresponding pixel of the celestial sphere
by a factor that is a monotonous function of $(1 + z)^{- 1}$. We found
that a satisfactory function was a power law of $(1 + z)^{- 1}$ for
negative $z$ (i.e., blueshift) and an exponential of $- (1 + z)$ for
positive $z$ (i.e., redshift).

\section{Drawing the celestial sphere -- Sophisticated version}
\label{sec_cel2}

In the Section above, the number of geodesic equations we have to
solve is equal to the number of pixels of the screen and can therefore
easily reach several millions and severely affect the computational
time for each picture. The situation is worsened by the fact that some
smoothing may be necessary when computing the image, i.e., one may
need to split one pixel into several subpixels, compute the color of
all of them and average the result accordingly, see
Ref.~\cite{thorne15}.

Such a large number of integrations of the geodesic equation is
actually not necessary because, as is well known, is a spherically
symmetric metric, null geodesics are described by a single parameter,
which can be taken to be the impact parameter $b = L / E$. Moreover,
we are considering a situation where the only information about the
geodesics we are interested in is the amount of deflection the photon
trajectory experiences between the direction it travels at the
observer's position and the direction it was traveling at
infinity. Therefore, we adopt the following procedure:

\begin{itemize}

\item Before computing the image, we solve the geodesics equation for
  a freely falling observer (or the one that defines our reference
  tetrad, with four-velocity $T^\mu$) for all possible angles $\delta$
  between the geodesic and the radial direction $R^\mu$. This amounts
  to sample, in some appropriate way, all the values of the geodesic
  impact parameter.

\item For each of these geodesics we obtain the angle $\delta - \chi$
  by which the geodesics has been deflected before reaching the
  observer.

\item We sample this quantity sufficiently by computing a moderately
  large number of geodesics so that the both functions $\chi(\delta)$
  and its inverse $\delta(\chi)$ are well sampled. 

\item Then, for each pixel of the image, compute the angle $\delta$
  between the corresponding null geodesic and the radial direction,
  and identify the plane into which the photon trajectory lies.

\item We rotate within this plane the observer position by an angle
  $\chi + \pi$ so as to identify the point of the celestial sphere the
  photon originates from.

\item We compute the RGB coordinates of the pixel one has to draw
  following the selected assumptions for redshift rendering, see
  \S\ref{ssec_rgb}.

\end{itemize}

\begin{figure}[htbp]
\begin{center}
\includegraphics*[angle=270,width=3.2in]{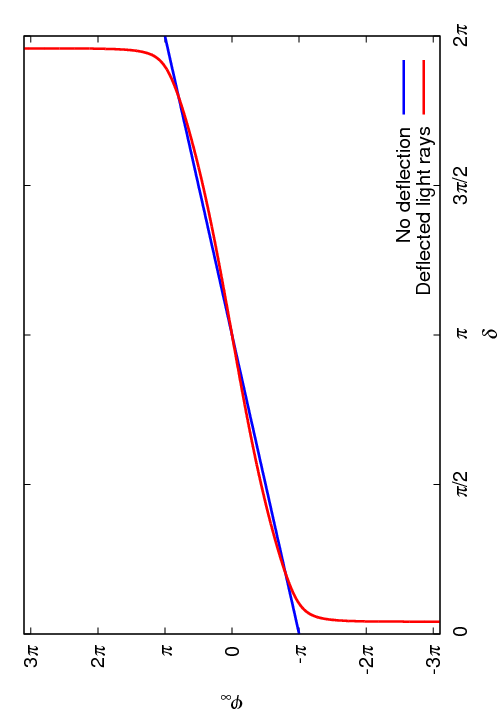}
\caption{An example of the deviation function
  $\varphi_\infty(\delta)$. The one shown here was computed for a
  freely falling observer on a Schwarzschild black hole at coordinate
  distance $r = 30 M$.  In the absence of any relativistic effect, the
  function $\varphi_\infty(\delta)$ would trivially reduce to
  $\varphi_\infty = \delta - \pi$, shown in dashed lines. But the
  adjunction of aberration (since the observer is freely falling) and
  light deflection modify it, only slightly away from the black hole
  ($\delta$ close to $\pi$), or much more importantly toward the black
  hole ($\delta$ close to $0$ or $2 \pi$). The black hole angular
  radius is given by the first value of $\delta$ for which
  $\varphi_\infty$ diverges. Here, the function was sampled so as to
  include values of $|\varphi_\infty|$ slightly larger than $5 \pi$,
  which is sufficient to include any visible ghost images of stars in
  most situations.}
\label{fig_dev}
\end{center}
\end{figure}

The angle $\delta$ we use here is defined with respect to our
reference tetrad~(\ref{tetrad_start}--\ref{tetrad_end}). This means
that it is given by the formula
\begin{equation}
\cos \delta = \frac{- k^\mu R_\mu}{k^\nu T_\nu} ,
\end{equation}
or, equivalently, this means that we will integrate a bunch of
geodesics whose wavevector $k^\mu$ is defined as
\begin{equation}
\label{def_k_prec}
k^\mu \propto 
   T^\mu
 + \cos \delta R^\mu
 + \sin \delta (\cos \varpi \Phi^\mu + \sin \varpi \Theta^\mu) .
\end{equation}
We can of course choose, without loss of generality, $\varpi = 0$ and
have only yo sample $\delta$.  With this definition, a radial geodesic
originating from the black hole past horizon is described by an angle
$\delta = 0$, and a radial geodesic originating from past null
infinity is described by $\delta = \pi$.  Because of spherical
symmetry, when we compute the $\chi(\delta)$ function, we can place
ourselves in the equatorial plane at the same $r$ coordinate as the
observer and by considering equatorial geodesics only that start from
the $x$ axis (i.e., our freely falling fiducial observer lies at
coordinates $(t, r, \pi / 2, 0)$ and $\varpi = 0$ in
Eq.~(\ref{def_k_prec})). From this position and this wavevector, we
propagate the geodesic backward in time till it reaches very large
values of $r$ (typically larger than $10^{10} M$, but the exact value
does not matter). We then obtain the azimuthal angle $\varphi_\infty$
with respect to the observer of the corresponding origin of the
geodesic (for example, $\varphi_\infty = 0$ for $\delta = \pi$, and
the angle $\chi$ is then given by
\begin{equation}
\chi = \varphi_\infty + \pi - \delta .
\end{equation}
Should there be no aberration nor deflection of light, then one would
just have $\chi = 0$, i.e., $\varphi_\infty = \delta - \pi$, regardless
of the observer's position and velocity\footnote{We implicitly neglect
  any parallax effect here.}.  An example of deviation function
$\chi(\delta)$ (or, in fact, $\varphi_\infty (\delta)$) is given in
Figure~\ref{fig_dev}.
The main characteristic of this function is that it diverges for the
extreme values of $\delta$ for which the function is defined, a fact
that is recalled in Appendix~\ref{app_geod} and that is usually
called the zoom-and-whirl effect~\cite{giampedakis02}.

Note that the two angles $\delta$ and $\chi$ are associated to
somewhat different contexts: $\delta$ is an angle defined by a freely
falling observer in the Schwarzschild metric, whereas $\chi$
corresponds to an angle measured by a static observer in Minkowski
space whose origin and orientation matches that of the Schwarzschild
metric.

In practice, sampling this $\chi(\delta)$ function necessitates a few
thousands of geodesics to be computed, i.e. a factor between $10^2$
and $10^3$ less than the brute force computation of the previous
section assuming million pixel sized images. This situation is also
enhanced by the fact that it suffices to perform the computation in
the equatorial plane, so that, in practice, there is no need to
consider the $\theta$ and $k^\theta$ variable in the ODE
system~(\ref{deb_ode1}--\ref{fin_ode1},\ref{def_ode2}).

Then, once this deflection function is computed, all what we need to
do rely on very simple trigonometric operations for all each pixel of
the screen and some search in the deflection function $\chi (\delta)$
pre-computed array. More importantly, the deviation function will
further be used in the next section (through its inverse) in order to
perform a very clean rendering of the stars.

\section{Drawing the stars}
\label{sec_stars}

The main drawback of the sky rendering above is that it is not well
suited to include point sources. A point source such as a star will
never appear as perfectly pointlike. Because of diffraction, a point
source will always appears as possessing a small but non zero angular
size typically of circular shape (with possibly diffraction patterns)
and can, in principle, belong to the pixellized version of the
celestial sphere. However, when distorting the image of such a source,
its appearance will also be distorted. As is well known, amplification
due to gravitational lensing goes with a large amount of shear, so
that an initially circular pattern will become quite elongated. This
is not the way this source should appear because given the actual
smallness of a star angular size, even in a strong lensing regime it
should be considered a pointlike and its visual finite angular size
would only result in the optical distortions caused by the observation
apparatus. Consequently, it is not possible to consider stars as
``points'' that would be ``impainted'' on the celestial sphere, and
stars (or any almost pointlike sources) should be processed using a
different procedure.

\subsection{Direct ray tracing from interpolation}

Since we work in a Schwarzschild metric, any direction on the
celestial sphere will possess an infinity of images seen from any
observer point of view, even though most of these images will be
extremely faint and close to the edge of the black hole silhouette
(see, e.g., \cite{chandrasekhar83}). Since the metric is spherically
symmetric, any geodesic is planar and all the geodesics starting from
the same point of the celestial sphere and reaching the observer
belong to the same plane.

If we have some star catalog, what we know about a given star is its
position $\boldsymbol{n}^*$ on the celestial sphere. We also know the
observer position within the metric, and consequently we know the
plane spanned by the observer, the black hole and the star, as well as
the angle between the observer radial position and and star position,
$\varphi^*_\infty$. We can choose the orientation of the
star-observer-black hole plane so that $\varphi_\infty^*$ lies between
$0$ (observer is between the star and the black hole) and $\pi$ (black
hole is between the observer and the star). We can now invert the
procedure outlined in the previous section, with the extra
complication that for a given $\delta_0$ such that
$\varphi_\infty(\delta_0) = \varphi_\infty^*$, there exists other
values of $\delta$ that we shall note $\delta_k$ such that
\begin{equation}
\varphi_\infty(\delta_k) = \varphi_\infty^* + 2 k \pi . 
\end{equation} 
Those are the ghost images of the star. There exists an infinity of
such images as long as the function $\varphi_\infty(\delta)$ diverges
(see Fig.~\ref{fig_dev}), which is the case for the Schwarzschild
metric. These ghost images are not difficult to find numerically as
long as the function $\varphi_\infty(\delta)$ is accurately sampled. We
therefore can find without difficulty the apparent position of any
ghost image of any star.

\subsection{Amplification}
\label{ss_ampl}

This being done, we know the direction under which we see a given
image of the star. However, because of both aberration and lensing,
the actual (microscopic) angular area of the star will be different,
and hence its luminosity. In order to take account for this, we shall
define two directions very close to that where the star (or, in fact,
one of its image) is seen.  Since we work here in the observer's
frames, the directions we are talking about can be expanded in term of
the spacelike vectors $X^\mu$, $Y^\mu$, $Z^\mu$ and can be considered
as Euclidean three-vectors which we shall write in bold notations and
the direction into which the (image of the) star is seen will be
written $\boldsymbol{n}$. In order to define two directions
infinitesimally close to $\boldsymbol{n}$, we choose one direction
$\boldsymbol{n}'$ that is different from $\boldsymbol{n}$. This can be
for example either $\boldsymbol{x}$ or $\boldsymbol{z}$ (one may
choose the one among these two which has the smallest dot product with
$\boldsymbol{n}$). Then, we compute
\begin{equation}
\boldsymbol{n}^\perp_1 = \frac{\boldsymbol{n} \wedge \boldsymbol{n}'}
                              {|\boldsymbol{n} \wedge \boldsymbol{n}'|},
\end{equation}
and then
\begin{equation}
\boldsymbol{n}^\perp_2 = \boldsymbol{n} \wedge \boldsymbol{n}_1^\perp.
\end{equation}
We then choose two small quantities $\delta_1$ and $\delta_2$ and define
\begin{eqnarray}
\boldsymbol{n}_1 & = &   \boldsymbol{n} + \boldsymbol{\delta n}_1^\perp
                   =     \boldsymbol{n} + \delta_1 \boldsymbol{n}_1^\perp , \\
\boldsymbol{n}_2 & = &   \boldsymbol{n} + \boldsymbol{\delta n}_2^\perp
                   =     \boldsymbol{n} + \delta_2 \boldsymbol{n}_2^\perp .
\end{eqnarray}
Up to $O(\delta_{1,2}^2)$ terms, these two vectors are units vectors
that are very close to $\boldsymbol{n}$. Now, it is obvious that the
solid angle $\Omega$ spanned by three directions $\boldsymbol{n}$,
$\boldsymbol{n}_1 = \boldsymbol{n} + \boldsymbol{\delta n}_1^\perp$
and $\boldsymbol{n}_2 = \boldsymbol{n} + \boldsymbol{\delta
  n}_2^\perp$ is given by the formula
\begin{equation}
  \Omega = \boldsymbol{n} \cdot
           (\boldsymbol{\delta n}_1^\perp \wedge \boldsymbol{\delta n}_2^\perp)
 = \boldsymbol{n} \cdot (\boldsymbol{n}_1 \wedge \boldsymbol{n}_2 ).
\end{equation}
Equivalently, if we propagate the null geodesics which originate from
these three direction till the celestial sphere, we obtain three
direction on the celestial sphere, $\boldsymbol{n}^*$,
$\boldsymbol{n}^*_1$ and $\boldsymbol{n}_2^*$ which span a solid angle
$\Omega^*$ given by
\begin{equation}
  \Omega^*
 = \boldsymbol{n}^* \cdot (\boldsymbol{n}_1^* \wedge \boldsymbol{n}_2^* ).
\end{equation}
With these notations, the amplification or de-amplification factor $f$
induced both by lensing and aberration is simply written as
\begin{equation}
\label{eq_ampl}
f = \frac{\Omega}{\Omega^*} .
\end{equation}
(This is, of course, nothing more than solving in this particular
context the relevant part of the optical scalar equation, i.e.,
convergence, see, e.g., Ref.~\cite{seitz94}.)  Computing this factor
numerically is in fact not necessary. There exists an analytical
formula for amplification due to aberration, and the deflection
function $\chi(\delta)$ (or, in fact, the derivative of its inverse)
allows after some algebra to address the lensing part. However in
practice this does not allow to perform significant enhancement in
term of CPU time so that we won't address this here but rather keep it
for future work.

\subsection{Effective drawing of the star}
\label{sec_stars_eff}

In the two last subsections, we explained how to determine the
position of (the image of) a star on the observer's screen, in
addition to its redshift and the amplification of its light because of
aberration and lensing. For simplicity, we shall assume here that
stars have a black-body type emission whose temperature is given by
their spectral type (O and A being hotter, K and M cooler).

If one wants to compute the colour perceived by human eye a given
light source (in term of, say, its RGB coordinate of a computer
screen), one has to know the eye perception for each visible
monochromatic frequency. These data are called the spectral
tristimulus values are have been tabulated since a long time by the
dedicated authority, the International Commission on Illumination
(CIE)~\cite{cie}. For simplicity, we assume that eye response do not
depend on light intensity, that is, we somewhat questionably assume
that the observer's vision always works in diurnal (photopic) mode
rather than in low brightness (scotopic) environment. This allows more
colorful hues for stars than what we are used to. In any case, if one
starts from a light source with a given spectrum, one can compute the
RGB values of this source by using the trichromatic tables delivered
by the CIE.

In practice, with the assumptions we make about the star spectra, the
procedure is the following:
\begin{enumerate}

\item Prior to launching the code, we have processed our star catalog
  in order to transform each star's spectral type and magnitude in the
  V-band into a bolometric magnitude $m_*$ and a surface temperature
  $T_*$.

\item For each image of each star, we compute the corresponding
  redshift $z$ and amplification factor $f$. 

\item We deduce that the star image possesses an apparent (in the
  sense of observer-dependent) bolometric magnitude and temperature
  given by
\begin{eqnarray}
m_*^\OBS & = & m_* + 4 \log(1 + z) - \log f , \\
T_*^\OBS & = & \frac{T_*}{1 + z}
\end{eqnarray}

\item With these new values, we compute the RGB values of this
  image of the star.

\end{enumerate}

This being done, we have to decide how to simulate a pointlike light
source with these RGB values. Since the source is supposed to the
pointlike, the most natural choice is to add to the pixel where the
star image is seen the RGB values of the star to that of the
already computed background. However, this naive assumption quickly
leads to difficulties. The reason is that there is nothing that
guarantees that the RGB values are smaller than 1, i.e., that
adding the star to this pixel will not saturate it. If they are not,
putting all the luminosity of the star into a single pixel will
truncate its true luminosity to the maximum that a pixel can draw and
many bright star will have, in practice, identical magnitude because
of the limitations of a computer screen. Therefore, in order to allow
recovering the whole luminosity range of observable stars, we draw
them as extended blobs which are several pixel wide. We have found
that a pleasant rendering is obtained if our blob intensity profile
looks like a truncated Gaussians both in R, G, and B colors. This
means that we impaint the already computed distorted celestial sphere
by those blobs, imposing that the Gaussians are centered on the actual
position of the star (not necessarily at the center of the pixel they
belong to) and by choosing their size so that their integrated
flux corresponds to the one that has been computed. In other
words, the brighter the star, the larger the blob that represents
it. This procedure is inspired by the beautiful pictures made by
famous amateur astronomer Akira Fuji~(see, e.g.,~\cite{fuji}) which
are obtain by putting a diffusing filter in front of the camera so as
to artificially spread a bright star images on some extended area of
the picture in order to more faithfully reproduce the luminosity
contrast between faint and bright stars.

In order to reproduce the most satisfactory rendering of the stars,
some cooking is necessary here. For example, its appears less
artificial to saturate the centre of the blobs that represent bright
stars, so that its color is less saturated than the edge of the blob
which reproduces more faithfully the star color. Also, it appears that
when performing videos, a faint stars slowly crossing the screen
appears more aesthetic if one imposes that its blob is always at least
a few pixel wide in diameter even if the center of the blob is then
not saturated.

Even if there are obviously some ``artistic'' choices that are made
here (apologizing in advance that not every reader will agree with
them!), we insist on the fact that the really physically significant
quantities that are needed, $m_*^\OBS$ and $T_*^\OBS$ are computed as
accurately as possible, so that we have very carefully split the
problem into its physical content ($m_*^\OBS$ and $T_*^\OBS$) and its
representational content (how to associate a colored blob to these two
quantities).

\section{Data that are actually used to produce images}
\label{sec_data}

\subsection{Background sky}
\label{ssec_bckgsky}

For pedagogical purposes, a celestial sphere made of a coordinate grid
is by far sufficient. In what follows we have defined a celestial
sphere under the form of a checkerboard structure of $5$~degrees both
in celestial latitude and longitude. The two type squares (``light'')
and (``dark'') all correspond to black-body type emission but with same
temperature ($8000\;{\rm K}$) but different intensity (a factor $4$
between light and dark squares). In order to avoid thinner and almost
triangular squares toward the pole, the polar regions are covered by
discs of $5$~degrees in diameter which are both redder
($T = 2000\;{\rm K}$) and brighter.

For aesthetic rendering and/or astronomical outreach use, it is by far
better to use a celestial sphere that looks like a real one. The
simple way to obtain this is by using actual pictures of the sky seen
from Earth. Many such high resolution pictures of this exist. One of
the most famous is the one made by A.~Mellinger some time
ago~\cite{mellinger09}, at the same epoch as the one made by
S.~Brunier~\cite{brunier09}. Unsurprisingly, the latter was much
advertised in French speaking countries, whereas the former was mostly
known in the rest of the world. Independently however of the
astonishing quality of these two pictures, both suffer from the fact
that stars are part of the pictures in the sense that their appear to
be impainted on the celestial sphere. This is not satisfactory for the
reasons given in Sec.~\ref{sec_stars}.

For this reason, it is better to look for a full sky picture of the
celestial sphere whose individual bright stars have been removed. This
could presumably be done with some dedicated software such as
SExtractor~\cite{bertin96}, however this could also be very simply
achieved in a rather clean way by the 2MASS collaboration which
produced a starless picture of the celestial sphere~\cite{2mass}. It
seems that the pictures is not actually ``starless'', but that the
luminosity of each pixel was computed by averaging the luminosity of
the stars over some windows function.  The picture obtained in this
way is of moderate size ($2400 \times 4700$ pixels) corresponding
(once borders are removed) to a resolution of around $4.8'$ per
pixel. This is a factor $\sim 5$ coarser that the natural resolution
of human eye, however, for practical purposes, what is of interest is
the ratio between the digital resolution of the produced images and
that of the images that are used for rendering. If one considers
normal (for modern screen standards) pictures of around $1280$ pixel
wide corresponding to a view computed with an opening angle of
$90$~degrees, then the theoretical resolution of such a picture is
around~$4.2'$ which is only marginally better than the 2MASS picture,
which is therefore is sufficient for many purposes.

This picture suffers however from small but noticeable problems. The
most obvious one is that the lower border is missing on a width of a
few pixels, so that one has to complete the missing part by some
(rather arbitrary) cosmetic procedure. The second one is that the
2MASS project has observed the sky in the infrared bands and the
structure of the Milky way is somewhat affected by this. The most
noticeable difference comes from the fact that there is far less
absorption, especially in the direction of the Galactic centre which
appears much more regular and more symmetric with respect to the
Galactic plane than in optical images. Also, since this picture is
unavoidably made in false colors, the overall hue does not correspond
to visible image, notable the Milky way band which is both brighter a
has much more yellowish hue than what human eye is used to. Finally,
since the image is already given in term of a planar projection and
its pixellization follows this projection\footnote{Regarding this, it
  has to be noted that the 2MASS website claims that it is an Aitoff
  projection but following the nomenclature of Ref.~\cite{proj}, it
  rather seems to be an Aitoff-Hammer projection.}. When viewed in a
spherical context, then the underlying pixellization of the image
appears more or less elongated depending on the direction of
observation. This is particularly true at the joining of the two
edges of the pictures some quite visible and unaesthetic patterns are
visible on the bare image. However, again, when adding a star catalog,
those issues are barely noticeable.

\subsection{Stars}

Regarding the stars, visual inspection of simulated images shows that
the more stars, the more spectacular the result is. We therefore need
a large, uniform and magnitude limited catalog comprising at least
several $10^4$ stars.

We have chosen to use the electronic version of the Henry Draper
catalogue with its extension~\cite{henry_draper}. This catalog
comprises more than 250,000 stars.  For high magnitudes, it is not
uniform, some regions of the sky having a deeper coverage (the
Galactic anticenter, among others). We therefore truncate up to some
magnitude around 9. After this, some handmade changes have to be
performed because the magnitude of some binary star systems are not
given, something which happens for some easily noticeable stars such
as $\beta$~Lyrae.

For very high resolution images, we also have used the ASCC catalogue
of around 2.5~million stars~\cite{kharchenko01}.

\subsection{CPU issues}

With the described implementation, the typical computational time for
a single high resolution image is usually split in equal proportion
between the few thousand geodesic calculation sampling the deflection
function $\chi(\delta)$, and drawing the stars pixel by pixel along
the lines described above. Typical images of one million pixel
initially computed at a twice larger resolution and then smoothed
needed in the first versions of our code needed one minute of single
CPU time to be computed. Assuming one makes a movies at 25 of 30
frames per second, one second of the movie can be computed in less
than half an hour, thus allowing to obtain a one minute long movie in
less than one CPU day.

In some situations however, most notably when large portions of the
sky experience a large blueshift (for example, a freely falling
observer within the horizon and approaching the singularity or static
observer close to the horizon), the CPU time is dominated by the star
drawing, something which is far from being optimized, thus reducing the
length of the movie that could be made in one CPU day. However, we did
not meet any critical CPU issues here. Moreover, since any frame
can be computed independently of the others, we did not have any need of
parallelization. It was in practice simpler to launch by hand our code
to compute a few hundreds of frame per computer, and then to
split the not so numerous hard-to-compute frames after they had been
identified. 

\subsection{An example}

We have computed several images in various context, so as to highlight
this or that special or general relativistic effect. They are not
essential to the discussion here, but since they at least are of
obvious pedagogical interest, we give several examples in
Appendix~\ref{app_img}. We shall here in Fig.~\ref{APOD} give only one
example of a high resolution picture including a realistic background
celestial sphere and a deep star catalog.
\begin{figure}[htbp]
\begin{center}
\includegraphics*[width=3.2in]{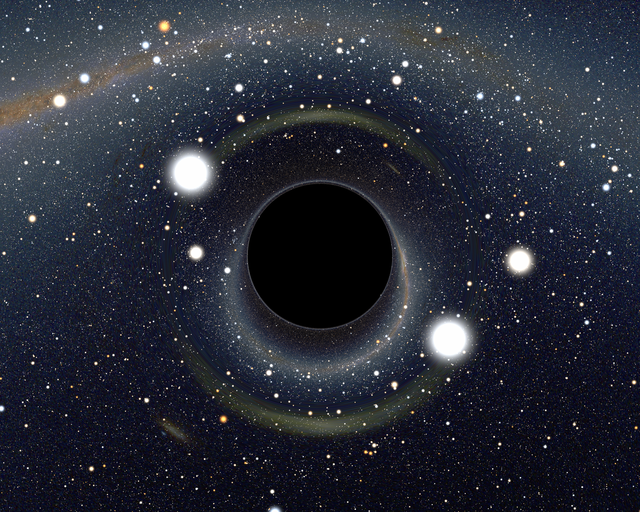}
\caption{A example of an image computed using the methods outlined
  above, using the false color 2MASS full sky starless picture as well
  as the Henry Draper catalogue of stars. The black hole is put in
  front of the Large Magellanic Cloud, and specifically aligned with a
  7.5~mag star (HD~49359) which is strongly lensed and appears as a
  two stars at equal distance but opposite position with respect to
  the black hole.}
\label{APOD}
\end{center}

\end{figure}

\section{Adapting simulations for planetariums}
\label{sec_plane}

In addition to their obvious mathematical/physical interest, an
obvious use of astronomical simulations is for popular science shows,
especially for planetarium since those are since more than a decade
built upon a fully digital projection system. Since special and
general relativistic effects are more spectacular close to the black
hole horizon, which is therefore sustained by a large angular
diameter, hemispherical projection are most naturally adapted to
full-dome projection. For the purpose, we need to provide still frame
or movies using the image format that is widely used in this field,
the Domemaster format. Those are high definition (typically
4k$\times$4k, or even 8k$\times$8k square images whose largest
incircle correspond to a half sphere. pixel distance with respect to
the center of the image is proportional to colatitude and angle
between a given ray and a downward half-line starting from the center
of the square represent latitude. The pixel that lies at the middle of
the inferior side of the square is in front of the audience and the
pixel in the middle of the upper side of the square in behind
it. Although the audience can look at any point of the planetarium
screen, the most comfortable part to look at extends till around
60~degrees away from point of $60^\circ$ colatitude and $0^\circ$
longitude.

Regarding the rendering of the celestial sphere explained
in\S\ref{sec_cel2}, the procedure here is exactly the same, except
that Eqns~(\ref{pix_i}, \ref{pix_j}) have to be changed according to
the above description of the Domemaster format, which amounts to
change the relation between the pixel distance from the center of the
screen to a given pixel with the angular separation between their
associated directions (i.e., $Z^\mu$ and $N^\mu$ with the notations
defined above). Those two equations are therefore rewritten as
\begin{eqnarray}
\label{pix_i_geode}
i & = &   \frac{S + 1}{2} 
        - \frac{S}{\pi} \frac{\theta}{\sin \theta} X_\mu  N^\mu , \\
\label{pix_j_geode}
j & = &   \frac{S + 1}{2} 
        - \frac{S}{\pi} \frac{\theta}{\sin \theta} Y_\mu N^\mu ,
\end{eqnarray}
where $S$ is the number of pixel of any side of the square
image (typically 4000 or 8000), and $\theta$ is the angle between
$N^\mu$ and $Z^\mu$, i.e., $\theta = \arccos (- N_\mu Z^\mu)$. Note
that this transform does not only work for pixels belonging to the
square incircle, but to all the pixels of the square image. Inverse
transform follows immediately from
Eqns~(\ref{pix_i_geode}--\ref{pix_j_geode}).

Drawing the stars involves a slightly tweak with respect to
\S\ref{sec_stars} an more specifically \S\ref{sec_stars_eff}. This
time, we want that the star appears as a circular blob with respect to
the sky coordinates and not the pixel coordinates of the
picture. Therefore, we proceed using a supplementary step:
\begin{enumerate}

\item From the star overall luminosity computed including the Doppler,
  amplification and lensing effect described above, we define the
  angular size of the blob that represents the star, after having
  chosen a normalization for this (i.e., a 1\SST magnitude
  star is represented by a blob of given angular size).

\item Then, we compute the position of the star in pixel space as before.

\item Further, we focus on the square region in pixel space centered
  on the star position and including all the pixels with angular
  separation (in real space) smaller than the assumed star size.

\item Finally, for each of the pixels in this region, we compute the
  angular separation (in real space) between the pixel center and the
  star position, and we draw the star according to its luminosity
  profile we have chosen (truncated Gaussian, or whatever else).

\end{enumerate}

\begin{figure}[htbp]
\begin{center}
\includegraphics*[width=3.2in]{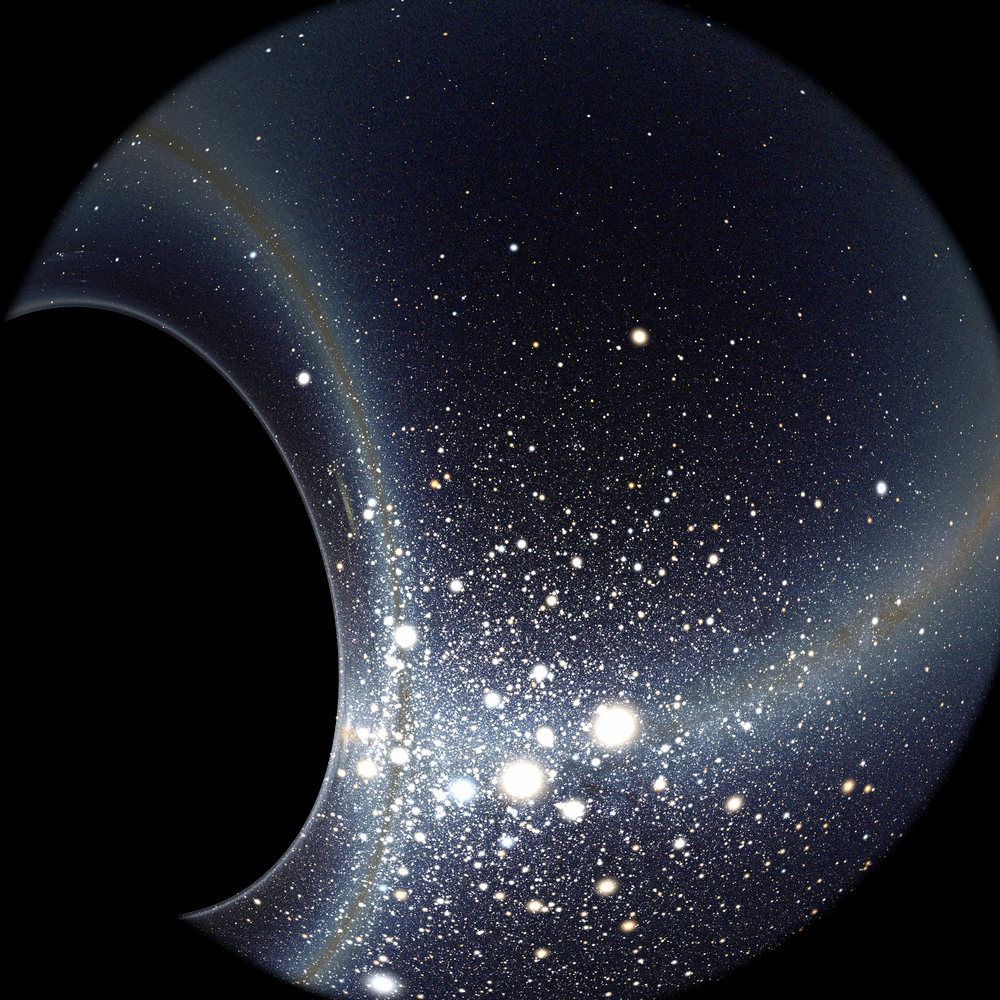}
\caption{An example of a 4k$\times$4k picture computed for planetarium
  projection. It shows the view of an observer at the last stable
  orbit around a Schwarzschild black hole (i.e., $r = 6 M$), looking at
  the front direction, corresponding to 45~degrees above the
  horizontal edge of the planetarium, which here corresponds to the
  pixel (rounded down) $(2000, 3000)$.  The curious reader is
  encouraged to search for usually easy to recognize stars or
  constellations such as Ursa Major, Orion, $\alpha$~Boo, $\alpha$~Sco
  and $\alpha$~CMa, but it is quite involving due to the high velocity
  ($c / 2$) causing significant luminosity and colour distortion to
  the familiar star background, as well as strong aberration and light
  deflection which severely affect the angular distance between stars.}
\label{LSO_plane}
\end{center}

\end{figure}

When looking at picture on a flat screen, stars will appear as
elongated orthogonally to the radial direction, but when projected
into a planetarium, they will appear as a circular blobs for observers
sitting close to the center of the hemisphere. Observers sitting
closer to the edges of the planetarium will experience some
distortions, but this is the case for any picture projected that
way. An example of picture computed for planetarium is given in
Fig.~\ref{LSO_plane}.

\section{Delineating and crossing the horizon}
\label{sec_hor}

As long as the observer lies outside the horizon, any calculation can
be done in the standard Schwarzschild coordinates, although this is
not necessarily what we do in practice. Such a coordinate choice is
no longer possible when the observer is within the horizon since all
the null geodesics he or she intersects have crossed the horizon and
consequently have locally necessitated to use another coordinate
system, see \S\ref{ssec_max}. However, apart from this, the procedure
is the same: we compute the deflection function, and then we perform
the drawing of the celestial sphere and of the stars.

\begin{figure}[htbp]
\begin{center}
\includegraphics*[width=3.2in]{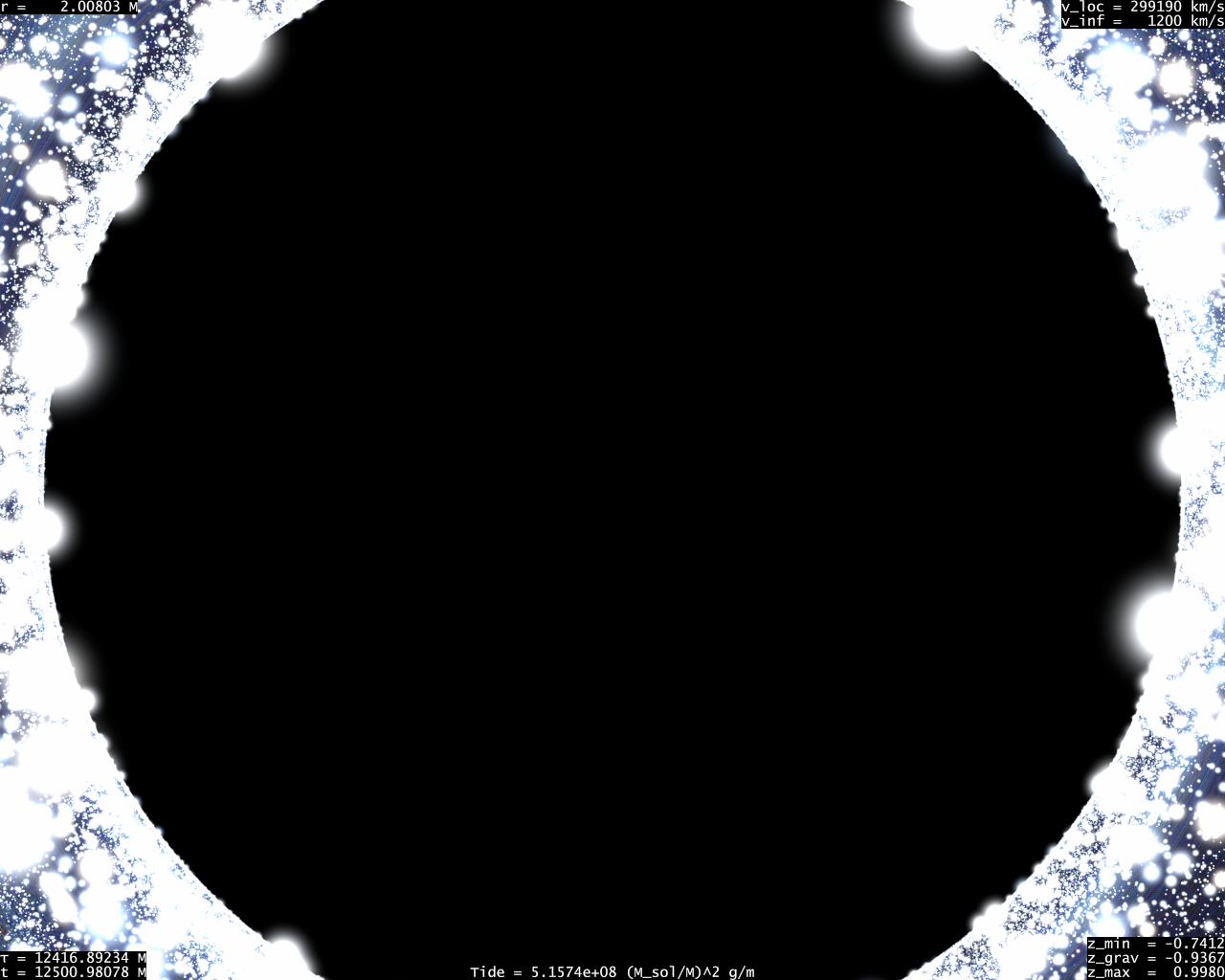}
\vskip 0.12cm
\includegraphics*[width=3.2in]{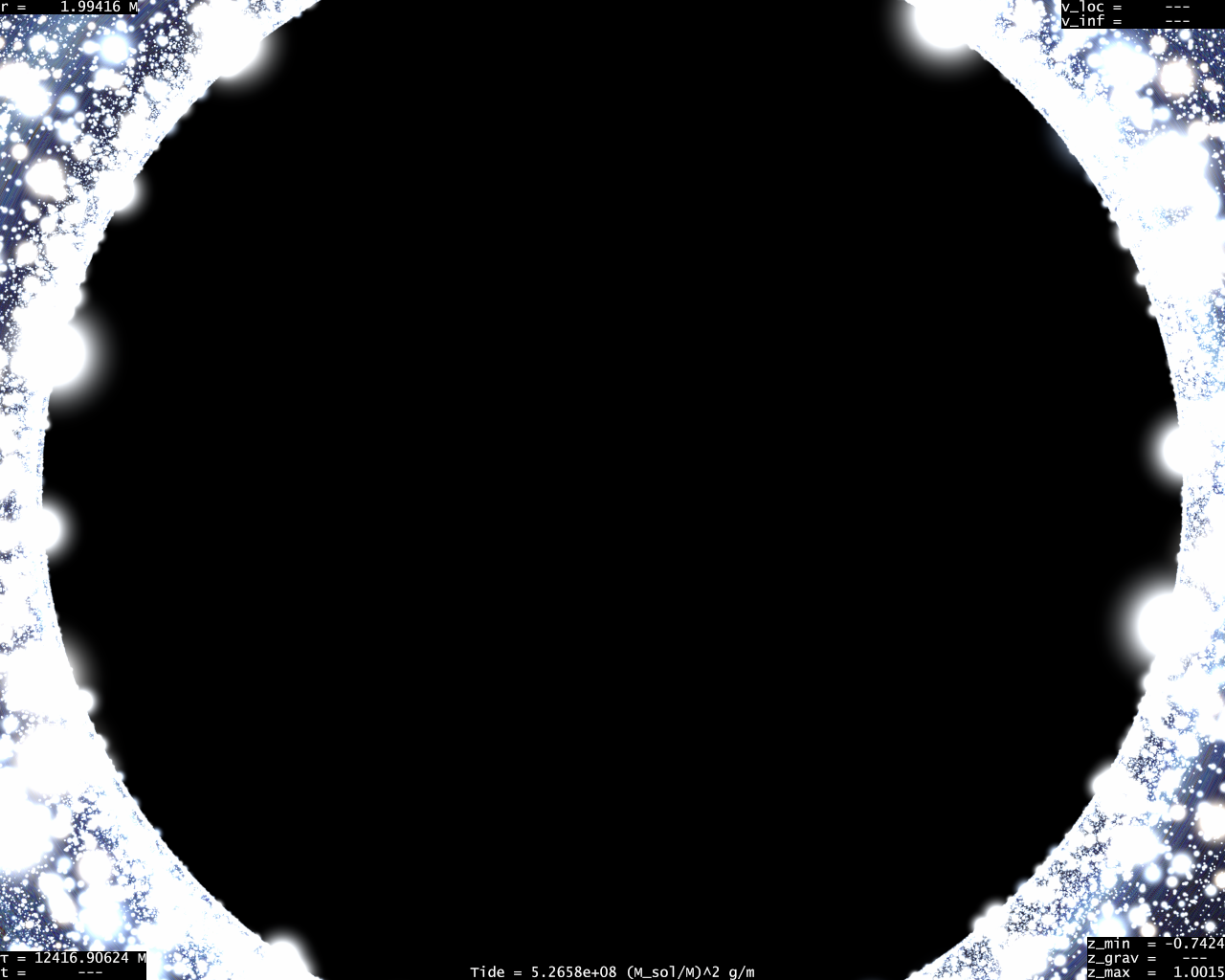}
\vskip 0.12cm
\includegraphics*[width=3.2in]{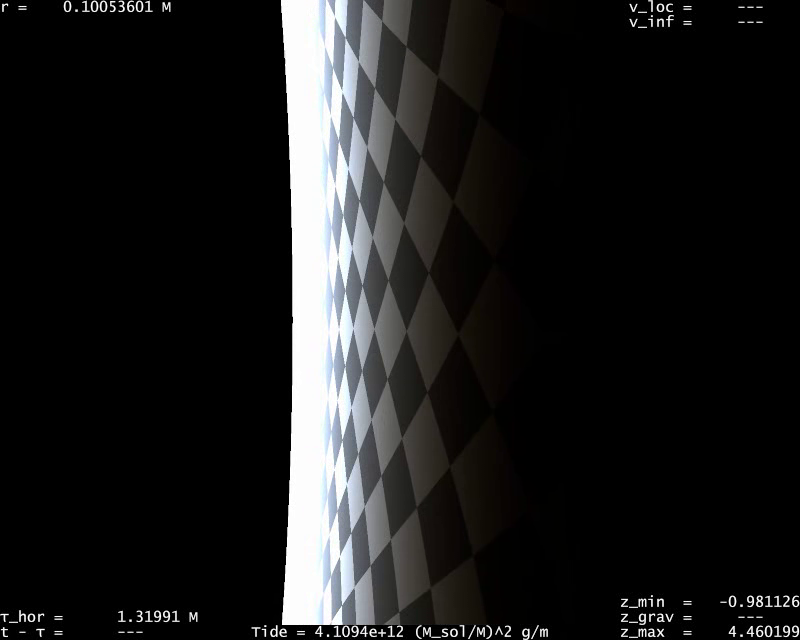}
\caption{Three views of a radial geodesic trajectory plunging into the
  black hole. Top and middle image show the front view just before and
  just after horizon crossing. As expected, no visual hint allows to
  decide easily whether or not the observer has crossed the
  horizon. Bottom image shows the side view soon before hitting the
  singularity ($r \sim 0.1 M$). We provide for this last picture a
  view with a coordinate grid only, so as to better emphasize the
  thinness of the blueshifted region. This representation also
  highlights that our celestial sphere grid, which lies at infinity,
  actually appears to be at decreasing, finite distance, falling on
  the observer.}
\label{fighor}
\end{center}
\end{figure}

We present in Fig.~\ref{fighor} three views of a radial geodesic
trajectory starting from infinity and plunging into the black
hole. Several interesting features seen in those images deserve an
explanation:

\begin{itemize}

\item Firstly, if we consider the natural case of a freely falling
  observer with zero angular momentum and zero velocity at infinity
  (so that this observer's velocity is $T_{\rm I}^\mu$), the angular
  size of the black hole is rather small at horizon crossing. This
  comes from the fact that an observer who is about to cross the
  horizon has a very large relative velocity with respect to a static
  observer close to the horizon. This means that the view seen by the
  former experiences a very strong aberration phenomenon with respect
  to the view seen by the latter, thus drastically reducing the black
  hole angular size. In order to determine the angular size of the
  black hole, one needs to consider a geodesic endowed with the
  critical impact parameter $|L / E| = 3 \sqrt 3 M$. As long as
  $r > 3 M$, the geodesic that delineates the horizon must have the
  above mentioned critical parameter, and must have almost reached
  the $r = 3 M$ region before bouncing back toward the
  observer. Therefore, it must be an outgoing geodesic, i.e., their
  $k^r$ component must be positive. For $r < 3 M$, the geodesics that
  start from null infinity and that reach the observer must be
  ingoing, i.e., their $k^r$ component must be negative. Without loss
  of generality, we can define a null geodesic under the form,
  following Eq.~(\ref{def_k_prec}):
  \begin{equation}
    \label{def_k_prec_2}
    k^\mu = \omega \left( T^\mu +  \cos \delta R^\mu 
                        + \sin \delta \Theta^\mu\right) ,
  \end{equation}
  $\omega$ corresponding to the observed wavenumber of the associated
  wave.  Without loss of generality, we can choose the $k^\theta$
  component to be positive, so that $0 \leq \delta \leq
  \pi$.
  Moreover, any geodesic starting for the observer's initial region
  past null infinity must have a positive frequency, which amounts to
  say that
  \begin{equation}
    \label{E_k_crit}
    \frac{E}{\omega} = 1 - \cos \delta \sqrt{2 M / r} > 0 .
  \end{equation}
  This constraint is always true when $r > 2 M$, but has to be checked
  when $r < 2 M$, something we will address in the next
  Section. Finally, the $k^r$ component of the
  geodesic is
  \begin{equation}
    k^r = \omega (\cos \delta - \sqrt{2 M / r}) .
  \end{equation}
  Therefore, following the previous discussion, we shall choose check
  that the sign of $k^r$ is compatible with the above criteria. These
  constraints being set, we need to compute the general solution of
  the equation $L / E = 3 \sqrt3 M$ for the null geodesic we consider
  here. With the above definition of $k^\mu$, this amounts to solve
  equation
  \begin{equation}
    \label{L_E_ff}
    \frac{r \sin \delta}
         {1 -  \sqrt{2 M / r} \cos \delta } 
     = 3 \sqrt{3} M .
  \end{equation}
  Defining $u \equiv r / M$ and solving the
  equation of the variable $\tan (\delta / 2)$, we find
  \begin{equation}
    \label{tan_delta}
    \tan (\delta_\eta / 2)
    = \frac{u + \eta  \sqrt{u^2 - 27 \left(1 - \frac{2}{u} \right)}}
    {3 \sqrt{3} \left(1 + \sqrt{\frac{2}{u}} \right)  } ,
  \end{equation}
  where $\eta = \pm 1$. The term within the square root in the
  numerator of Eq.~(\ref{tan_delta}) is bigger than $u$ when $u$ is
  smaller than $2$, so that the numerator is positive when $\eta > 0$
  whichever value of $u$, and is negative when $u < 2$ and $\eta < 0$.
  Regarding the denominator, it is always positive. Therefore,
  $\tan (\delta_\eta / 2)$ is negative when both $u < 2$ and
  $\eta < 0$, so that this part of the solution must be
  discarded. Further, one can show (or check after plotting the
  functions as calculations are less straightforward) that the
  solution $\delta_{+}$ is not outgoing when $r > 3 M$ and are
  therefore excluded, and so is $\delta_{-}$ when $r < 3 M$ as it is
  not ingoing. Further, Eq.~(\ref{E_k_crit}) is always valid for all
  remaining solutions, so that, taking into account the fact that the
  term under the upper square root of Eq.~(\ref{tan_delta}) can be
  factored by $(u - 3)^2$, the allowed solution reduces to
  \begin{equation}
    \label{tan_delta_final}
    \tan (\delta / 2)
    = \frac{u -(u - 3)  \sqrt{1 + \frac{6}{u}}}
    {3 \sqrt{3} (1 + \sqrt{\frac{2}{u}} ) } .
  \end{equation}
  The physical interpretation of $\delta$ is simple.  From an
  infalling observer point of view, a null geodesic of wavevector
  $k^\mu$ has an angular separation of $\delta$ with respect to the
  radial outgoing direction $R^\mu$. At horizon crossing, $u = 2$ and
  \begin{equation}
    \tan (\delta_{\rm hor} / 2) = \frac{2}{3 \sqrt{3}} ,
  \end{equation}
  so that
  \begin{equation}
    \cos \delta_{\rm hor} = \frac{23}{31} ,
  \end{equation}
  leading to an angular diameter of the black hole silhouette of
  $2 \arccos_{\rm hor} \delta \sim 84.2$ degrees. Such angular
  diameter would correspond, in Euclidean space, to that of a sphere
  seen at a altitude of $1 / \sin \delta_{\rm hor} - 1 \sim 0,\!49$
  times its radius. Coincidentally, such ``altitude'' in the
  Schwarzschild metric ($r = 3 M$) corresponds to that where a static
  observer sees the black hole silhouette encompass exactly the half
  of the celestial sphere, just as if one had landed on the perfectly
  dark surface, see middle part of Fig.\ref{fig_blue} in
  Appendix~\ref{app_img}.

\item Secondly, soon before hitting the singularity, the same
  calculation gives the simple result
  \begin{equation}
    \cos \delta_{\rm sing} = 0 ,
  \end{equation}
  leading to an angular diameter of 180~degrees. In other words,
  hitting the singularity happens when the black hole silhouette fills
  exactly half of the celestial sphere, just as what would happen in a
  Euclidean space should one land on the surface of the spherical
  body.

\item Thirdly, the celestial sphere is never very dark before horizon
  crossing. In order to see this, one simply has to notice that, as
  already stated, the scalar product $T_\mu k^\mu = \omega$ correspond
  to the observed wavenumber, whereas the constant of motion $E$
  corresponds to the wavenumber measured by a static observer at
  infinity. Therefore, the shift between the observed and the initial
  frequency is
\begin{equation}
\frac{1}{1 + z}
 = \frac{\omega}{E}
 = \frac{1}{1 - \cos \delta \sqrt{2 M / r}} .
\end{equation}
The redshift therefore takes the simple expression
  \begin{equation}
    z = - \cos \delta \sqrt{\frac{2 M}{r}} .
  \end{equation}
  Whatever value of $r$ one considers, the maximum redshift is
  unsurprisingly obtained to radial ingoing radiation
  ($\delta = - \pi$, i.e., $\cos \delta = -1$ and is 1 at horizon
  crossing. The redshift of the radiation decreases to 0 in the
  orthogonal direction $\cos \delta = 0$) and any direction closer to
  the black hole shows blueshifted radiation, the most blueshifted
  being the closest to the black hole silhouette, i.e., for
  $\delta = \delta_{\rm hor}$, for which $z = -23 / 31$.

  It is only when the observer is deep within the horizon that the
  maximum redshift significantly increases, taking the value
  $z = \sqrt{2 M / r}$ for radial ingoing radiation. However,
  orthogonal directions still exhibit a zero redshift, but the region
  where redshift is moderate (between 0 and 1, say) becomes
  increasingly small as one approaches the singularity since it is
  bound to the shell of inner and outer radius $\pi / 2$ and
  $\pi / 2 + \arcsin\sqrt{\frac{r}{2 M}}$. The maximum blueshift is
  again obtain for the largest allowed value of $\cos \delta$, which
  is the one for which $ L / E = 3 \sqrt{3} M$ The value of $\delta$
  for which this holds tends to $\pi / 2$, which allows to approximate
  Eq.~(\ref{L_E_ff}) as
  \begin{equation}
    3 \sqrt{3} M 
    \sim \frac{r}{1 - \cos \delta \sqrt\frac{2 M}{r}} = \frac{r}{1 + z} ,
  \end{equation}
  which gives 
  \begin{equation}
\label{zmin_I}
   1 + z_\MIN \sim  \frac{r}{3 \sqrt{3} M} . 
  \end{equation}
  Therefore, soon before hitting the singularity, only a very thin
  ring of directions shows a blueshift, but the maximal blueshift
  diverges. Consequently, the last sight of the Universe seen by a
  freely falling observer is a very thin but also very bright ring of
  light, whose angular diameter is approximately $\pi$. Within this
  ring, everything is perfectly dark as it corresponds to the black
  hole silhouette, and outside this ring, the sky appears to be very
  dark, being highly redshifted.

\end{itemize}

\section{Exploring the maximal analytic extension of the metric}
\label{sec_Kruskal}

\subsection{Looking at the other asymptotic region (region~III)}

From the Carter-Penrose diagram, it is obvious that an observer in
region~I cannot see anything of region~III because the latter is not
in the past lightcone of the former. The situation changes when the
observer enters into region~II because parts of both regions~I and III
are then in his/her past lightcone. A first obvious question one might
ask is how this region looks like from such observer. For the sake of
simplicity, we shall consider the simpler case of a freely-falling
observer starting from region~I with zero velocity at infinity.

The above framework can be adapted almost without much modification to
the maximal extension of the metric. Indeed, once one is able to
integrate backward a geodesic reaching an observer in region~II and
that had started from past null infinity of region~I, there is no
difficulty in doing so with geodesics originating from past null
infinity of region~III. Actually, any null geodesic that penetrates
into region~II can be cast under the form of Eq.~(\ref{def_k_prec_2}),
and it will originate from region~III if two conditions are satisfied:
(i) its constant of motion $E$ must be negative since in region~III,
$t$ is a past oriented timelike coordinate, so that
$k^t = \ddd t / \ddd p < 0$, and, (ii) that its impact parameter
$| L / E |$ is smaller than the critical value of $3 \sqrt {3} M$
since this condition is always mandatory for a geodesics originating
from past null infinity of any asymptotically flat region to be
absorbed by the black hole. The edge of region~III is therefore, as in
the case of region~I, delineated by geodesics of impact parameter
equal to $3 \sqrt{3} M$. We can, without loss of generality, impose
that $L / E = 3 \sqrt{3} M$, which imposes to take $L < 0$ since $E$
is negative. Therefore, this amount to solve the same equation as
previously, but with $- \pi \leq \delta \leq 0$, for which one has
$\tan \delta / 2 \leq 0$ and further take the opposite value of
$\delta$ so as to recover the case where $0 \leq \delta \leq \pi$.
Following the previous discussion, this amounts to only consider the
above part of the solution $- \delta_{-}$ when $u < 2$, a result that
makes sense since region~III is seen only when one enters into
region~II. In the end, we have
\begin{equation}
\tan (\delta_{\rm III} / 2)
 = \frac{1}{3 \sqrt{3} }
   \frac{- u - (u - 3) \sqrt{1 + \frac{6}{u}}}{1 + \sqrt{\frac{2}{u}}} ,
\end{equation}
a value which ensures that, as requested,
$\cos \delta_{\rm III} > \sqrt{u / 2}$ whenever $u < 2$.  In order to
check this, let us define $\cos \mu \equiv \sqrt{u / 2}$. The
requested inequality, $\cos \delta_{\rm III} > \sqrt{u / 2}$, is
equivalent to $\tan^2 (\delta_{\rm III} / 2) < \tan^2 (\mu / 2)$.
Given the second order equation verified by
$\tan^2 (\delta_{\rm III} / 2)$, we have
$(1 + \sqrt{2 / u}) \tan^2 (\delta_{\rm III} / 2) = \sqrt{2 / u} - 1 -
(2 u / 3 \sqrt{3}) \tan (\delta_{\rm III} / 2) \geq \sqrt{2 / u} - 1
$.
But we have also $ \tan^2 (\mu /2) = (1 - \cos \mu) / (1 + \cos \mu)$,
which is equivalent to
$(1 + \sqrt{2 / u}) \tan^2 (\mu / 2) = \sqrt{2 / u} - 1$, a value that
is precisely the upper bound of
$(1 + \sqrt{2 / u}) \tan^2 (\delta_{\rm III}/ 2)$, which completes the
proof.

\begin{figure}[htbp]
\begin{center}
\includegraphics*[angle=270,width=3.2in]{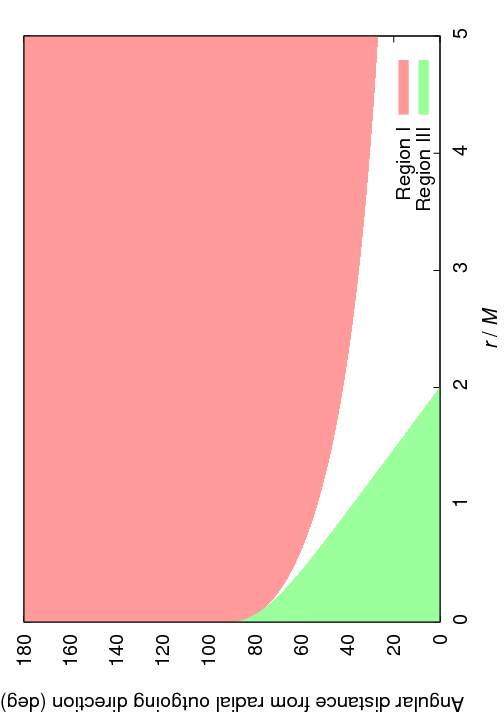}
\caption{Diagram showing which of region~I, III or black hole
  silhouette is seen as a function of angular separation between the
  front direction of a freely falling observer with zero velocity at
  infinity. Region~I is always seen and decreases in size till it
  occupies half of the field of view at the singularity.  Region~III
  is, as expected, invisible outside the horizon and steadily grows
  afterward, till it reaches the half of the field of view. Black hole
  silhouette, or, in this context, geodesics originating from past
  singularity (region~II) forms a shell between the two region, a
  shell that becomes infinitely thin as one approaches the
  singularity.}
\label{fig_delta_ee}
\end{center}
\end{figure}

Now, when $\delta = 0$, $L$ is also equal to $0$, so that region~III
correspond to any value of $\delta$ smaller than $\delta_{\rm III}$.
The above equation immediately tells that
$\tan (\delta_{\rm III} / 2) = 0$ at horizon crossing ($r = 2 M$),
which amounts to say region~III reduces to a single point when it
becomes observable just after horizon crossing. Then,
$\tan (\delta_{\rm III} / 2)$ grows almost linearly as $r$ decreases
(expanding the above relation gives
$\delta_{\rm III} = 3 \sqrt{3} (2 - u) / 8 + O ((2 - u)^2)$ and
reaches $1$ at $r = 0$, which means that region III now encompasses a
circular region of angular radius $\pi / 2$. In other words, both
regions I and III seem to join each other when the observer reaches
the singularity, and the remaining part, which corresponds to
geodesics originating from region~II occupies a narrow shell which
roughly is a great circle in the sky. The angular size of regions~I
and III are shown in Fig.~\ref{fig_delta_ee}.

\begin{figure}[htbp]
\begin{center}
\includegraphics*[angle=270,width=3.2in]{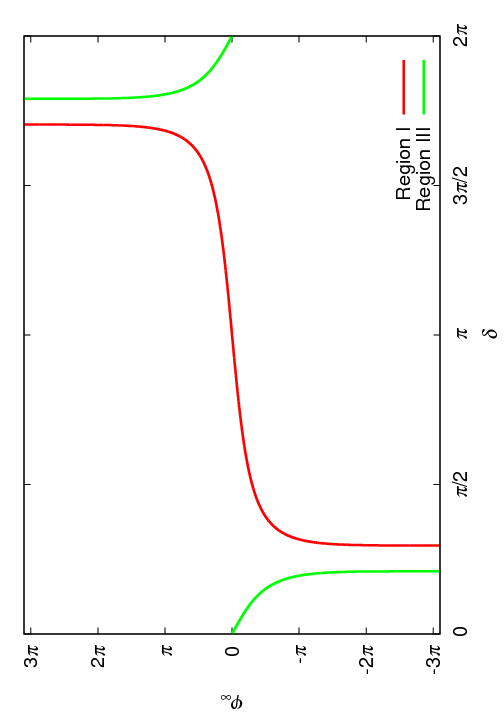}
\caption{An example of both deviation functions
  $\varphi_\infty(\delta)$ of regions~I and III as seen by an
  infalling observer at $r = M$. The server is still far, in term of
  coordinate distance, from the singularity, so that both regions have
  significantly different size, however, the deflection function are
  already quite similar.}
\label{dev_III}
\end{center}
\end{figure}

Within what is seen from region~III the angular distortion have a
similar behaviour to those of region~I. The deflection function is zero
in the front direction, and strongly increases as one approaches the
edges of region~III, leading to an infinity of closely packed multiple
images of the celestial sphere. An example of the deflection functions
of both region~I and III is show in Fig.~\ref{dev_III}.

An even less intuitive feature is how the redshift of region~III
evolves, both as a function of the angular distance to the central
direction and of the coordinate distance. In this case, a slight
difference arises with respect to null trajectories originating from
region~I. Indeed, if one assumes that a photon is sent from a static
observer of four-velocity $u_{\rm stat, III}^\mu$ at past null
infinity of region~III, the photon wavenumber will be given by
$\omega_{\rm III} = u_{\rm stat, III}^\mu g_{\mu\nu}k^\nu$. Assuming
that this distant observer is static amounts to say that the only non
zero component of its four-velocity is the time component, so that we
have $\omega_{\rm III} = u_{\rm stat, III}^t E$. Moreover, since in
region~III, $t$ is a past-oriented timelike coordinate,
$u_{\rm stat, III}^t = -1$, so that the wavenumber is
\begin{equation}
\omega_{\rm III} = - E .
\end{equation}
Now, according to Eq.~(\ref{def_k_prec_2}), $\omega$ corresponds to
the observed wavenumber from the point of view of the infalling
observer originating from region~I, so that the frequency shift is
$\omega / \omega_{\rm III} = \omega / (-E)$. Consequently, the
redshift or blueshift of the radiation is given by
\begin{equation}
z = \cos \delta \sqrt{\frac{2 M}{r}} - 2 .
\end{equation}
If we consider first the "front" direction $\delta = 0$, there is an
infinite blueshift at horizon crossing since $z = - 1$. Further, when
$r$ decreases, $z$ increases, becomes equal to 0 at $r = M / 2$ and
tends to infinity afterward, a situation that qualitatively mimics
what happens for radial ingoing radiation coming from
region~I. Moreover, since $\cos \delta > 0$, the redshift decreases as
$\delta$ grows, so that region~III appears as a disk whose center is
dimmer than the edge. The center of the disk has a diverging redshift
as $r$ goes to 0, whereas its edge experiences a blueshift, which,
when $\delta_{\rm III}$ approaches $\pi/ 2$ approaches
$1 + z \sim u / 3 \sqrt{3}$, a value that, again, matches what the
same observer sees with region~I (see Eq.~\ref{zmin_I}).

\begin{figure}[htbp]
\begin{center}
\includegraphics*[width=3.2in]{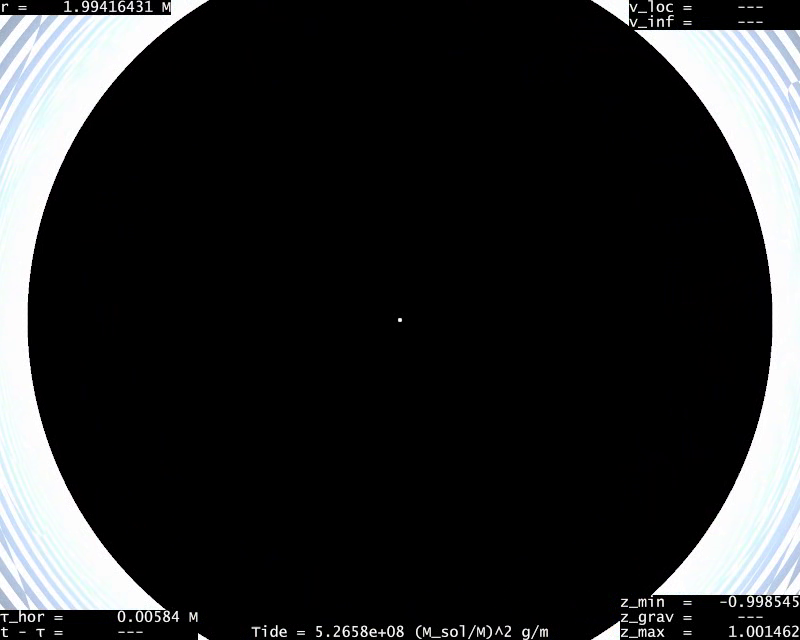}
\vskip 0.12cm
\includegraphics*[width=3.2in]{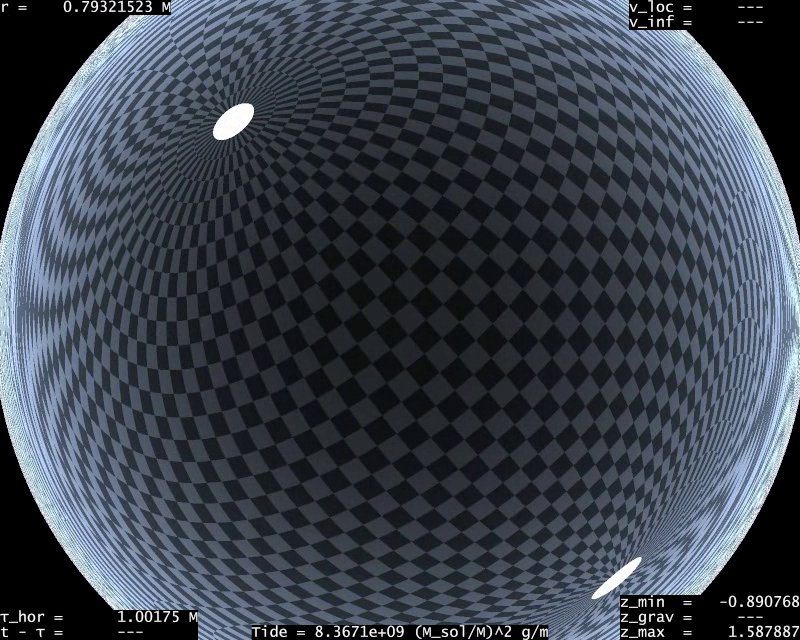}
\vskip 0.12cm
\includegraphics*[width=3.2in]{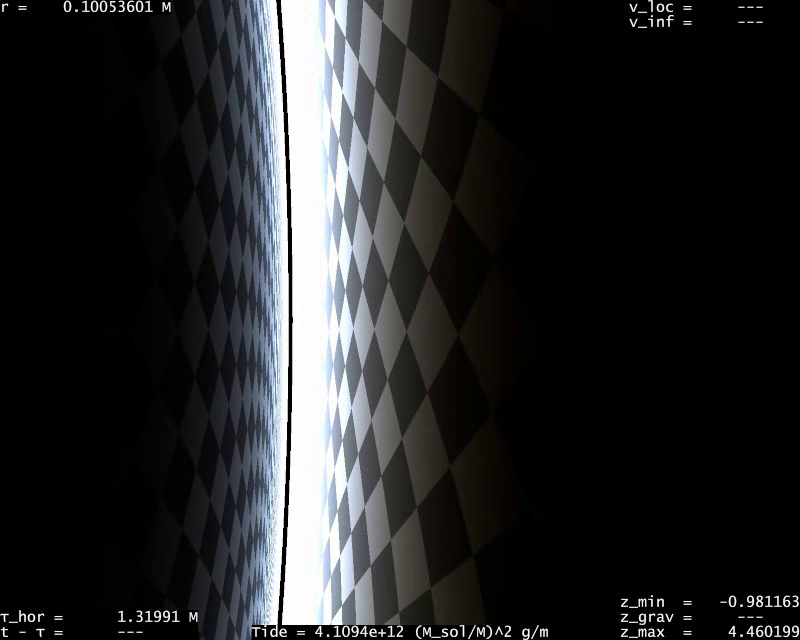}
\caption{Three views of the maximal analytic extension of the
  metric. Top view shows the appearance of region~III just after
  horizon crossing~($r \sim 1.994 M$). Region~I and region~III look
  very different, as region~III is an almost pointlike, infinitely
  blueshifted point, whereas region~I, that surrounds the silhouette of
  the black hole is of large but finite blueshift on its edge and
  moderate redshift in its center (behind observer, not shown).
  Middle image show the view of region~III at $r \sim 0.8 M$, where
  darkening of central part of region~III is visible, whereas its edge
  are blueshifted. Bottom image is a side view seen by an infalling
  observer close to the singularity ($r \sim 0.1 M$). Region~I, from
  which the observer originates, is on the right, and is identical
  with bottom view of Fig.~\ref{fighor}, whereas region~III is on the
  left. These two regions appear more and more identical as the
  observer approaches the singularity~(see text). Celestial spheres of
  both regions were taken to be a coordinate grid so as to better
  illustrate the fact that they look more and more the same (in term
  of distortion) as the observer is moved toward future singularity.}
\label{fighor2}
\end{center}
\end{figure}

It might seem unexpected that regions~I and III are seen almost
identically from an infalling observer originating from region~I,
however, there is a rather simple explanation to this. First, what we
have found here is identical to what a freely falling observer from
region~III would see, after exchanging in our results region~I and
region~III. If we note by $T^\mu_{\rm III}$ this new freely falling
observer's velocity, then $T^\mu_{\rm III}$ has the same components as
$T^\mu_{\rm I}$ except for the $t$ one, which is of opposite sign
since $t$ is a past directed timelike coordinate in
region~III. Consequently, the dot product between the two infalling
observers is
\begin{equation}
T^{\rm I}_\mu T^\mu_{\rm III}
 = \frac{ \frac{2 M}{r} + 1}{\frac{2 M}{r} - 1}.
\end{equation}
This quantity can be seen as the Lorentz factor of a boost one must
perform to go from one velocity to the other. This Lorentz factor is
infinite at horizon crossing. This explains why region~III appears
infinitely blueshifted from the point of view of observer
$T^\mu_{\rm I}$, whereas is it moderately redshifted for observer
$T^\mu_{\rm III}$~(such observer sees region~III at horizon crossing
exactly the same way observer $T^\mu_{\rm I}$ sees his own region~I at
horizon crossing). Now, as both observers move ahead within the
horizon, their relative Lorenz factor decreases toward 1, which means
that the two observers have a more and more similar perception of the
two regions. Consequently, region~III seen by observer $T^\mu_{\rm I}$
is very similar to the same region seen by observer $T^\mu_{\rm III}$,
which by definition is seen in the same way as region~I by observer
$T^\mu_{\rm I}$, which explains why the two regions look more and more
identical by this observer, as exemplified in Fig.~\ref{fighor2}.

So far, regions~I and III are seen in an asymmetric way since we dealt
with a freely falling observer coming from one of those regions. A
natural question that arises is what happens in the most symmetric
case, that is when the $t$ component of the observer four-velocity is
0 once in region~II. This amount to consider an observer in region~II
with a four-velocity given by
\begin{equation}
T_{\rm II}^\mu = \left(\begin{array}{c}
0 \\ -\sqrt{\frac{2M}{r} - 1} \\ 0  \\ 0 
                     \end{array}\right) .
\end{equation}
We then consider the tetrad with this four-vector, the vectors
$\Theta^\mu$ and $\Phi^\mu$ already defined in Eq.~(\ref{ThetaPhi})
and the ``radial'' vector
\begin{equation}
R_{\rm II}^\mu = \left( \begin{array}{c} 
\frac{1}{\sqrt{\frac{2 M}{r} - 1}}\\ 0 \\ 0 \\ 0 
                      \end{array}\right) .
\end{equation}
We can, as before, define a null vector crossing this new observer's
worldline as
$k^\mu = \omega (T^\mu_{\rm II} + \cos \delta R^\mu_{\rm II} + \sin
\delta \Theta^\mu$.
$\delta = 0, \pi$ correspond to the radial null trajectories
originating from regions~I and III, respectively. The most off-radial
null trajectories originating from those two regions are those of
impact parameter $3 \sqrt{3} M$ which correspond to an angle given by
\begin{equation}
\frac{u \tan \delta}{\sqrt{\frac{2}{u} - 1}} = 3 \sqrt{3} .
\end{equation}
Therefore, at the edge of region~II (i.e., $r$ close to $2 M$), both
regions I and III appear as a point whose redshift is given by
$1 / (1 + z) = 1 / (\sqrt{2 M / r - 1}|\cos \delta|)$, i.e., an
infinite blueshift. Further, as this observer goes toward the
singularity, both regions~I and III experience the same behaviour as
previously described for infalling observers from regions~I and~III.

\subsection{Looking at the ``white hole'' (region IV)}

\subsubsection{From outside the horizon}

When considering the maximal analytic extension of the Schwarzschild
metric, one usually assumes that nothing emerges from the past
singularity, however, nothing prevents from computing geodesics
originating from it and map what an observer can see from this
singularity. Since there are no bound null geodesics, an observer will
cross either (i) null geodesics originating from past null infinity of
either his/her own region (if the observer is outside the horizon), or
from any of the two asymptotic regions (if the observer has entered
into the black hole), or (ii) null geodesics originating from the past
singularity, i.e, region~IV. If we want to include such geodesics, we
simply need to assume that the past singularity emits some light whose
spectrum depends on both the direction $\theta, \varphi$ and the there
spacelike coordinate $t$.

\begin{figure}[htbp]
\begin{center}
\includegraphics*[angle=270,width=3.2in]{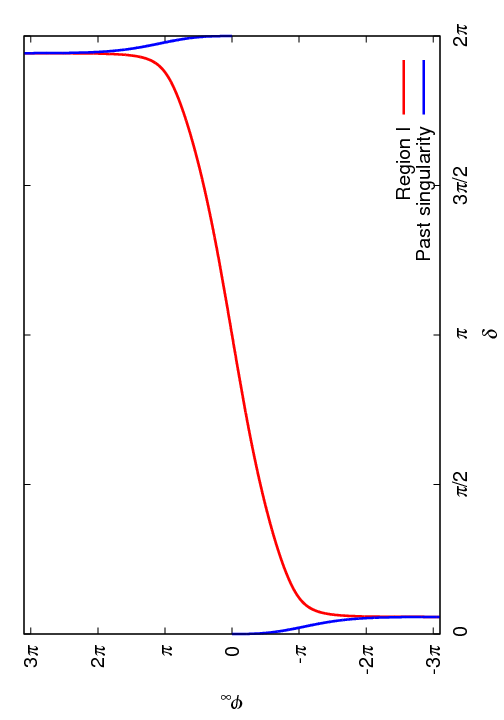}
\caption{An example of all deviation functions of region~IV (as well
  as I) seen from an observer outside the black hole, at $r = 20 M$.
  As explained in the text, there is an infinity of Einstein rings in
  what is seen of region~IV. Also, because the black hole silhouette
  is seen from far, the deviation function is rather steep. }
\label{dev_wh_0}
\end{center}
\end{figure}

\begin{figure}[htbp]
\begin{center}
\includegraphics*[width=3.2in]{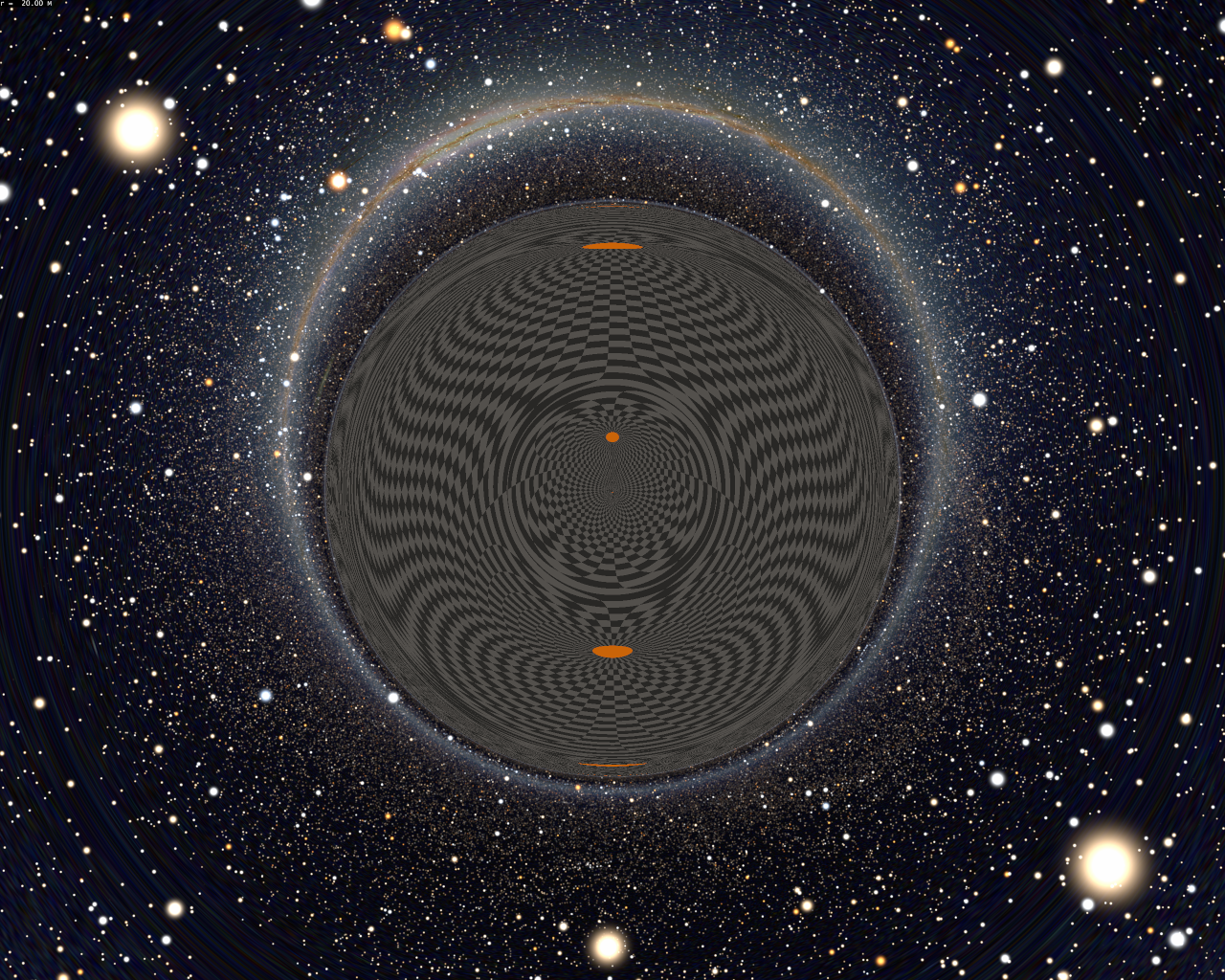}
\caption{A view of the white hole region seen from a static standard,
  observer at $r = 20 M$. Observer's asymptotic region is chosen to
  show the Milky Way, whereas white hole ``celestial sphere'' is
  chosen to be a coordinate grid (i.e., we do not assume any
  dependence with respect with the $t$ coordinate).  We also removed
  any redshift information from this region. The ``multiple images''
  in the white hole region are seen through the multiple views of the
  polar cap of the coordinate system. Note that since the deflection
  function is rather steep for close to radial outgoing trajectories,
  the coordinate grid in the center of the black hole silhouette is
  significantly shrunk.}
\label{fig_wh_1}
\end{center}
\end{figure}

If one neglects the $t$ dependence of the outgoing radiation, it is
rather easy to guess what an observer would see outside the past
horizon. Assuming an observer whose four-velocity is only along the
radial direction (i.e. it is equal to $T^\mu_{\rm I}$), the outgoing
radial null geodesic does not experience any deflection since it has
zero angular momentum. Therefore, this direction of the white hole
shows the radiation emitted from the same $\theta$ and $\varphi$ as
the observer lies. Conversely, close to the edge of the white hole,
when outgoing radiation has an impact parameter (measured at future
null infinity almost equal to, although smaller than $3 \sqrt{3} M$),
the outgoing radiation experienced an arbitrarily large deflection,
and the deflection decreases with the impact parameter. Consequently,
the observer will see a series of copy of the white hole emission
region, or, if one considers the $t$ dependence of the emission, a
series of rings of variable $t$. The deflection function together with a
view of the white hole region seen from an observer at $r = 20 M$ are
shown in Figs~\ref{dev_wh_0}--\ref{fig_wh_1}.

\subsubsection{From the future horizon (region~II)}

\begin{figure}[htbp]
\begin{center}
\includegraphics*[angle=270,width=3.2in]{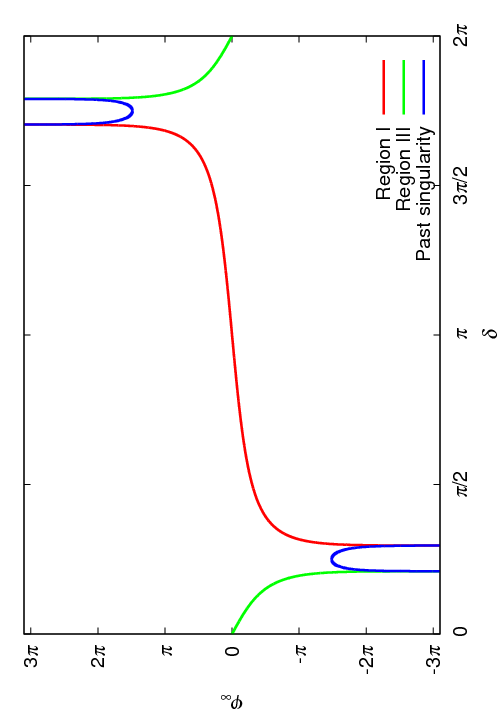}
\caption{An example of all deviation functions in the complete
  analytic extension of the metric, assuming that something emerges
  from the past singularity. As in Fig~\ref{dev_III}, the observer is
  situated at $r = M$.}
\label{dev_wh}
\end{center}
\end{figure}

\begin{figure}[htbp]
\begin{center}
\includegraphics*[width=3.2in]{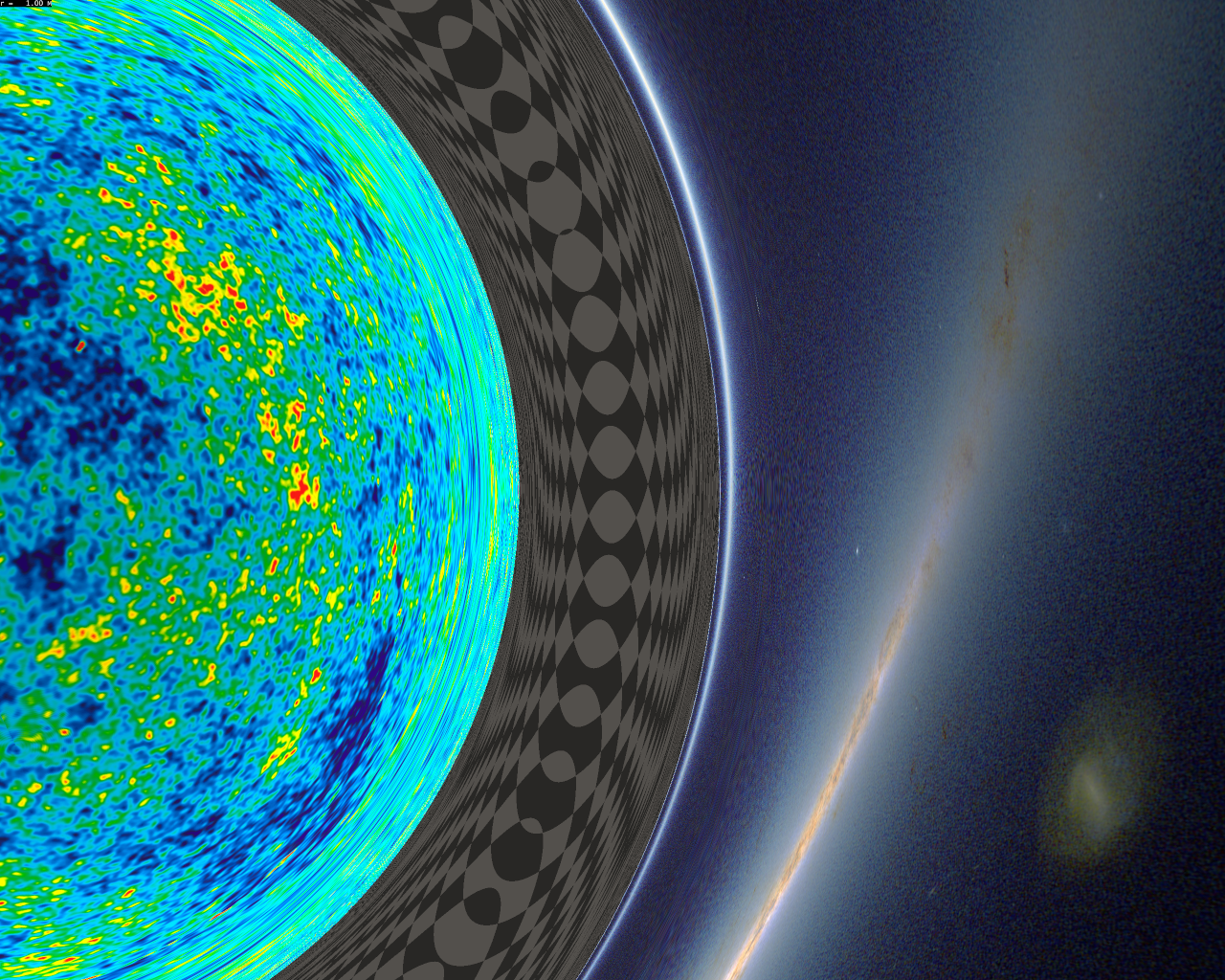}
\caption{A view of the white hole region seen from an infalling
  observer at $r = M$. The view is $45\circ$ away from radial
  direction. As for previous image, region~I, to the right, shows the
  Milky Way, whereas region~III, on the left shows (rather
  arbitrarily) a CMB map.}
\label{fig_wh_2}
\end{center}
\end{figure}

If one then considers an observer that has entered the future horizon,
then both regions~I, III and IV are visible. Region~IV fills the empty
space that existed between the two asymptotic regions. The deviation
function of region~IV behaves differently as compared to the previous
case. The reason is that outside the horizon, the observer can spot
the purely radial outgoing null geodesic.  This is not possible within
the future horizon as the observer cannot intersect this geodesic any
longer. Therefore, any geodesic originating from region~IV reaching
region~II is non radial and has experienced some amount of
deviation. Consequently, the deviation function has a local
extremum. On the edge of its domain of definition, the deviation
function still diverges as it corresponds to geodesics which have
exited the past horizon (either entering into region~I or III and
which then bounced back at $r = 3 M^-$ after having experienced an
arbitrarily large deviation. An example of the deviation function is
shown in Fig.~\ref{dev_wh} and the corresponding view is shown in
Fig.~\ref{fig_wh_2}.
*** DIRE PLUS

\subsubsection{From the past horizon}

Another question one may want to address is what happens in the
(rather academic) case of a very hypothetical observer originating
from the past singularity and still in region~IV. In this case, any
direction of observation shows null geodesics originating from the
past singularity since none of the other regions of the Kruskal
extension are in such an observer's past lightcone. Very close to the
singularity, so that we can keep only the leading terms in $1 / r^n$
in Eq.~(\ref{rdot2}), a null geodesics follows the simplified equation
\begin{equation}
\left( \frac{\ddd r}{\ddd p}\right)^2 \simeq \frac{2 M L^2}{r^3} ,
\end{equation}
where $p$ is an affine parameter of the geodesic. This amounts to say
that $r$ initially varies as $p^\frac{2}{5}$. From the constancy of
$L = r^2 \ddd \theta / \ddd p$, we have
$\theta - \theta_0 \propto p^\frac{1}{5}$, and applying the same
procedure with the other constant of motion $E$, we have
$t - t_0 \propto p^\frac{7}{5}$, where in both cases the $0$
subscripts denote quantities evaluated at the beginning of the
geodesic. Therefore, an observer close to the singularity will have a
past light cone that intersect only a small portion of the singularity,
both in term of the angular and ``time'' dependence.

In order to use our formalism in the region, we need to redefine
vectors $T^\mu$ and $R^\mu$, whose most natural choice is then
\begin{equation}
\label{TR_IV}
T_{\rm IV}^\mu = \left(\begin{array}{c} 
0 \\ \sqrt{\frac{2M}{r} - 1} \\ 0  \\ 0 
                     \end{array}\right) \qquad,\qquad
R_{\rm IV}^\mu = \left(\begin{array}{c} 
\frac{1}{\sqrt{\frac{2 M}{r} - 1}}\\ 0 \\ 0 \\ 0 
                     \end{array}\right) .
\end{equation}
We then consider the tetrad with this four-vectors, as well as the
four-vectors $\Theta^\mu$ and $\Phi$. As before, we shall consider a
hypothetical observer with four-velocity $T_{\rm IV}^\mu$, and
consider null geodesics of four-wavevector $k^\mu$ defined in
Eqns.~(\ref{def_k_prec},\ref{def_k_prec_2}) crossing his/her
worldline

\begin{figure}[htbp]
\begin{center}
\includegraphics*[width=1.6in]{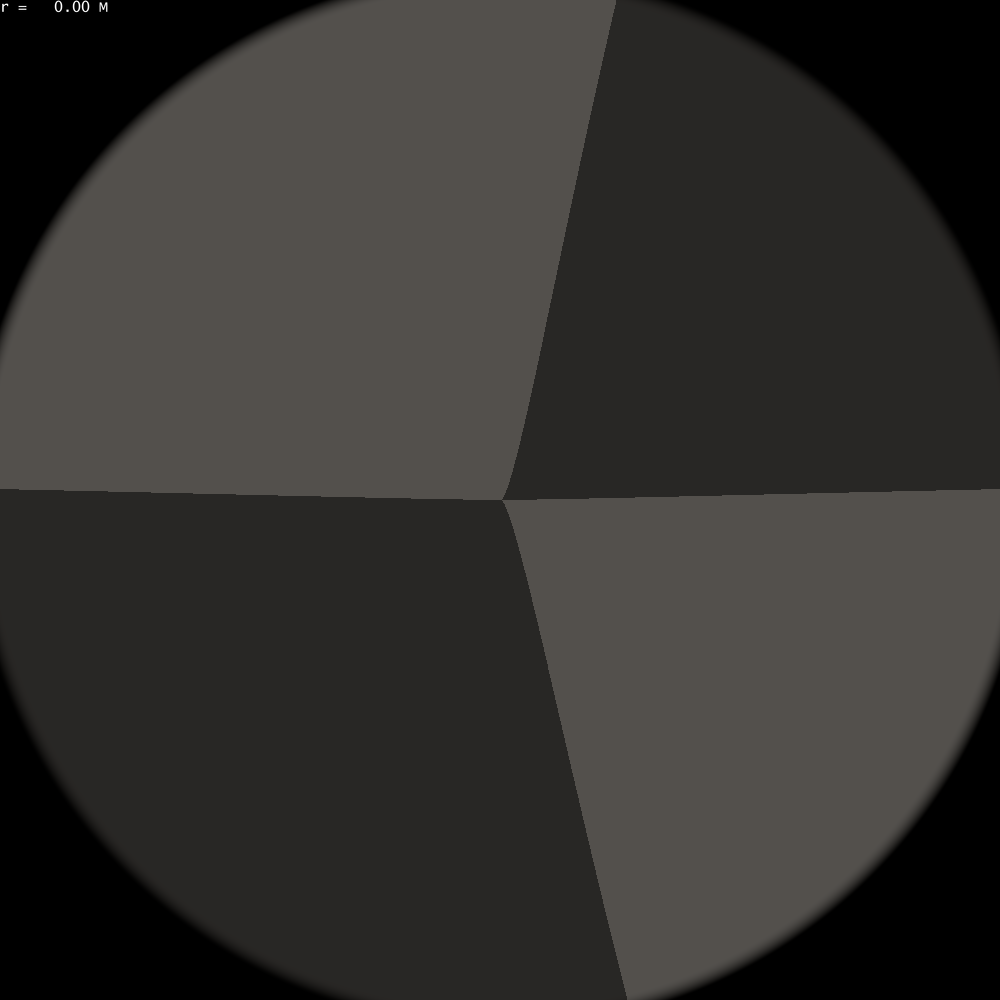}
\includegraphics*[width=1.6in]{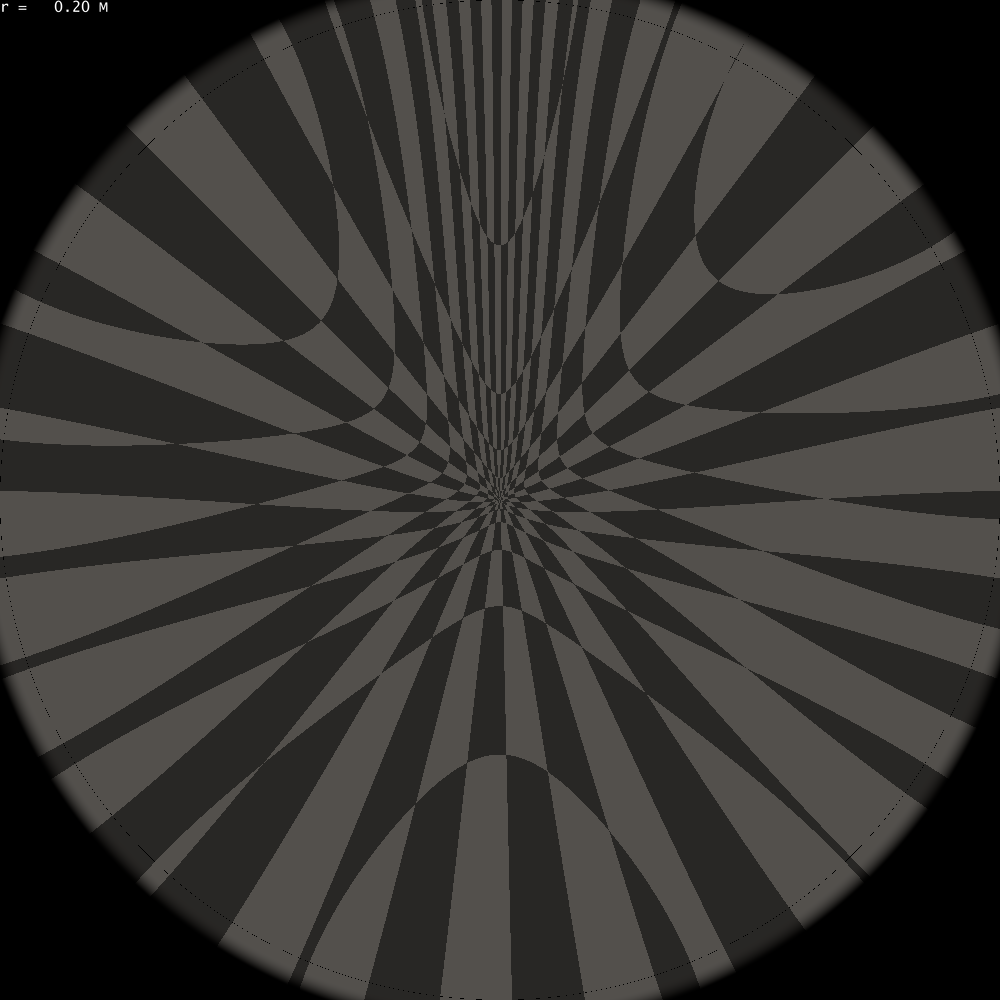}
\vskip 0.12cm
\includegraphics*[width=1.6in]{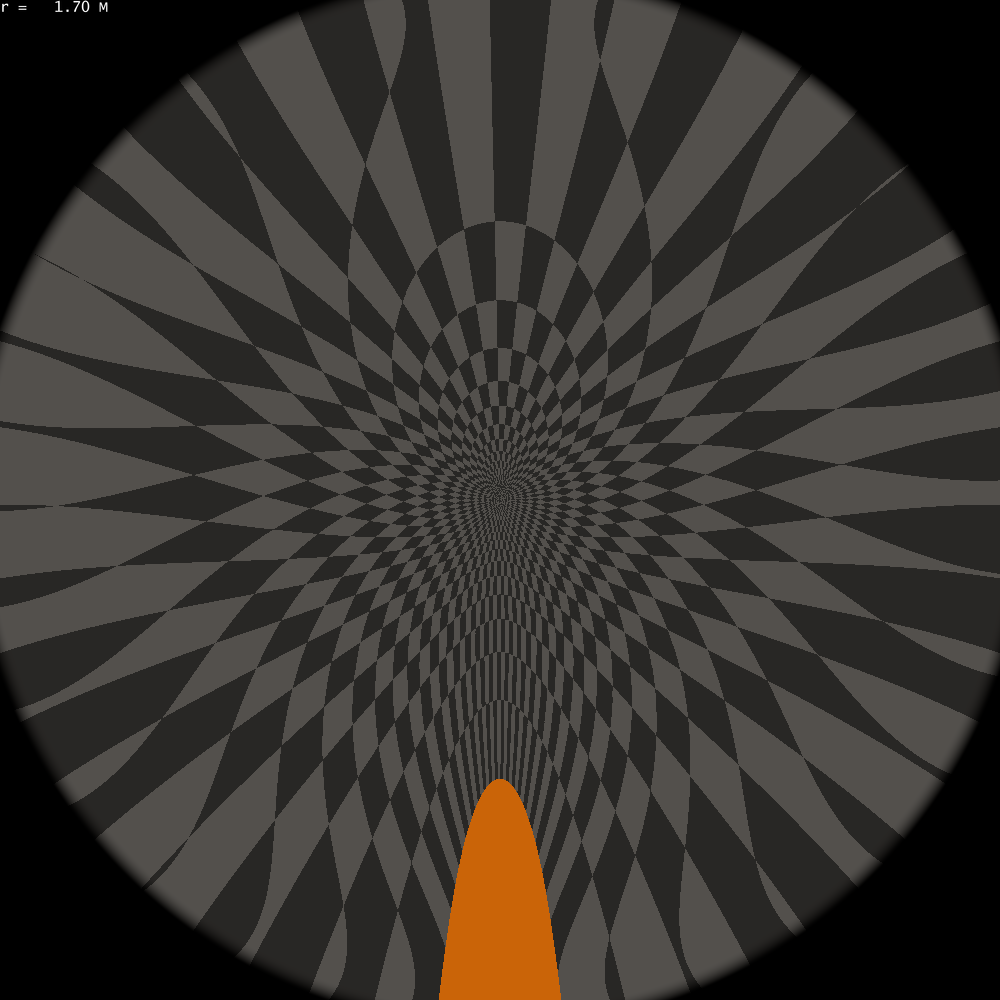}
\includegraphics*[width=1.6in]{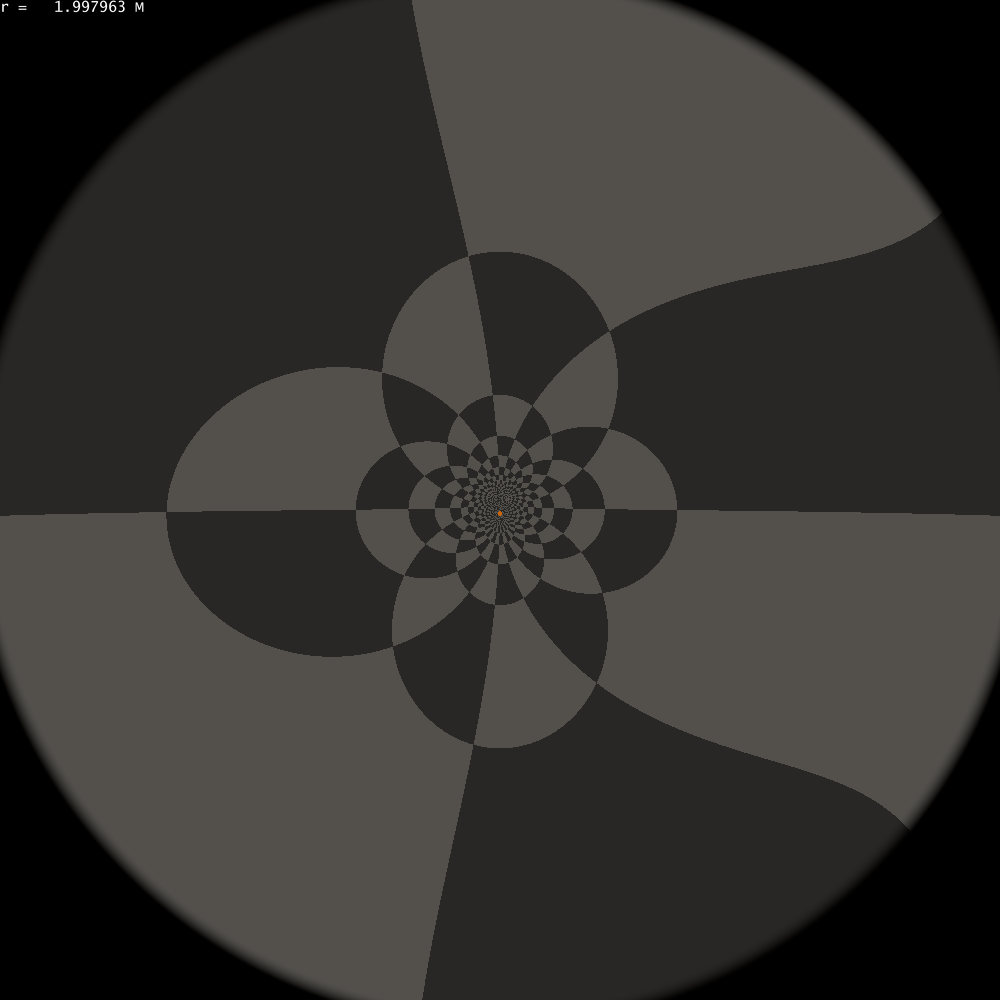}
\caption{Set of views of region~IV by an observer in region~IV. The
  four fisheye view are taken at $r = 0.001 M$, $r = 0.2 M$,
  $r = 1.7 M$ and $r = 1.998 M$ and toward the same direction. When
  the observer "takes off" from the singularity, only a narrow part of
  it is seen (bottom left picture). Further, when $'$ increases, the
  view of the singularity seems to become more and more elongated
  along the $\pm \partial / \partial t$ direction (opposite view is
  essentially identical to the one shown).}
\label{fig_IV_img}
\end{center}
\end{figure}

\begin{figure}[htbp]
\begin{center}
\includegraphics*[width=1.6in]{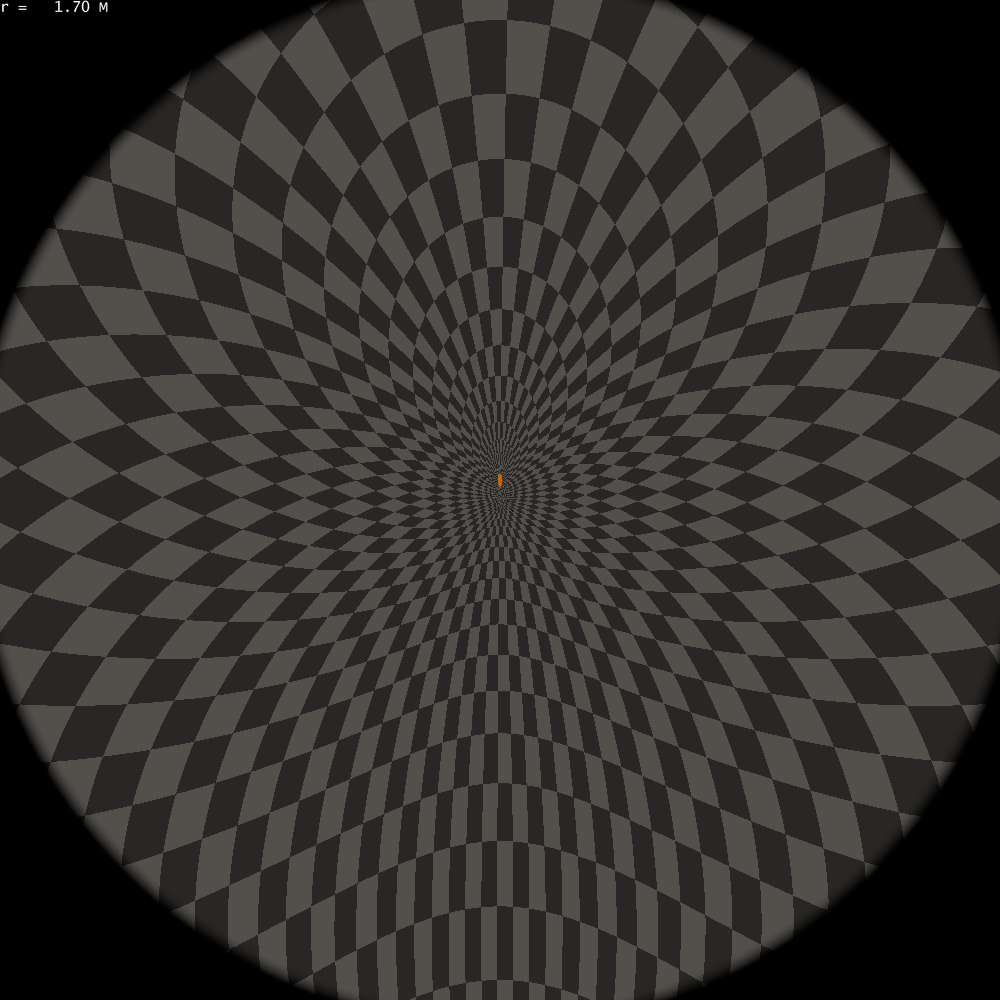}
\includegraphics*[width=1.6in]{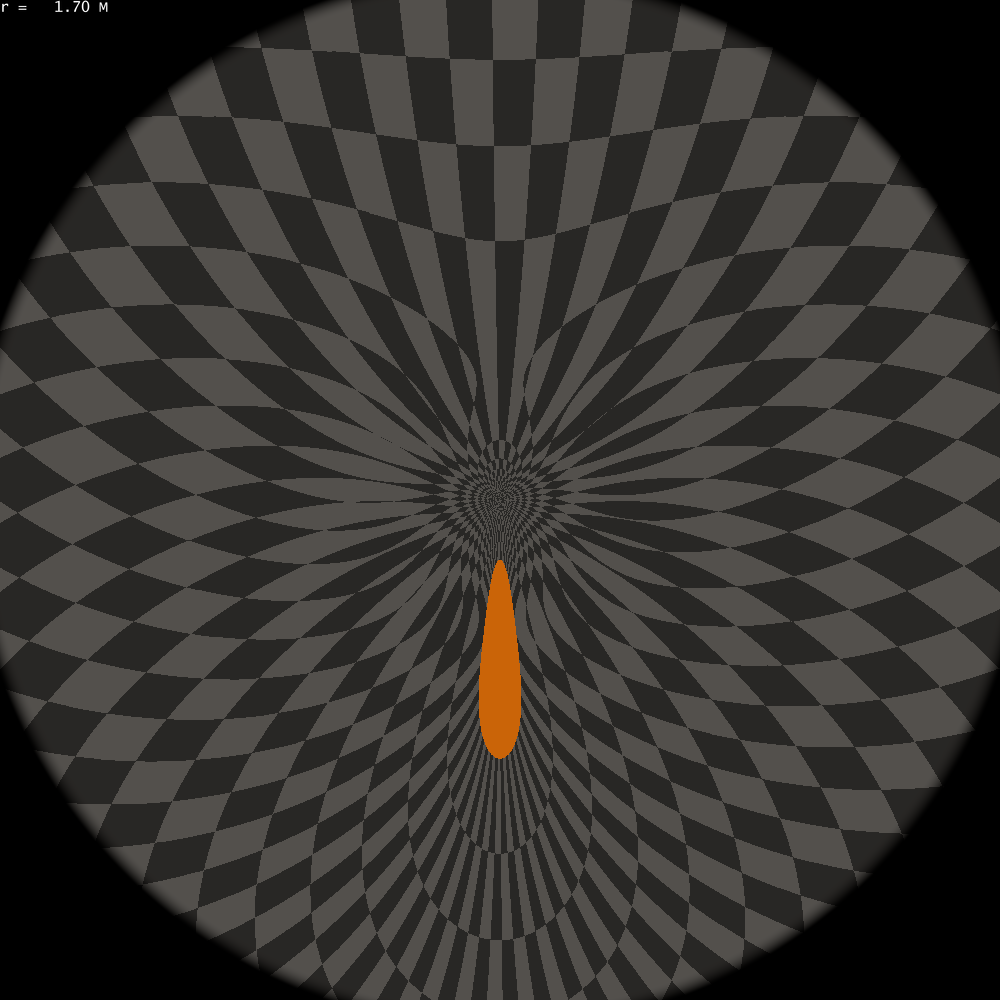}
\vskip 0.12cm
\includegraphics*[width=1.6in]{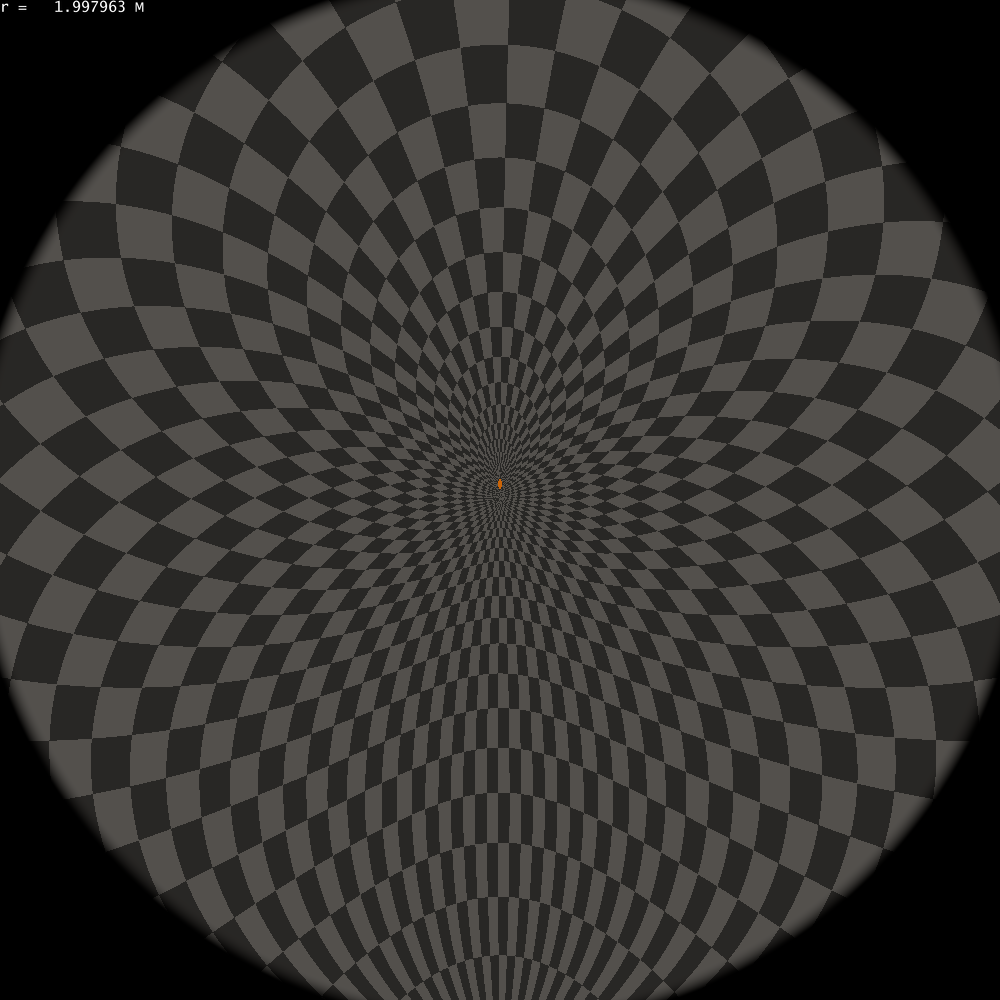}
\includegraphics*[width=1.6in]{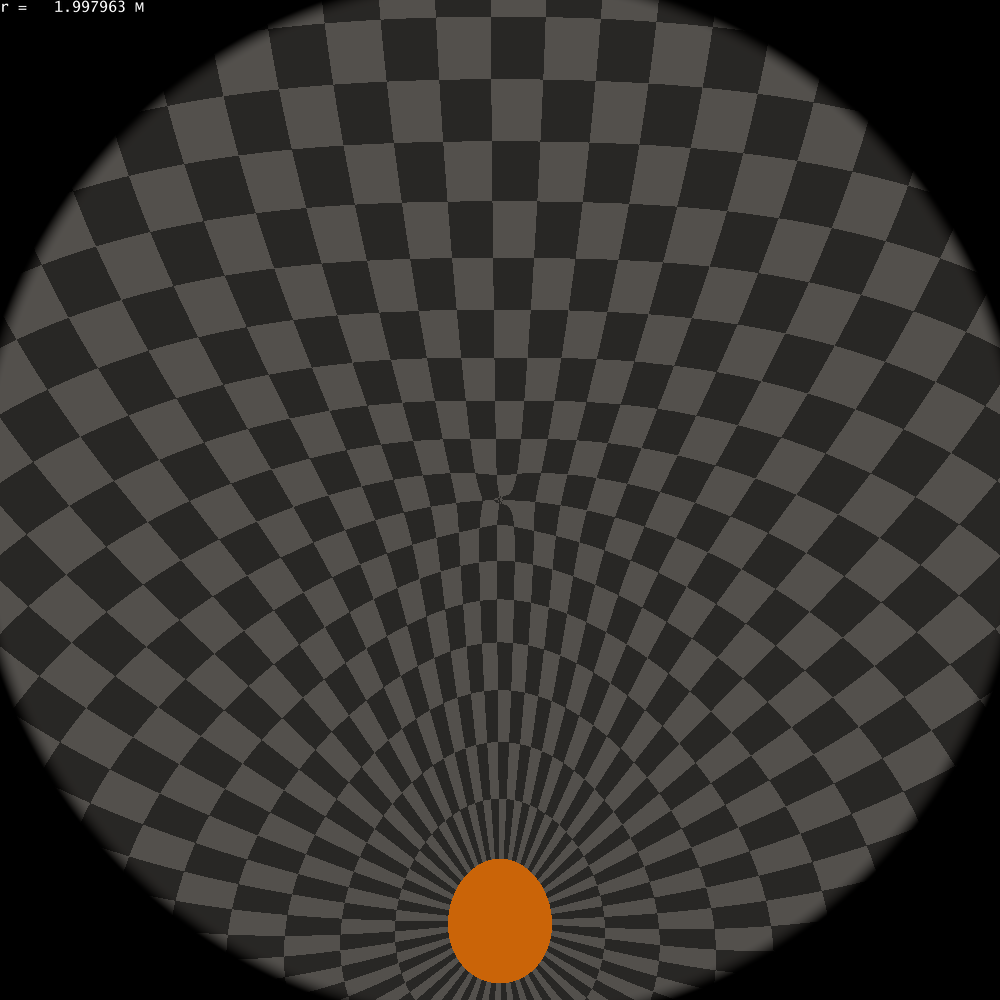}
\caption{Front (right) and rear (left) views at $r = 1.7 M$ (top) and
  $r = 1.998 M$ (bottom) by an observer about to exit region~IV to
  reach region~I and who will then become a ZAMO outgoing observer
  with zero velocity at infinity. The front view is to be compared
  with the bottom row of Fig.~\ref{fig_IV_img}: the field of view is
  significantly shrunk. The transition between the two is reasonably
  obvious for $r = 1.7 M$, as one can follow the displacement of the
  orange spot toward the center of the view. However, for
  $r = 1.998 M$ the view, that was already significantly shrunk toward
  the middle, is so much shrunk again whereas is the same time the
  rear view completely unfolds and encompasses the front view, so that
  one may think that there is only a minor distortion at the very
  center of the front view. }
\label{fig_IV_last_img}
\end{center}
\end{figure}

The amount of deflecting experienced by a null geodesic between the
singularity and an observer at some $r_\OBS < 2 M$ is given by the
formula
\begin{equation}
\Delta \varphi = \int_0^{r_\OBS} 
\frac{L \ddd r}
     {r^2 \sqrt{E^2 + \left(\frac{2 M}{r} - 1 \right) 
                      \frac{L^2}{r^2}}} , 
\end{equation}
where $E$ and $L$ are the null geodesics associated constants of
motion. Setting $u = r / M$ and defining the dimensionless impact
parameter $\bar b = L / E M$, we have
\begin{equation}
\Delta \varphi = {\rm sgn}(L)\int_0^{u_\OBS} 
\frac{\ddd u}{\sqrt{\frac{u^4}{\bar b^2} + 2 u - u^2 }}  . 
\end{equation}
When $\bar b = 0$, i.e., a radial trajectory (or what will become so
after exiting the horizon, the deflection is 0, and the absolute value
of the deflection grows as $\bar b$ grows since the denominator in the
integral is a decreasing function of $\bar b$. For an infinite value
of $\bar b$, which corresponds to a trajectory whose
$\ddd t / \ddd p = 0$, the above equation reduces to
\begin{equation}
\Delta \varphi = {\rm sgn}(L)\int_0^{u_\OBS} \frac{\ddd u}{\sqrt{2 u - u^2 }}  ,
\end{equation}
whose solution is
\begin{equation}
\Delta \varphi
 = {\rm sgn}(L) \left(\arcsin(u_\OBS - 1) + \frac{\pi}{2}\right)  ,
\end{equation}
a value which tends to $\pm \pi$ as $r$ approaches $2 M$. Now, the
relation between the viewing angle $\delta$ and the reduced impact
parameter $\bar b$ is, given our definitions of $T_{\rm IV}^\mu$ and
$R_{\rm IV}^\mu$ in Eq.~(\ref{TR_IV}), given by
\begin{equation}
\bar b = \frac{L}{E M}
       = \frac{u}{\tan \delta \sqrt{\frac{2}{u} - 1} } .
\end{equation}
For fixed $\tan \delta$, and hence a fixed observing direction, the
associated impact parameter decreases toward zero as $r$ approaches
$2 M$, so that the observer sees geodesics originating from almost the
same $\theta$ and $\varphi$ he/she is situated at, regardless on which
direction he/she is looking at. From a visual point of view, this
translates into the (rather unusual) features:
\begin{enumerate}
\item Very near the singularity ($r \ll 2 M$), the observer sees only
  a small patch of the singularity, something that is intuitive.

\item Far less intuitively, as $r$ grows, a large part of the field of
  view is occupied by neighbouring parts of the singularity, whereas
  two opposite direction ($\pm \partial / \partial t$) show a rapid
  variation of the region of the singularity that is seen. Along these
  two directions, the visual aspect of the angular coordinate grids
  looks like as if we had projected a sphere along a more and more
  elongated funnel.

\item When $r \lesssim 2 M$, almost all directions of observation show
  the same angular part of the singularity, and only two opposite
  directions show all the rest of the angular part of the singularity,
  which is confined within a decreasing angular size.

\end{enumerate} 
Four examples of deviation functions together with the corresponding
views are shown in Figs.~\ref{fig_IV_img},\ref{fig_IV_img}.
\begin{figure}[htbp]
\begin{center}
\includegraphics*[angle=270,width=3.2in]{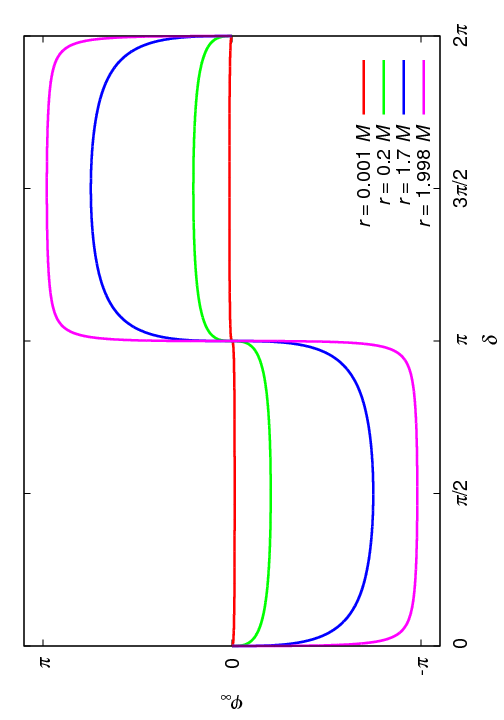}
\caption{Deviation function for region~IV seen from region~IV for four
  values of $r$. As explained in the text the deviation function tends
  to a step function as $r$ approaches $2 M$, which translates into
  the confusing visual aspect of next Figure.}
\label{fig_IV_dev}
\end{center}
\end{figure}
Of course, of choice of the observer's four-velocity $T_{\rm IV}^\mu$
is somewhat questionable because such observer's trajectory is the
only one that does does not exit to horizon, as it directly goes from
region~IV to region~II when when its trajectory reaches its ``apex'',
at $r= 2 M$. One may prefer to consider instead an analog of our
freely-falling observer $T_{\rm I}^\mu$, except that we want that this
new observer originates from the past horizon (and, hence, the past
singularity) instead of heading toward its future counterpart. This
amounts to change the $T_{\rm I}^\mu$ of Eq.~(\ref{tetrad_start}) into
\begin{equation}
T_{\rm I'}^\mu = \left(
\begin{array}{c}
\frac{1}{1 - \frac{2}{u}} \\ + \sqrt{\frac{2}{u}} \\ 0
\end{array}
\right) . 
\end{equation}
The scalar product between $T_{\rm I'}^\mu$ and $T_{\rm IV}^\mu$ gives
\begin{equation}
T_{\rm I'}^\mu T_{{\rm IV}, \mu} = \frac{1}{\sqrt{1 - \frac{u}{2}}} ,
\end{equation}
which means that the two observers have a larger and larger relative
velocity (equal to $\sqrt{u / 2}$). Therefore the views seen by
observer endowed with four-velocity $T_{\rm IV}^\mu$ and
$T_{\rm I'}^\mu$ become more and more different because of aberration
(and also Doppler shift, which we do not show in the views of
region~IV). An example of this is shown in
Figure~\ref{fig_IV_last_img}, to be compared to the bottom row of
Fig.~\ref{fig_IV_img}.

\section{Conclusion}

\label{sec_conc}

In this paper, we have outlined the main steps in order to produce a
correct rendering of relativistic ray tracing in the Schwarzschild
metric. Although this problem is not new, we have obtained a very
satisfactory (and, to our knowledge, new) way to simulate correctly
the rendering of stars without any significant increase of the CPU
time. This was made possible thanks to the spherically symmetric
character of the Schwarzschild metric. This method does not rely on
the specific projection scheme one adopts for the viewing screen and is
therefore very well adapted for digital planetarium hemispheric
projection.

One major drawback of the techniques presented here is that they are
still computationally intensive and do not allow real-time rendering
of the metric since even at moderate resolutions, the CPU time is of
several dozens of seconds. Real-time rendering thus necessitates to
reduce CPU time by a factor greater than $10^3$, which is a rather
ambitious goal, the attainability of which will be presented in a
future work.

Using these techniques, we performed a thorough exploration of the
Schwarzschild metric. Many results were already known, but we found
several novel effects when exploring the visual aspect beyond the
horizon, and more specifically when considering the maximal analytic
extension of the metric, i.e., the Kruskal-Szekeres extension. In
particular, we found that when one visualizes the other asymptotic
region, it appears at first infinitely blueshifted when the observer
crosses the horizon and starts seeing it, but further, it is
increasingly redshifted, except on its edges which are
blueshifted. Also, we studied the case of an observer originating from
the past singularity of the metric and who witnesses very unusual
visual effects, assuming that radiation originates also from the
singularity.

A natural follow-up of this work is to implement other spherically
symmetric metric, the most natural of which being the
Reissner-Nordstrom one, some result of which will be presented
elsewhere. Unfortunately, out fast ray tracing method is less suited
for the Kerr metric or any other non spherically symmetric metric such
as the Papapetrou-Majumdar one. Some results regarding these metrics
will also be presented elsewhere.

\section*{acknowledgments}

A.R. thanks Y.~Zolnierowski, G.~Esposito-Far\`ese, G.~Faye, B.~Fort,
S.~Prunet, \'E~Hivon, S.~Hocevar, J.~Weeks, D.~Monniaux and
M.~Bocquien for useful discussions during the implementation of the
numerical code that was used for these simulations, as well as
B.~Crowell for careful proofreading of part of this manuscript.

\appendix

\section{On the structure of null geodesics in the Schwarzschild
  metric}
\label{app_geod}

We briefly recall here the different types of null geodesics in the
Schwarzschild metric as well as the parameters that allow to
distinguish them.

A null geodesics described by wavevector $k^\mu$ is characterized by
the trivial equation
\begin{equation}
g_{\mu\nu} k^\mu k^\nu = 0 .
\end{equation}
Using the constants of motion $E$ and $L^2$ defined in
Eqns~(\ref{def_E},\ref{def_L2}), this can be rewritten
\begin{equation}
\label{rdot2}
E^2 - \dot r^2 = \frac{L^2}{r^2} \left(1 - \frac{2 M}{r} \right) ,
\end{equation}
where for simplicity $\dot r$ denotes the $r$ component of the
wavevector.

Such an equation can formally be seen as describing the motion of a one
dimensional particle along coordinate $r$, with a total energy $E^2 /
2$ and subject to a potential $V(r)$ given by
\begin{equation}
\label{def_pot}
V(r) = \frac{1}{2} \frac{L^2}{r^2} \left(1 - \frac{2 M}{r} \right) .
\end{equation}
This potential goes to $0$ at infinity and tends to minus infinity
when $r$ goes to $0$. It has a unique maximum at $r = 3 M$ (regardless
of the value of $L^2$) whose value is $V_\MAX = L^2 / 54 M^2$. 

Consequently, 
\begin{enumerate}

\item If $E^2 / 2 > V_\MAX$, or, equivalently $|L / E| < 3 \sqrt{3} M$,
  then the geodesic does not have any turning points along the $r$
  coordinate and therefore its endpoints are $r = 0$ and $r =
  \infty$. It originates from infinity if and only if $\dot r < 0$.

\item If $E^2 / 2 < V_\MAX$ then the geodesics has its both endpoints
  either at $r = 0$ or at $r = \infty$. In the first case, it always
  lies within the $[0, r_\MAX < 3 M]$ region and in the second case it
  always lies in the $[r_\MIN > 3 M, \infty[$ region. The values of
  $r_\MAX$ and $r_\MIN$ are computed by finding the positive roots of
  the third degree polynomial equation $E^2 / 2 = V(r)$, the solution
  of which can be expressed in term of moderately simple elementary
  functions which do not matter here since if one has $E^2 / 2 <
  V_\MAX$) a geodesic originates from infinity if and only if one lies
  at $r > 3 M$.

\item The critical case $|L / E| = 3 \sqrt{3} M$ corresponds to
  geodesics which are stuck at $r = 3 M$ or which indefinitely spiral
  toward this value, either originating from $r= 0$ of from $r =
  \infty$. In practice such geodesics never have to be taken care of
  since numerical round-off error prevent from having this exact value
  of the $L / E$ ratio even if one tries to.

\end{enumerate}
Moreover, the ratio $|L / E|$ can be interpreted as some apparent
impact parameter for the geodesic~\cite{wald84}. 

From the shape of the effective potential, it is easy to understand
that if the observer is at $r > 3 M$, outgoing null geodesics that
intersect his/her worldline can originate from past null infinity only
if their impact parameter is larger than the critical value
$b_\CRIT = 3 \sqrt{3} M$. From the observer's point of view, there
exists therefore a critical angle which separates outgoing geodesics
with $b > b_\CRIT$ and $b < b_\CRIT$. The former originate from past
null infinity whereas the latter originate from the black hole past
horizon. Therefore, the deviation function $\varphi_\infty (\delta)$
is bounded for the $\delta$ corresponding to such outgoing geodesic
with impact parameter larger than the critical value
$b_\CRIT$. Moreover, outgoing null geodesics with an impact parameter
close but above $b_\CRIT$ will experience a large deviation since,
given the effective potential, the value of $\ddd \varphi / \ddd r$ is
given by
\begin{equation}
  \frac{\ddd \varphi}{\ddd r}
  = \frac{1}{r} 
    \frac{1}{\sqrt{\frac{r^2}{b^2} - \left(1 - \frac{2 M}{r} \right)}} ,
\end{equation}
a value which is very large when $r \sim 3 M$, $b \sim b_\CRIT$,
because then geodesics on the potential hill when its slope is very
small. Consequently, geodesics which correspond to the edge of the
domain of definition of $\varphi_\infty (\delta)$ experience a
diverging deviation as shown in Fig.~\ref{fig_dev}. 

A similar reasoning also applies for an observer at $r < 3 M$ if we
now consider ingoing geodesics with impact parameter close to but
smaller than $b_\CRIT$.

\section{Some images}
\label{app_img}

The methods that we have outlined can now be implemented to compute
high quality images of black holes in various situations, and to
explore the unexpected variety of special and general relativistic
effects that one would visually experience close to a black hole. For
pedagogical purpose, we shall try to isolate each of those effects one
by one.

\subsection{Special relativistic effects only}

In this subsection, we shall consider Minkowski space only. 

Let us assume an observer traveling at a constant velocity
$\boldsymbol{v}$ with respect to a reference observer whose celestial
sphere is well-defined. We note $(t, {\boldsymbol{x}})$ a set of
Cartesian coordinates for the reference observer and $(t',
\boldsymbol{x}')$ those of the second one. Up to some unimportant
constant, the second set of coordinates is expressed in term of the
first one as:
\begin{eqnarray}
t' & = & \gamma(t - \boldsymbol{x} \cdot \boldsymbol{v} ) , \\
\boldsymbol{x'}_\perp & = & \boldsymbol{x}_\perp , \\
\boldsymbol{x'}_\parallel &
 = & \gamma(\boldsymbol{x}_\parallel - \boldsymbol{v} t ) ,
\end{eqnarray}
where $\boldsymbol{x}_\parallel$ and $\boldsymbol{x}_\perp$ represent
the parallel part and the perpendicular part of the spacelike
coordinates with respect to velocity $\boldsymbol{v}$. As well-known,
the last two equations can be rewritten into a single one:
\begin{equation}
\boldsymbol{x}'
 =   \boldsymbol{x} - \gamma \boldsymbol{v} t
   + \frac{\gamma^2}{\gamma + 1} 
     (\boldsymbol{v}\cdot \boldsymbol{x} ) \boldsymbol{v} . 
\end{equation}
The associated Lorentz transform can be seen as the matrix whose
components are $\Lambda^\mu_{\;\nu} = \partial x'^\mu / \partial
x^\nu$, which here can be written (see Eq.~(\ref{Lambda_cov}))

\begin{equation}
\Lambda^\mu_{\;\nu} = \left( 
\begin{array}{cc}
\gamma & - \gamma \boldsymbol{v} \\
- \gamma \boldsymbol{v} & {\rm Id} 
+ \frac{\gamma^2}{\gamma + 1} \boldsymbol{v} \otimes \boldsymbol{v}
\end{array}\right) 
\end{equation}
From the point of view of the reference observer, a wavevector $k^\mu$
associated to a direction $\boldsymbol{n}$ has components given by 
\begin{equation}
k^\mu = \omega \left(\begin{array}{c} 
               1 \\ 
               - \boldsymbol{n} \end{array} \right) .
\end{equation}
In the second observer frame, the wavevector has the new coordinates
\begin{eqnarray}
k'^\mu
 & = & \Lambda^\mu_{\;\nu} k^\nu \nonumber \\
 & = & \omega \left(\begin{array}{c} 
                      \gamma (1 + \boldsymbol{v} \cdot \boldsymbol{n}) \\ 
                    - \boldsymbol{n}
                    - \gamma \boldsymbol{v} 
                      \left(1 + \frac{\gamma}{\gamma + 1} 
                                (\boldsymbol{v} \cdot \boldsymbol{n}) 
                      \right) \end{array} \right) \\ 
\nonumber & \equiv &
                \omega' \left( \begin{array}{c} 
                               1 \\
                             - \boldsymbol{n}'\end{array} 
                        \right).
\end{eqnarray}
The direction $\boldsymbol{n}'$ and frequency $\omega'$ of the
corresponding photon seen from the second observer are therefore
\begin{eqnarray}
\label{nprime_aber}
\boldsymbol{n}'
 & = & \frac{  \boldsymbol{n}
             + \gamma \boldsymbol{v} 
               \left(1 + \frac{\gamma}{\gamma + 1} 
                         (\boldsymbol{v} \cdot \boldsymbol{n}) 
               \right)}
            {\gamma (1 + \boldsymbol{v} \cdot \boldsymbol{n})}, \\
\label{omprime_aber}
\omega' & = & \omega \gamma (1 + \boldsymbol{v} \cdot \boldsymbol{n}) .
\end{eqnarray}
Although it is not obvious at first sight, the vector $\boldsymbol{n}'$
has norm $1$ as expected.

\subsubsection{Aberration}

\begin{figure}[htbp]
\begin{center}
\includegraphics*[width=1.6in]{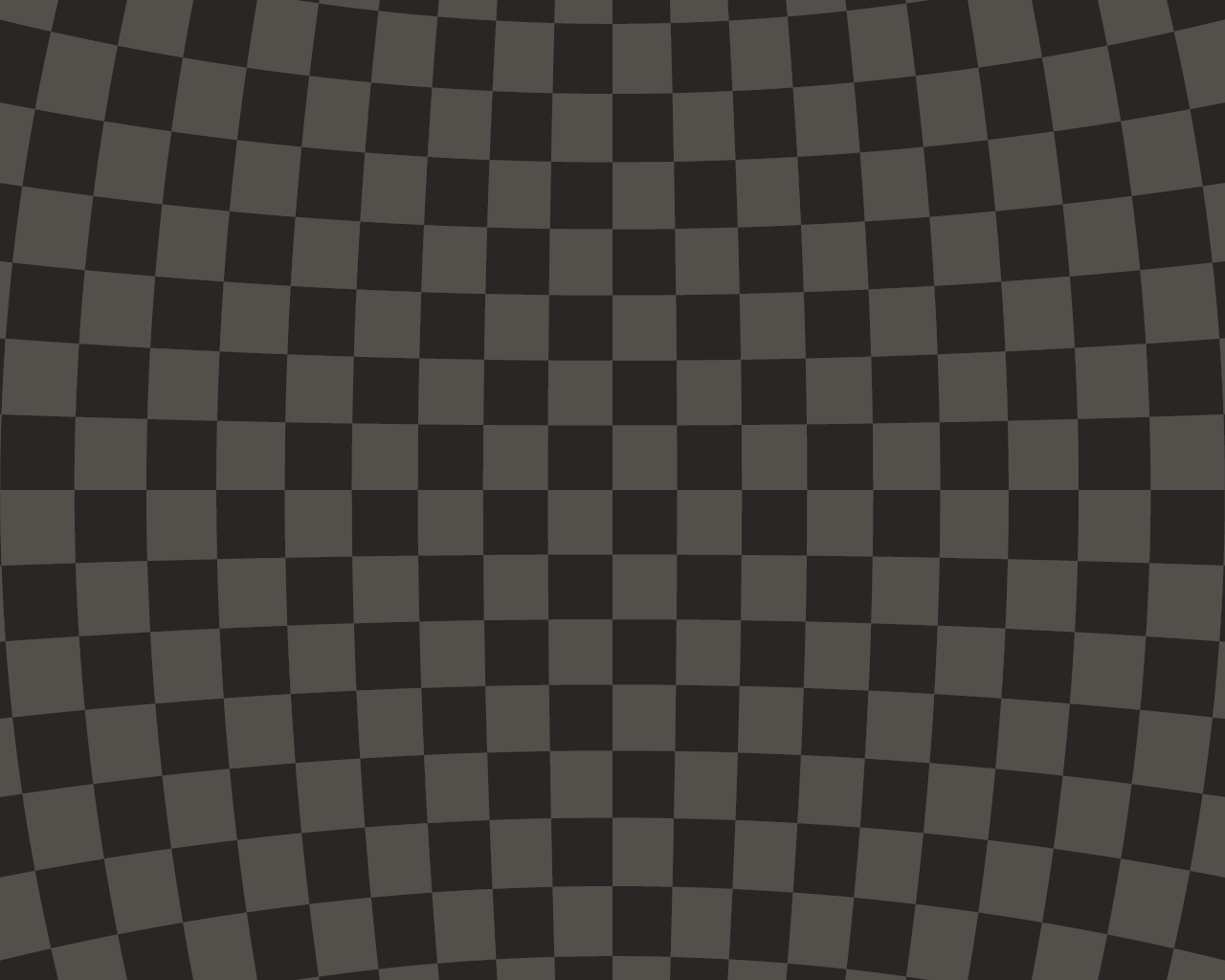}
\includegraphics*[width=1.6in]{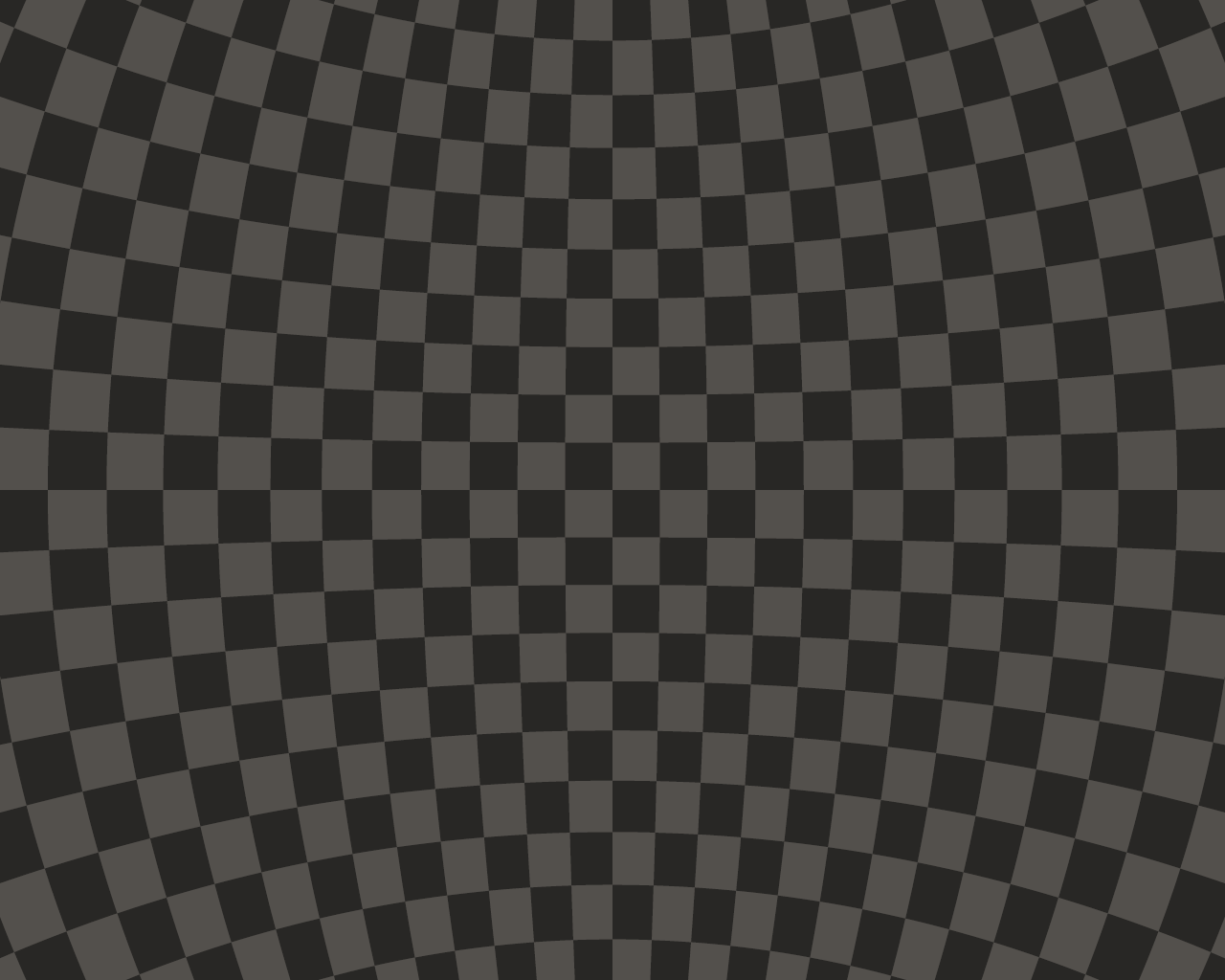}
\vskip 0.12cm
\includegraphics*[width=1.6in]{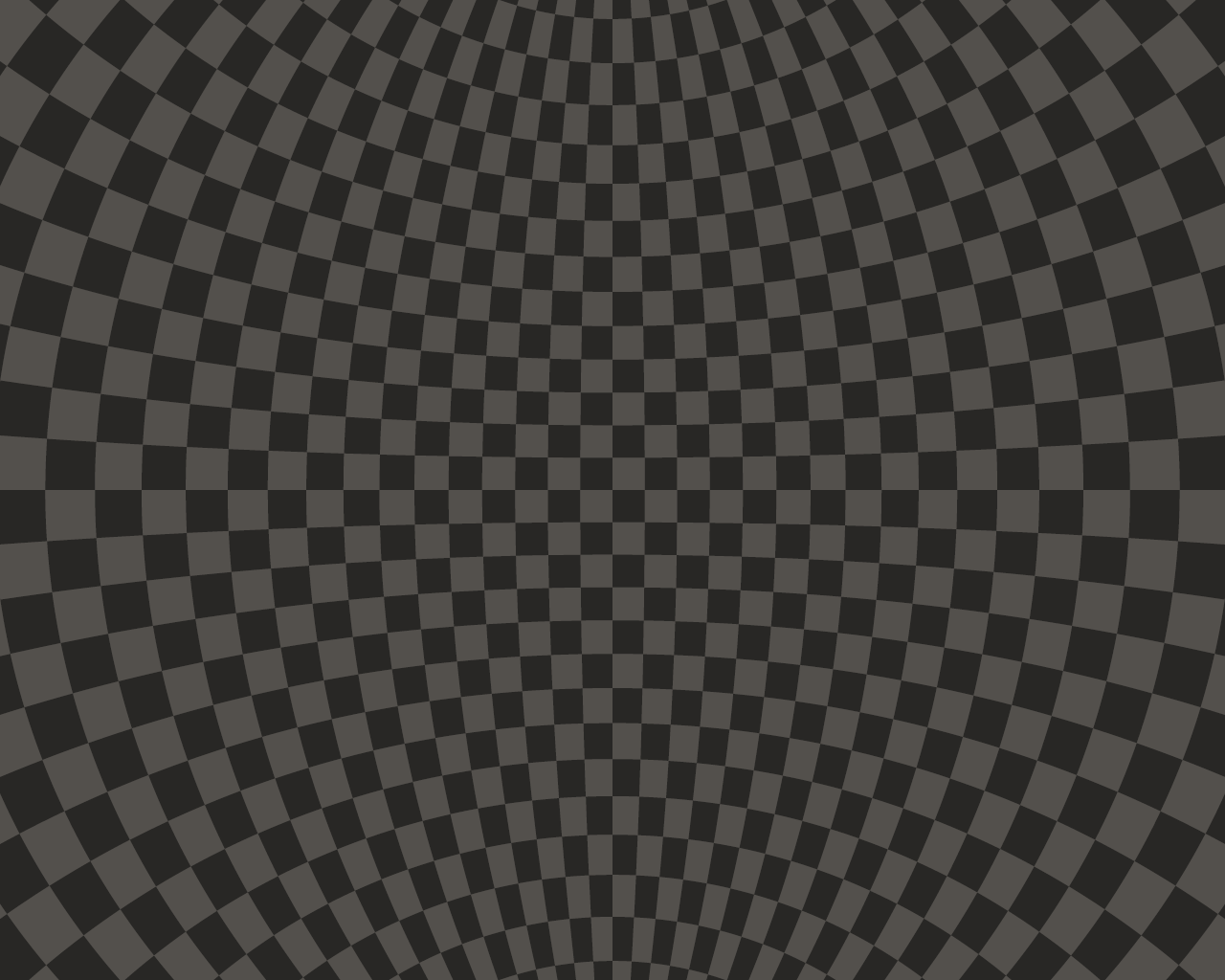}
\includegraphics*[width=1.6in]{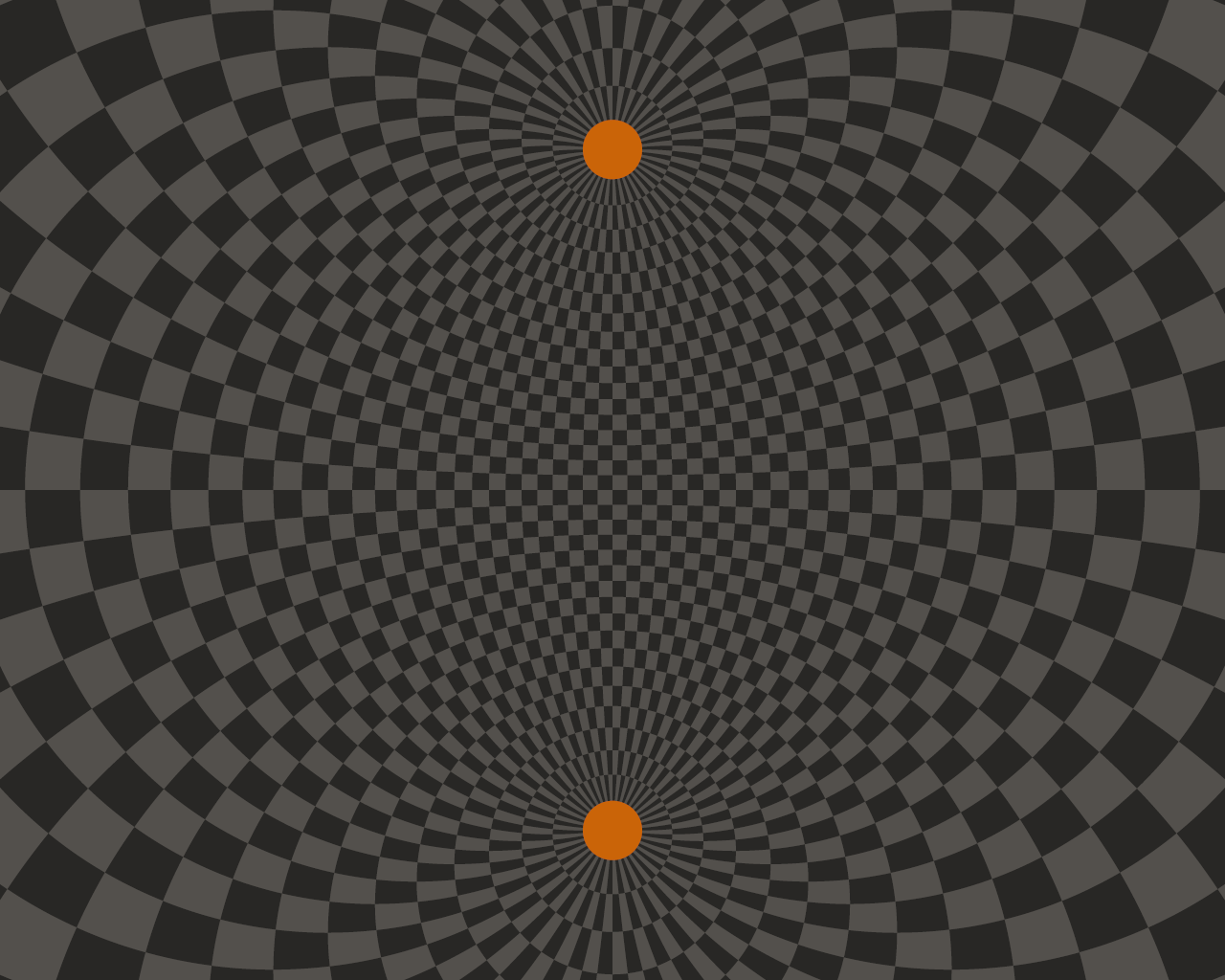}
\caption{Aberration seen by an observer starting from a static
  situation with respect to the celestial sphere (upper left image) to
  $v = 0.3$ (upper right image), $v = 0.6$ (lower left image), and $v
  = 0.9$ (lower right image). In all the pictures the field of view is
  $90$~degrees along the horizontal direction. The ``squares''
  delineating the celestial sphere are $5$~degrees wide both in
  latitude and longitude (see Section~\ref{ssec_bckgsky}).
  Eq.~(\ref{eq_inv_aberr}) can be checked by visual inspection of the
  number of squares along the central horizontal band. The initial
  angular separation between the central part of the vertical edges of
  the images is $90$, $~117.8$, $158.6$, $244.1$~degrees,
  respectively. The two poles of the celestial sphere, which by
  definition are 180~degrees apart for a static observer are now, from
  Eq.~(\ref{eq_aberr}), only $51.7$~degrees apart on the last image. }
\label{fig_aberr1}
\end{center}
\end{figure}

\begin{figure}[htbp]
\begin{center}
\includegraphics*[width=1.6in]{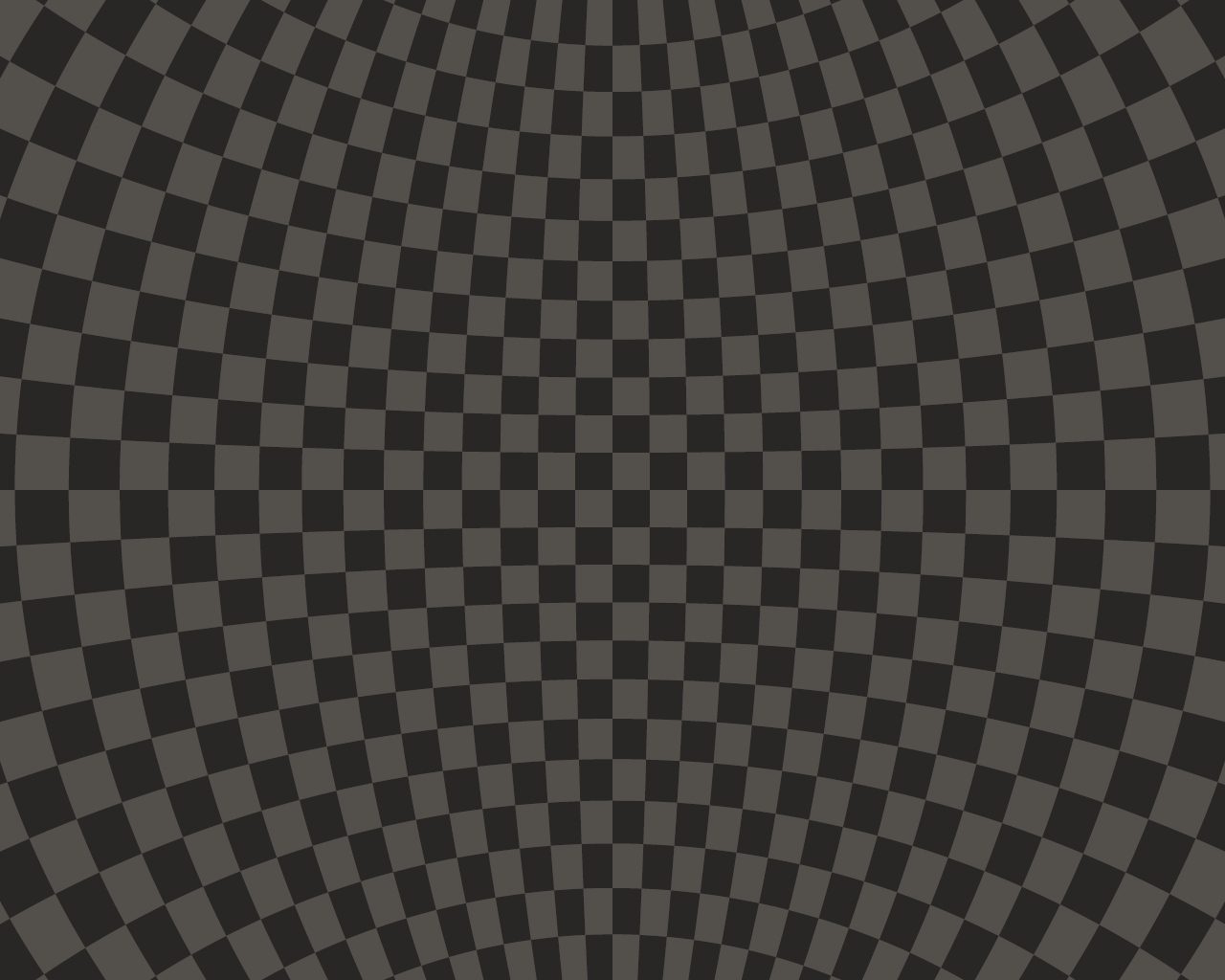}
\includegraphics*[width=1.6in]{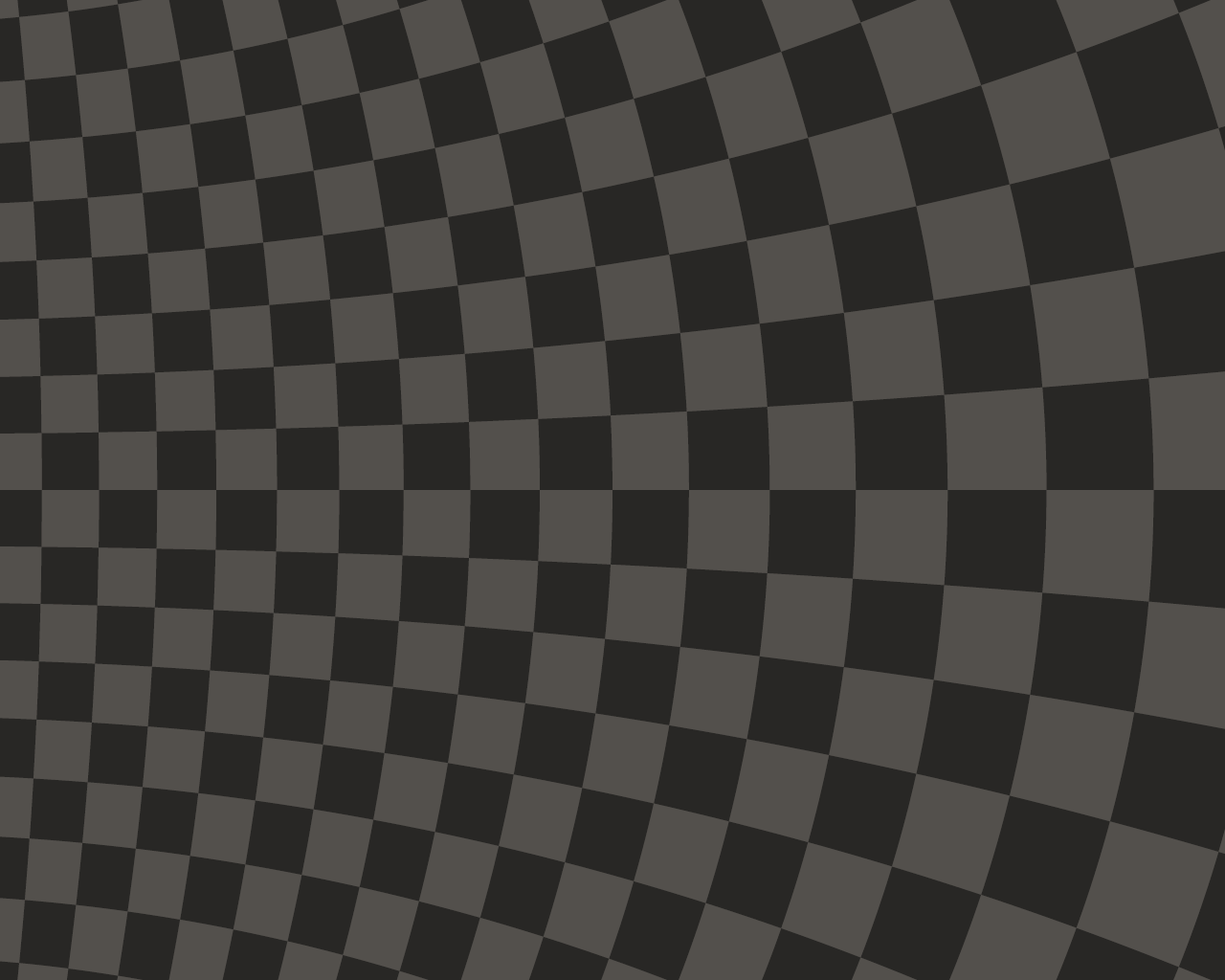}
\vskip 0.12cm
\includegraphics*[width=1.6in]{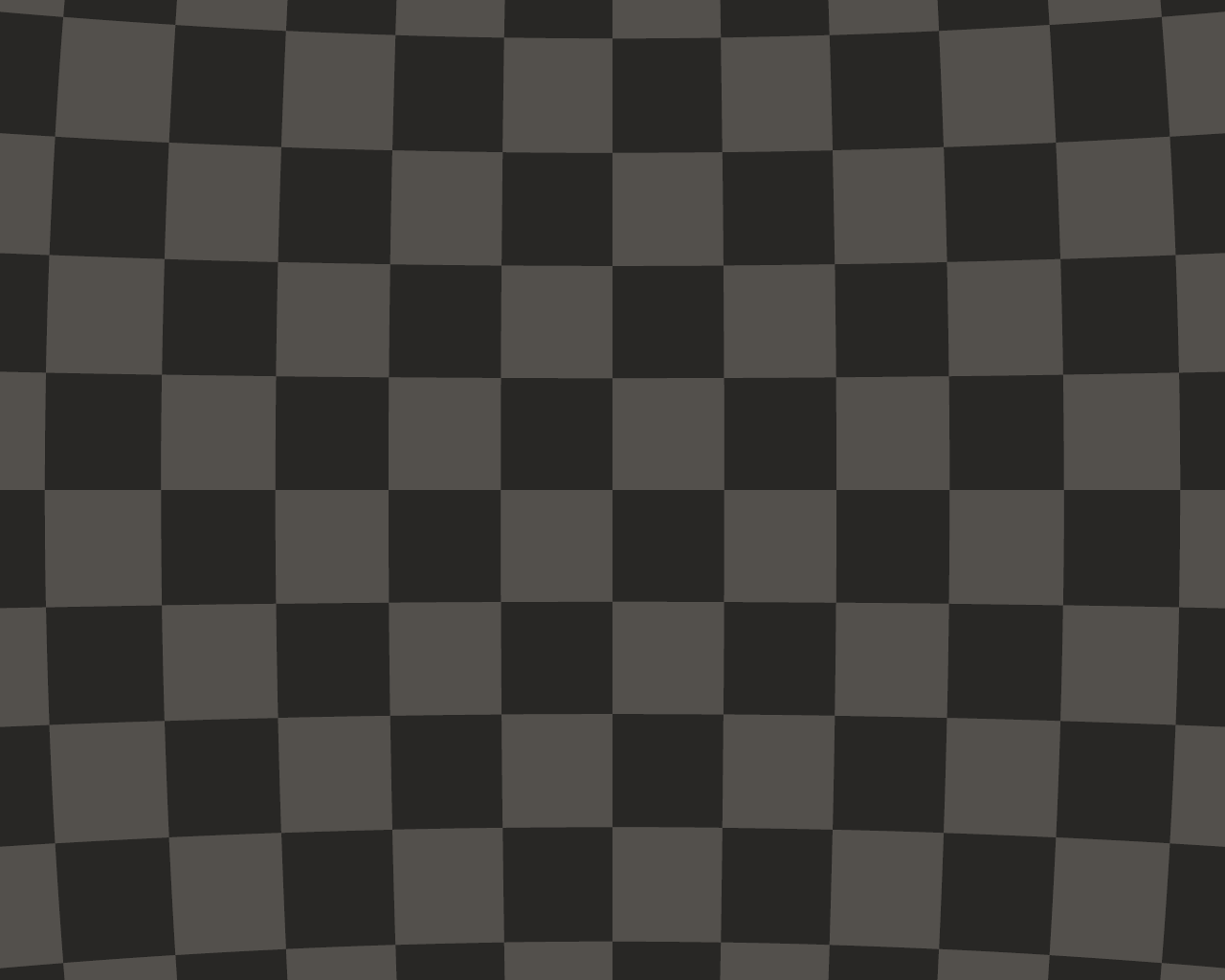}
\includegraphics*[width=1.6in]{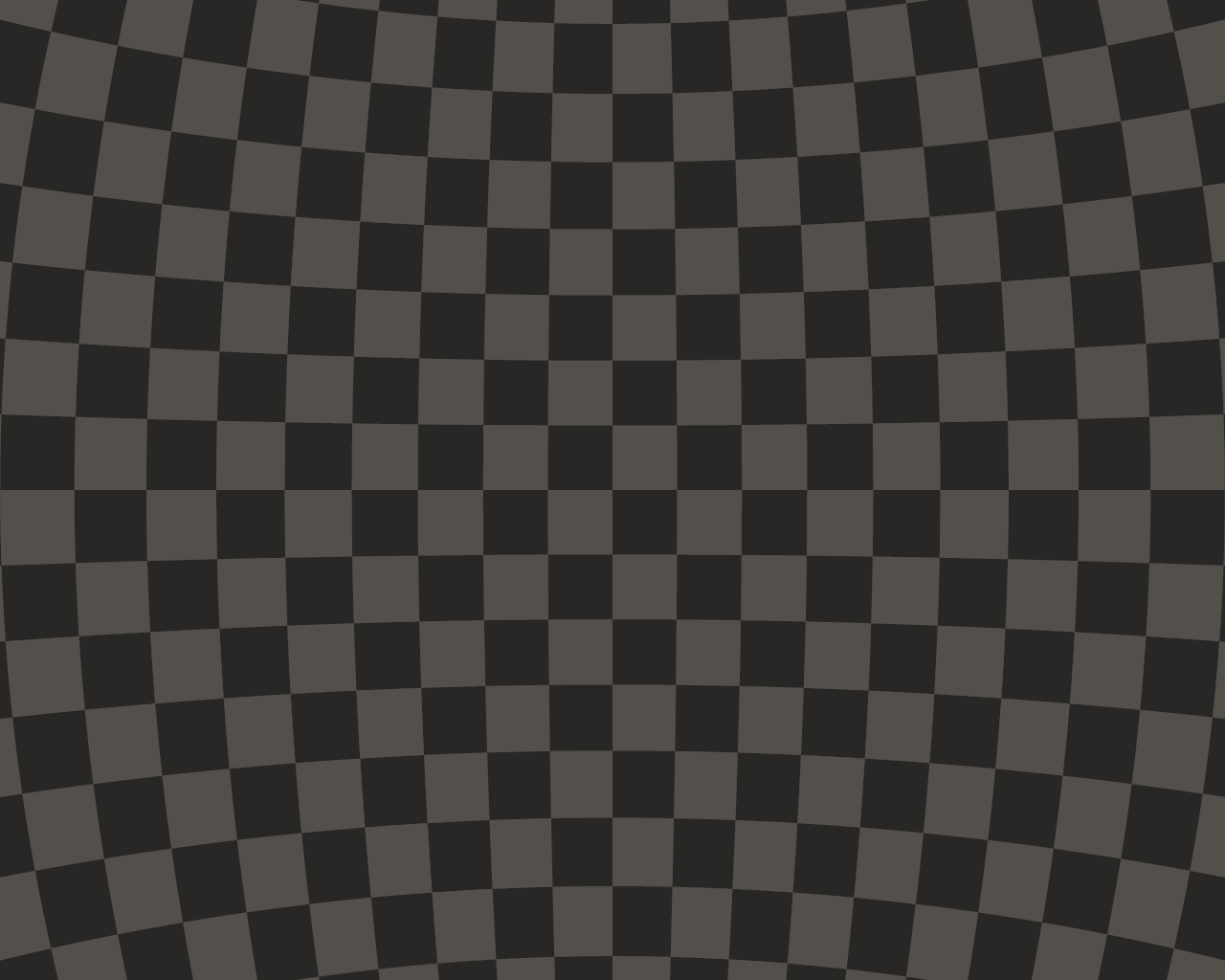}
\caption{Aberration seen by an observer travelling at $v = 0.5$ with
  respect to the frame where the celestial sphere is defined (see
  Sec.~\ref{ssec_bckgsky}). Upper left image shows the front view,
  followed by the right view (upper right) and rear view (lower
  left). Without aberration, these three views should match the static
  one, in the lower right corner.}
\label{fig_aberr2}
\end{center}
\end{figure}

Aberration describes the way the spacelike components of a null vector
are transformed during a Lorentz boost. Its main effect is that the
angle between some star and the direction one is travelling to
diminishes as velocity increases. Starting from
Eq.~(\ref{nprime_aber}) and denoting $\alpha$, $\alpha'$ the angle
between $\boldsymbol{n}$ and $\boldsymbol{v}$ on the one hand, and
$\boldsymbol{n'}$ and $\boldsymbol{v}$ on the other hand, one has
\begin{equation}
\label{eq_aberr}
\cos \alpha' = \frac{\cos \alpha + v}{1 + v \cos \alpha} ,
\end{equation}
so that, as expected,
\begin{equation}
\cos \alpha' - \cos \alpha = \frac{v \sin^2 \alpha}{1 + v \cos \alpha} > 0 .
\end{equation}
The relation between $\alpha$ and $\alpha'$ can be inverted, either by
direct algebra or by noticing that it suffices to change the sign of
$v$ in order to make the inversion make sense, so that
\begin{equation}
\label{eq_inv_aberr}
\cos \alpha = \frac{\cos \alpha' - v}{1 - v \cos \alpha'} .
\end{equation}
A few examples of the aberration effect are shown in
Figures~\ref{fig_aberr1},\ref{fig_aberr2}.

\subsubsection{Doppler}

\begin{figure}[htbp]
\begin{center}
\includegraphics*[width=1.6in]{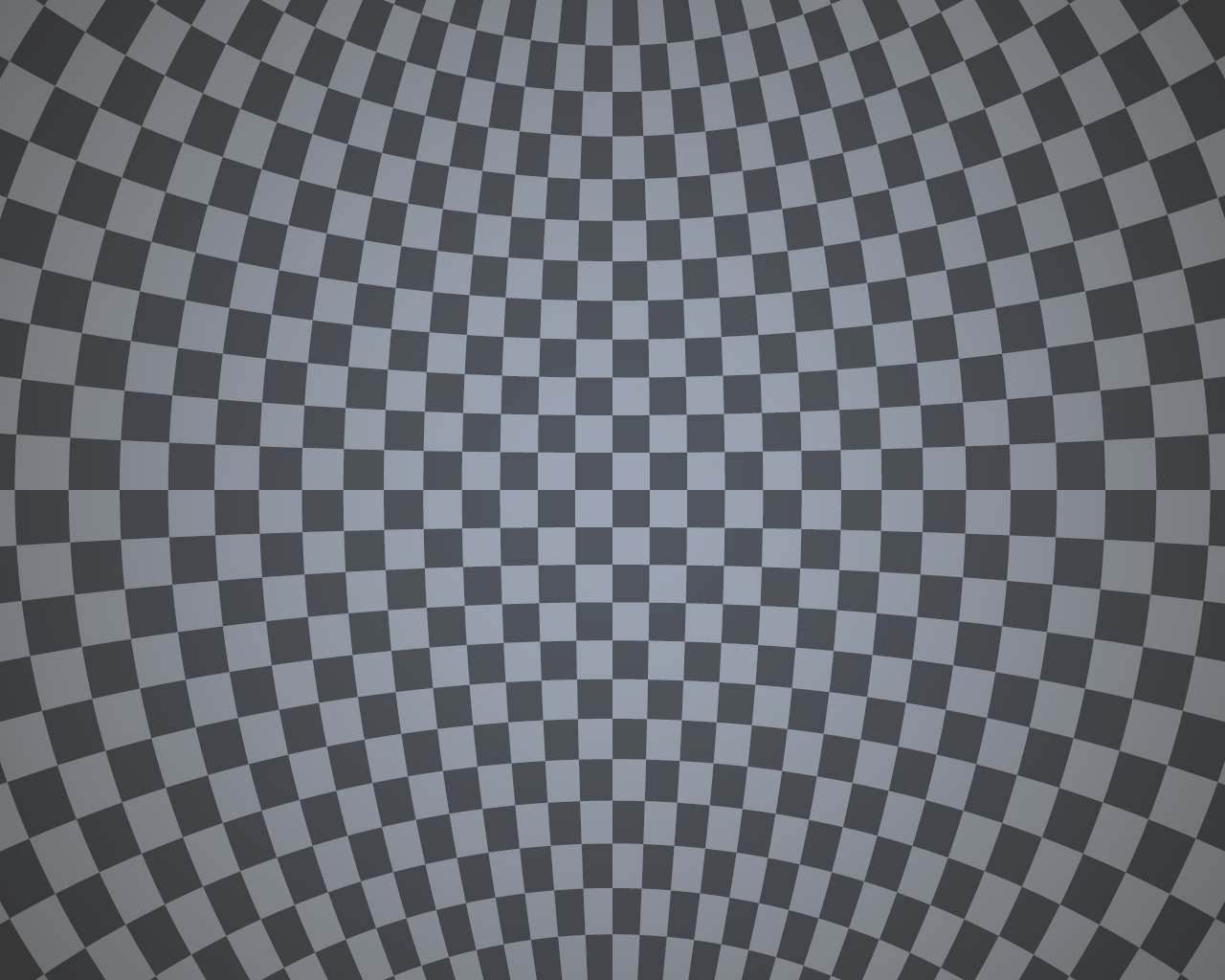}
\includegraphics*[width=1.6in]{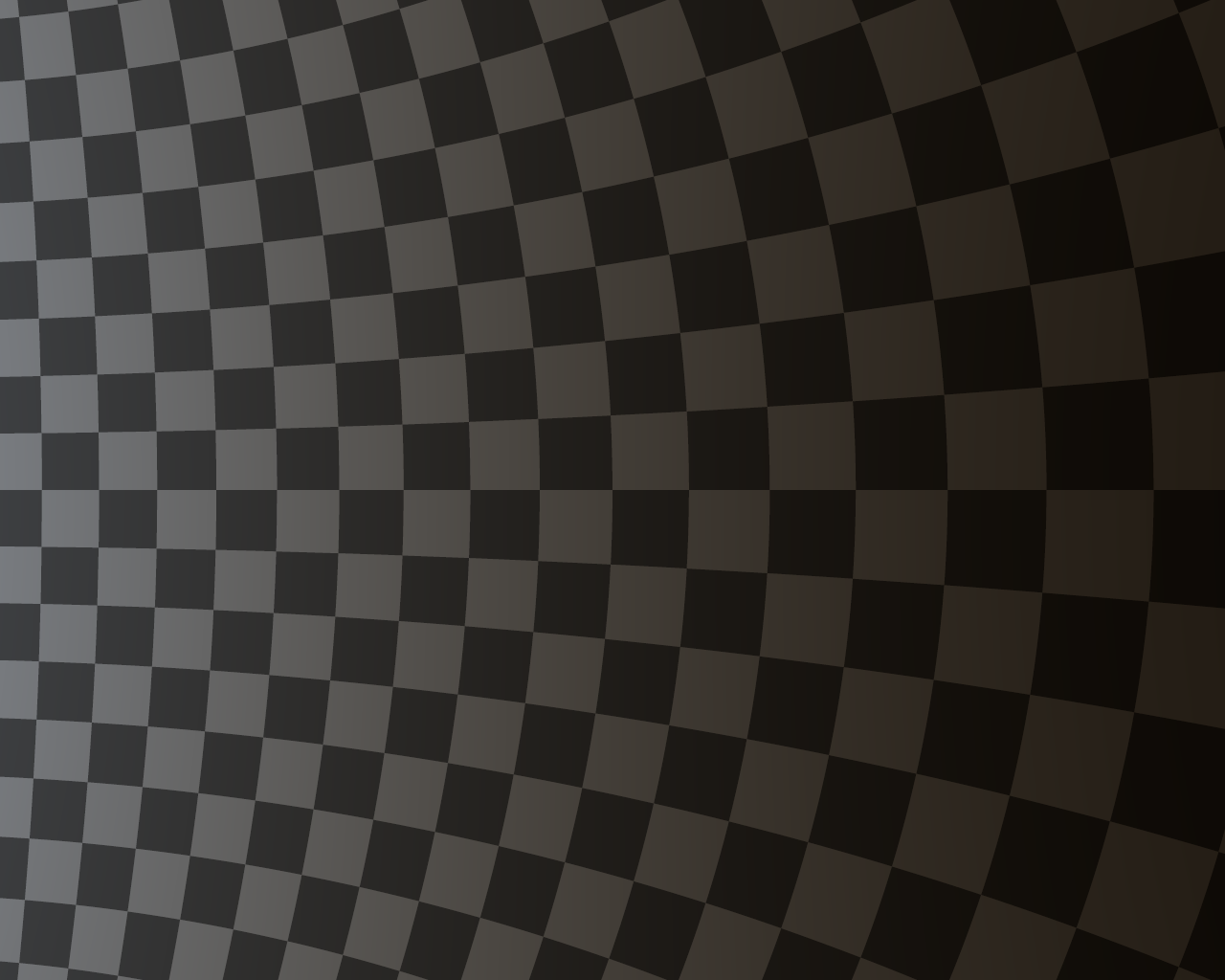}
\vskip 0.12cm
\includegraphics*[width=1.6in]{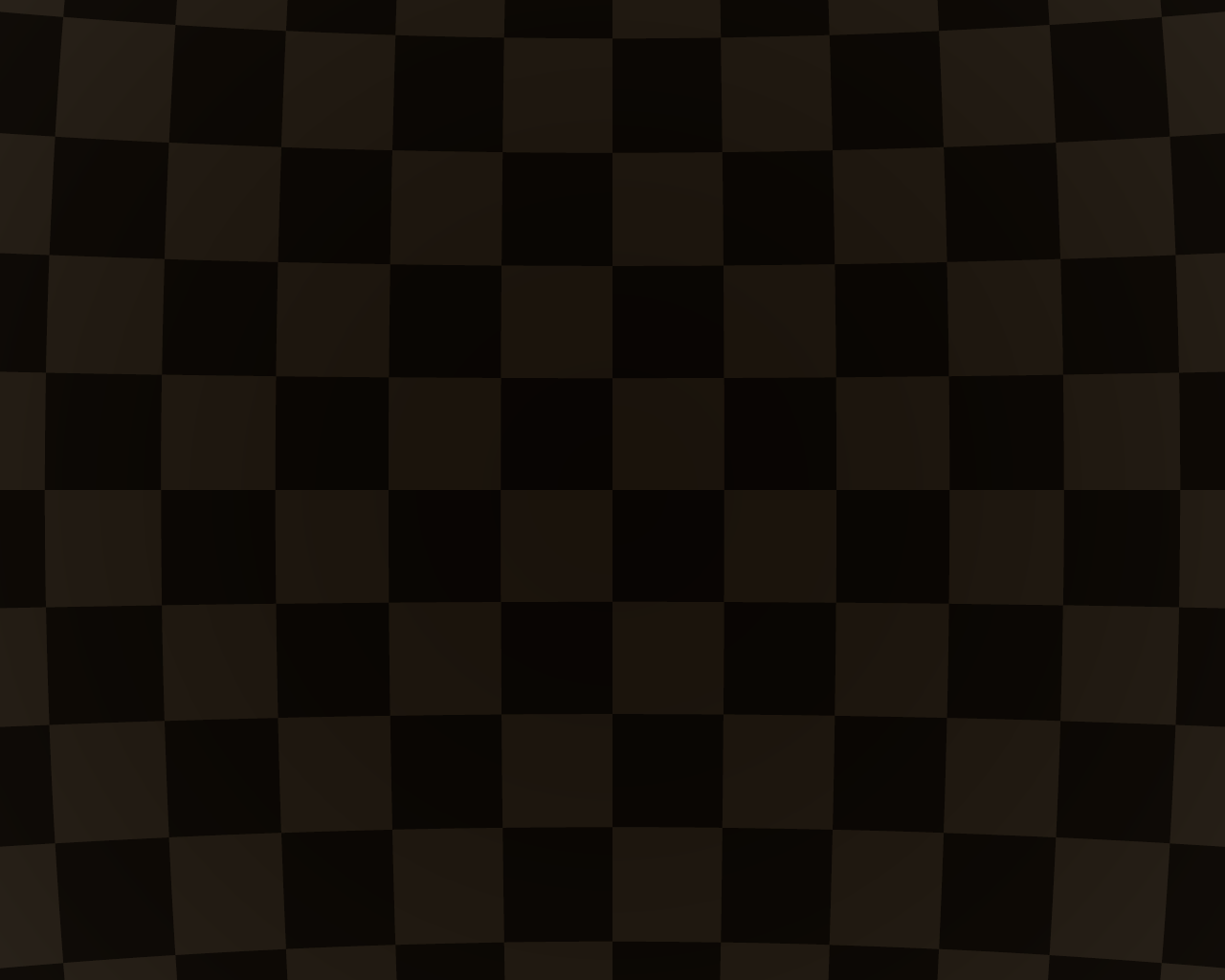}
\includegraphics*[width=1.6in]{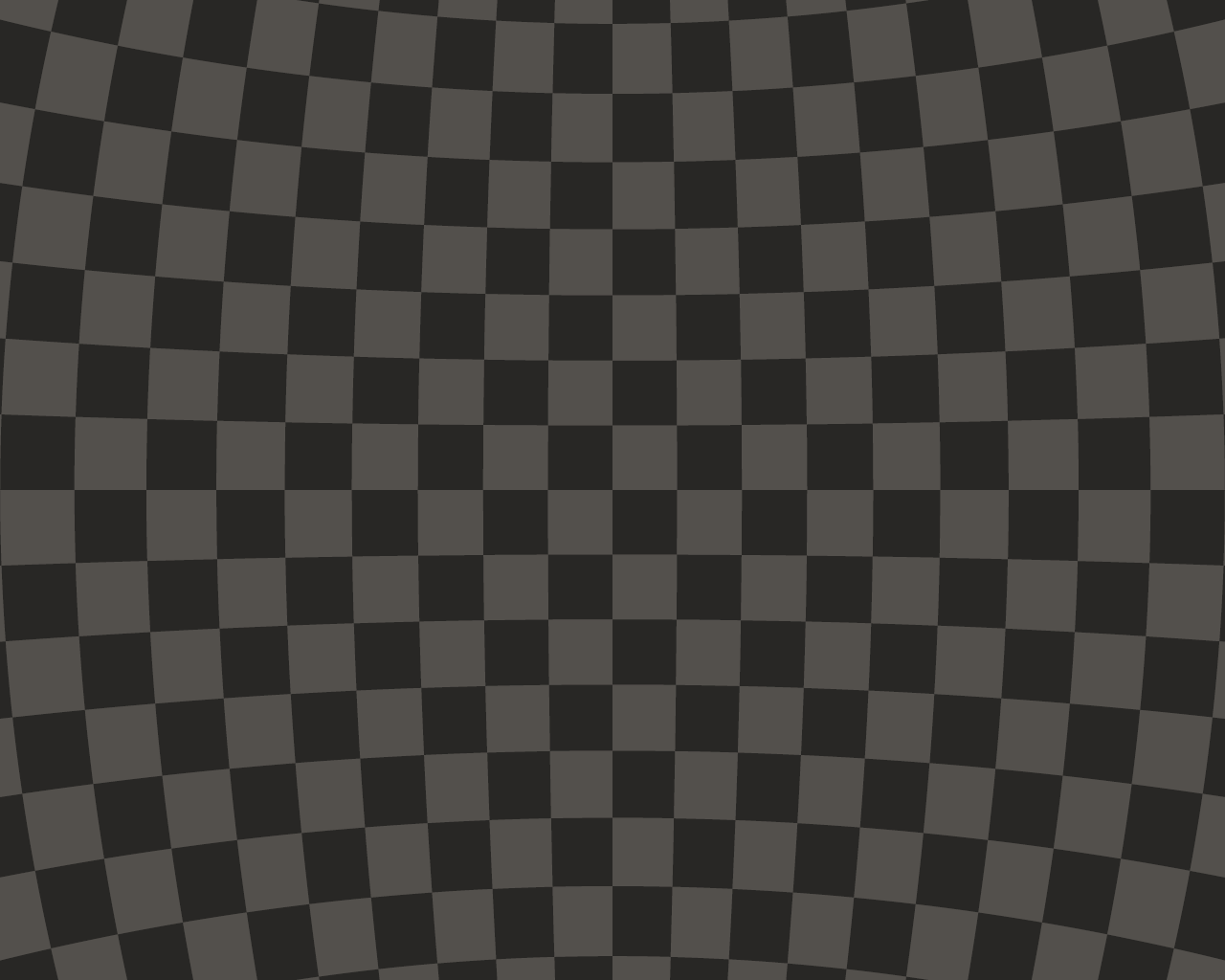}
\caption{Same as Fig.~\ref{fig_aberr2} above, but now including color
  change due to Doppler shift. The bolometric intensity of each pixel
  of the celestial sphere is kept unchanged, so that the intensity
  changes that are seen are due to the variation of sensitivity of the
  eye with respect to the variable temperature, the loss of
  sensitivity being largest in the rear direction, where the red
  spectrum peaks in the infrared domain and has an intensity in the
  visible band that is exponentially suppressed as redshift
  increases.}
\label{fig_dop}
\end{center}
\end{figure}

The frequency shift $\omega'$ given by Eq.~(\ref{omprime_aber}) is
more conveniently written in term of the angle $\alpha'$ as it is the
one that is actually observed. It reads
\begin{equation}
\label{zkin}
\omega' = \frac{\omega}{\gamma (1 - v \cos \alpha')} .
\end{equation}
One recovers the usual result that the frequency shift goes between
$\sqrt{(1 - v) / (1 + v)}$ and $\sqrt{(1 + v) / (1 - v)}$ when going
from the opposite direction to the direction of
$\boldsymbol{v}$. Along the perpendicular direction (i.e., $\cos
\alpha = 0$), the frequency shift is $\gamma^{- 1}$, i.e., there is an
observed redshift. The region along which there is a blueshift if the
one where $\omega' > \omega$, which corresponds to
\begin{equation}
\cos \alpha' > \frac{\gamma v}{\gamma + 1} ,
\end{equation}
which, in term of the angle $\alpha$ corresponds to
\begin{equation}
\cos \alpha > - \frac{\gamma}{\gamma + 1} .
\end{equation}
This means that when the velocity is large, the angular size of the
blueshifted region is increasingly smaller (the lower bound on $\cos
\alpha'$ increases), but corresponds to an initially increasingly
larger patch of the sky seen by the first observer (the lower bound on
$\cos \alpha$ increases). An example of the observed color change due
to the Doppler effect is shown in Fig.~\ref{fig_dop}.

\subsubsection{Intensity}

\begin{figure}[htbp]
\begin{center}
\includegraphics*[width=1.6in]{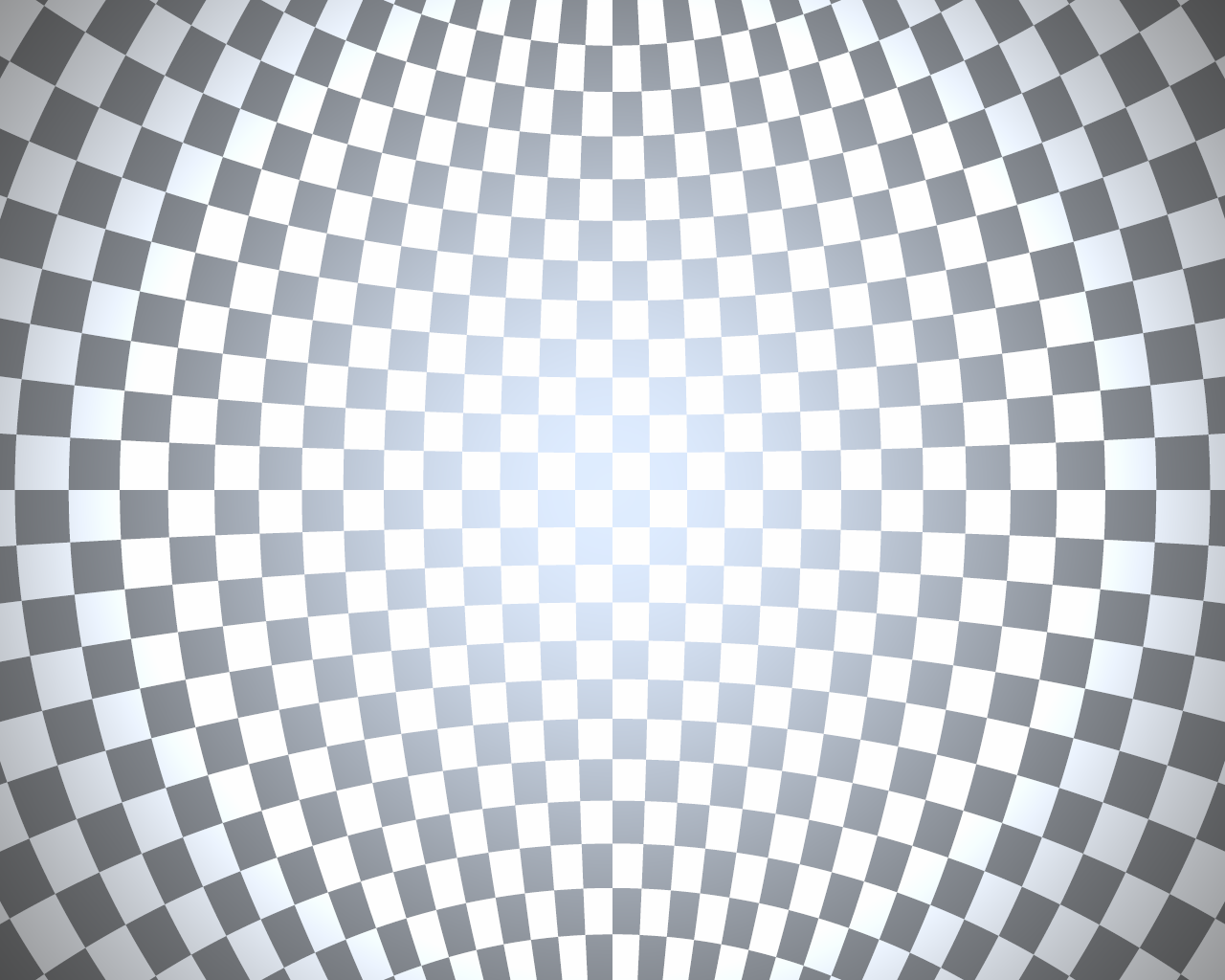}
\includegraphics*[width=1.6in]{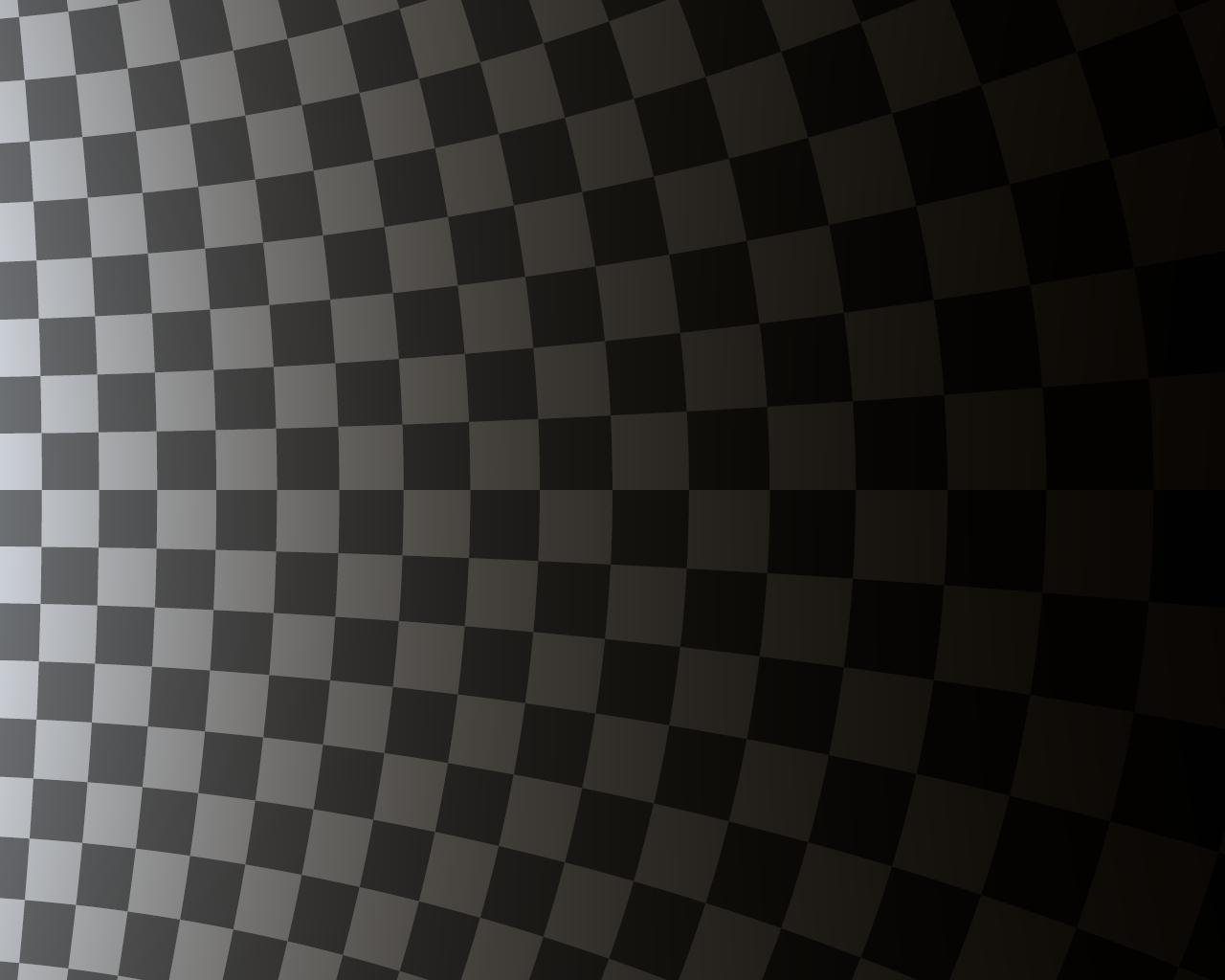}
\vskip 0.12cm
\includegraphics*[width=1.6in]{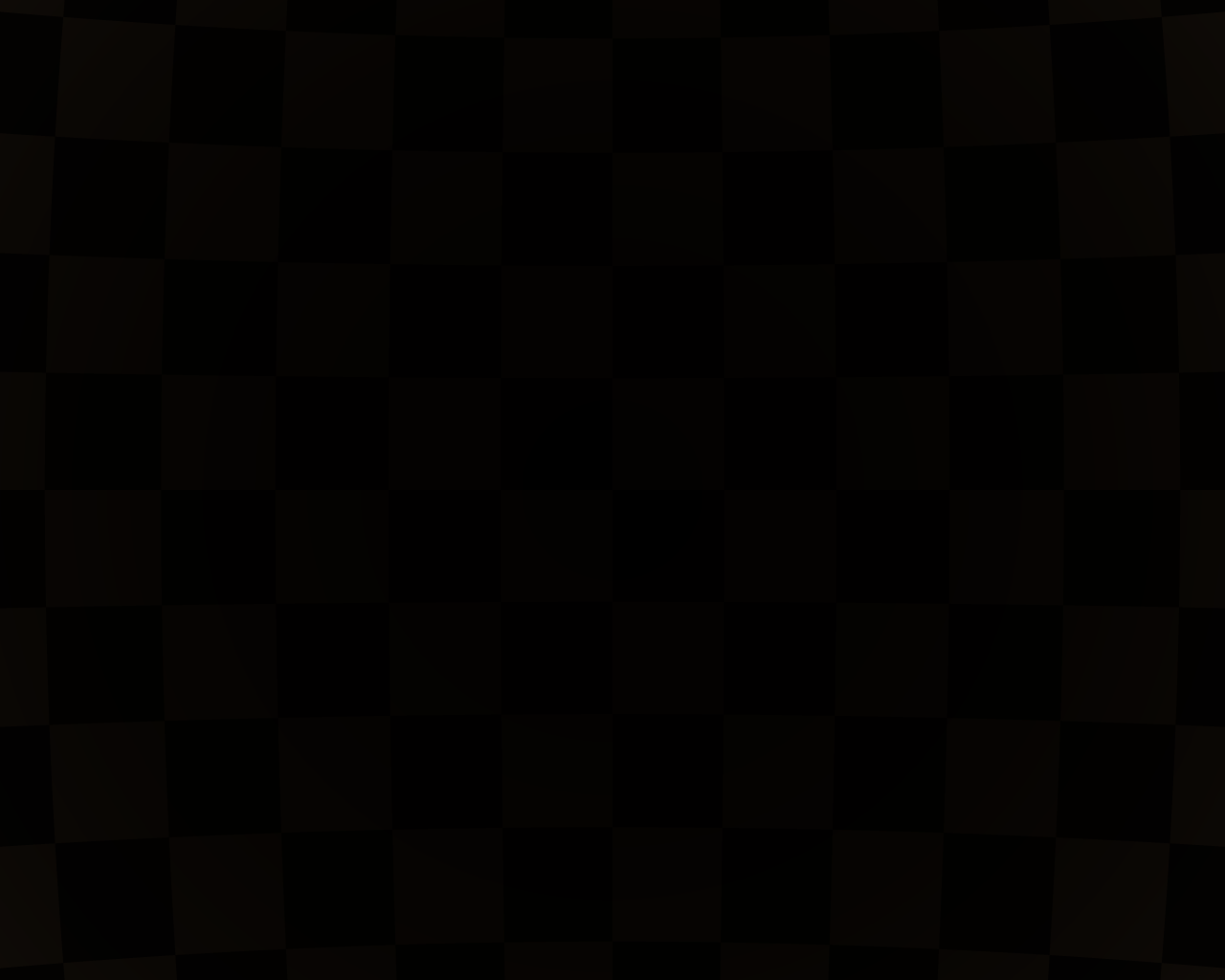}
\includegraphics*[width=1.6in]{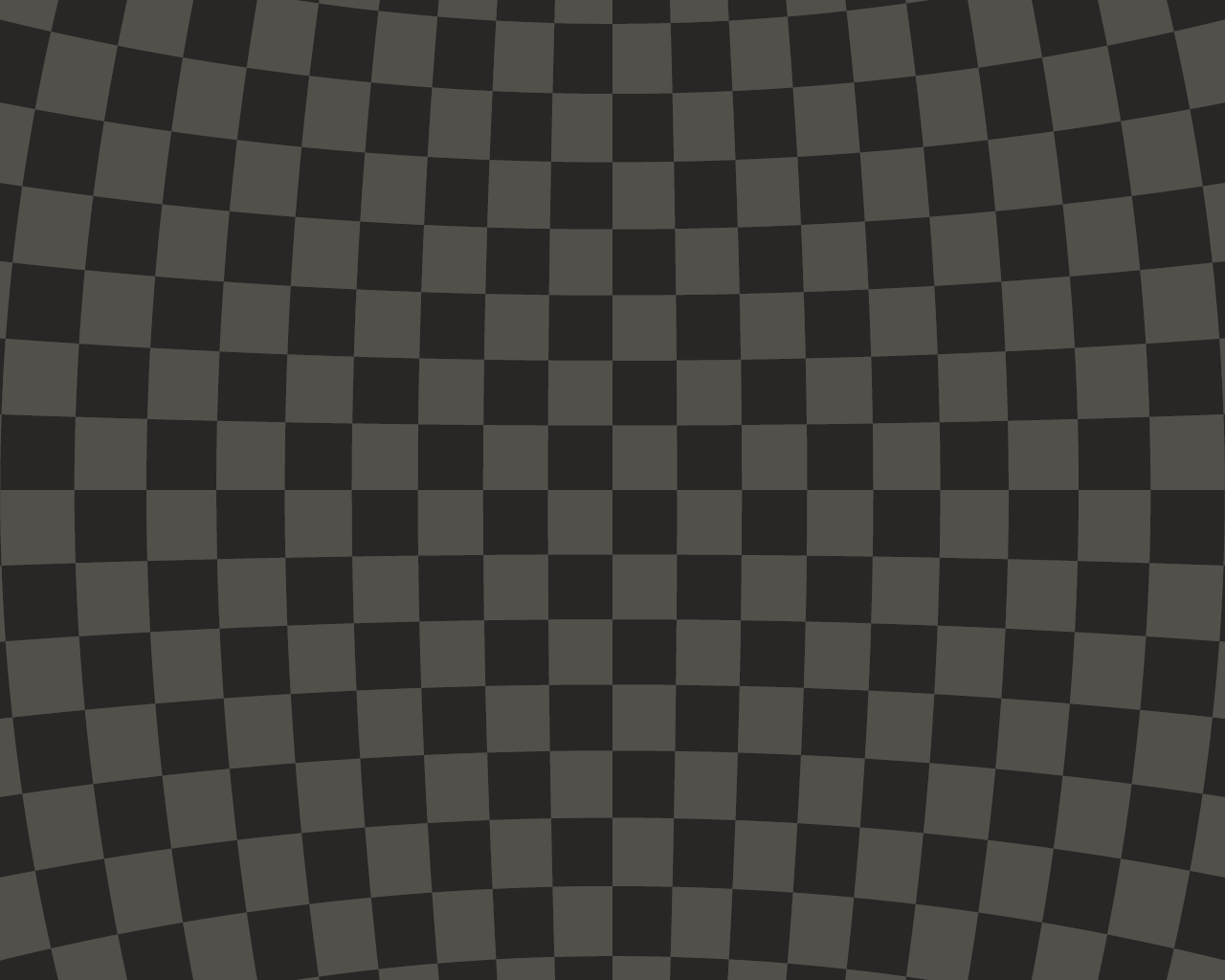}
\caption{Same as Fig.~\ref{fig_aberr2} and~\ref{fig_dop} above, but
  now including everything. The luminosity gradient as one goes from
  the front to rear direction is even larger than previously since, in
  addition to that of the previous figure, it is modulated by an extra
  $(1 + z)^{- 4}$ factor, which here, for $v = 0.5$, varies from $9$
  in the front direction to $1 / 9$ in the rear one.}
\label{fig_int}
\end{center}
\end{figure}

In addition to the frequency shift, the Doppler effect produces a
variation of the overall intensity of a light source. In the case
considered here, where we assume that our celestial sphere (and,
later, the stars) have a black-body emission, the bolometric luminosity
varies as $T^4$, where $T$ corresponds to the temperature of the
celestial sphere pixel or star of interest. Consequently, the
bolometric luminosity is modulated by a factor $(1 + z)^{- 4}$, where
the redshift $z$ is given by 
\begin{equation}
1 + z = \frac{\omega}{\omega'} .
\end{equation}
The intensity variation therefore varies of a factor $[(1 + v) / (1 -
v)]^2$ between the front and rear directions, a factor which can be
quite large for relativistic velocities ($\sim 360$ for $v =
0.9$), which make it difficult to represent on a computer screen since
with sRBG coordinates, the relative intensity between the brightest
pixel and the dimmest one (for a fixed hue) is $12.92 \times 255 =
3294.6$. Note however that such a factor is valid for the bolometric
luminosity only. Taking into account the eye response drastically
change this factor, although it does not improve much the situation: a
highly redshifted object become completely invisible not because its
bolometric luminosity decreases, to the $(1 + z)^{-4}$ factor but
because almost all its energy becomes radiated in the infrared domain
which is not visible at all to a human eye. An example of the
intensity effect is shown in Fig.~\ref{fig_int}.

\subsubsection{Amplification}

When considering our checkerboard-like celestial sphere, the total
intensity of a given square relies on the combination of its
temperature change because of the Doppler effect and its angular size
change because of aberration. This last part is computed implicitly
by the fact that the number of pixels which span the square changes
because of aberration. If we consider a star, then only the Doppler
term is known a priori. However, the star is an extended object,
although a tiny one, just as our checkerboard is. Therefore, we must
add to its intensity change the amplification factor whose computation
is outlined in Section~\ref{ss_ampl}.

\subsection{General relativistic effects only}

\subsubsection{Light bending}

\begin{figure}[htbp]
\begin{center}
\includegraphics*[width=3.2in]{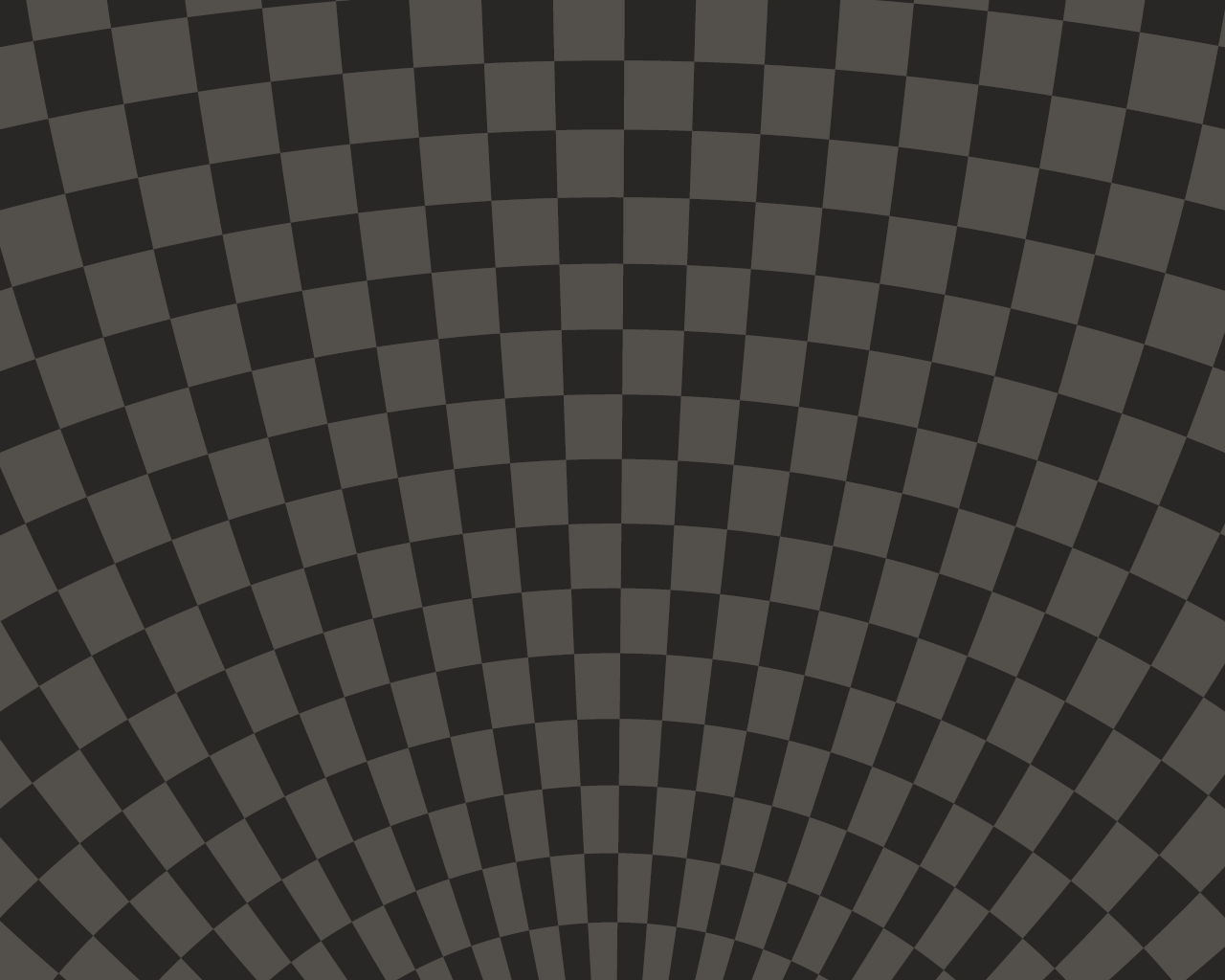}
\vskip 0.12cm
\includegraphics*[width=3.2in]{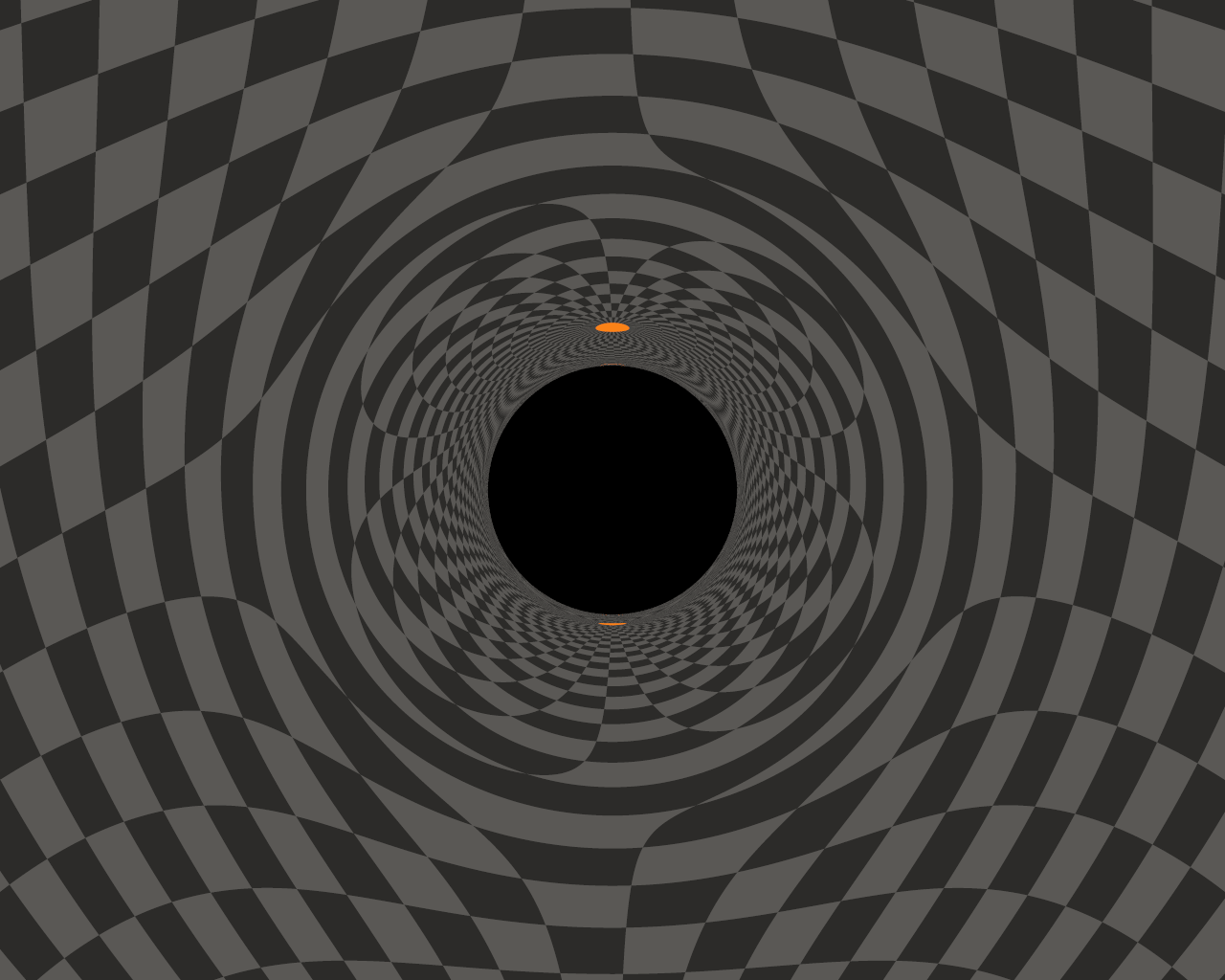}
\caption{Comparison of a view seen by a static observer with and
  without the presence of the black hole. In the latter case, the
  observer lies at coordinate $r = 30 M$. The black hole lies in front
  of the most central dark square of the top picture. This square is
  seen highly distorted surrounding the silhouette of the black hole
  in the second picture. The two poles of the coordinate grid are seen
  as secondary/ghost images between the silhouette and the distorted
  black square, the south pole ghost image appearing above and that of
  the north pole below.  Note that the second image is very slightly
  brighter than the first one because we have included the
  gravitational blueshift, which here takes the value $z = \sqrt{1 - 2
    M / r} - 1 \sim -0.0339$.  However strange it may look like, such
  picture is nothing more than the visual translation of the
  deflection function shown in Fig.~\ref{fig_dev}.}
\label{fig_dev2}
\end{center}
\end{figure}

Let us now add a black hole to our scenery. An example of the same
picture with and without the black hole is shown on
Figure~\ref{fig_dev2}. The observer is looking toward latitude
$-32.4\,{\rm deg}$ with respect to the coordinate grid we use. the
visualization parameters are the same as previously.
From a visual point of view, the central region of the picture, where
we put the black hole, is now scattered all around (and at some
distance) of the black hole silhouette. Therefore, the presence of the
black hole somehow pushes away the background image as compared to the
undistorted case. Between the distorted image of the celestial sphere
lies a series of ghost images which we will discuss below.

\subsubsection{Gravitational blueshift}

\begin{figure}[htbp]
\begin{center}
\includegraphics*[width=3.2in]{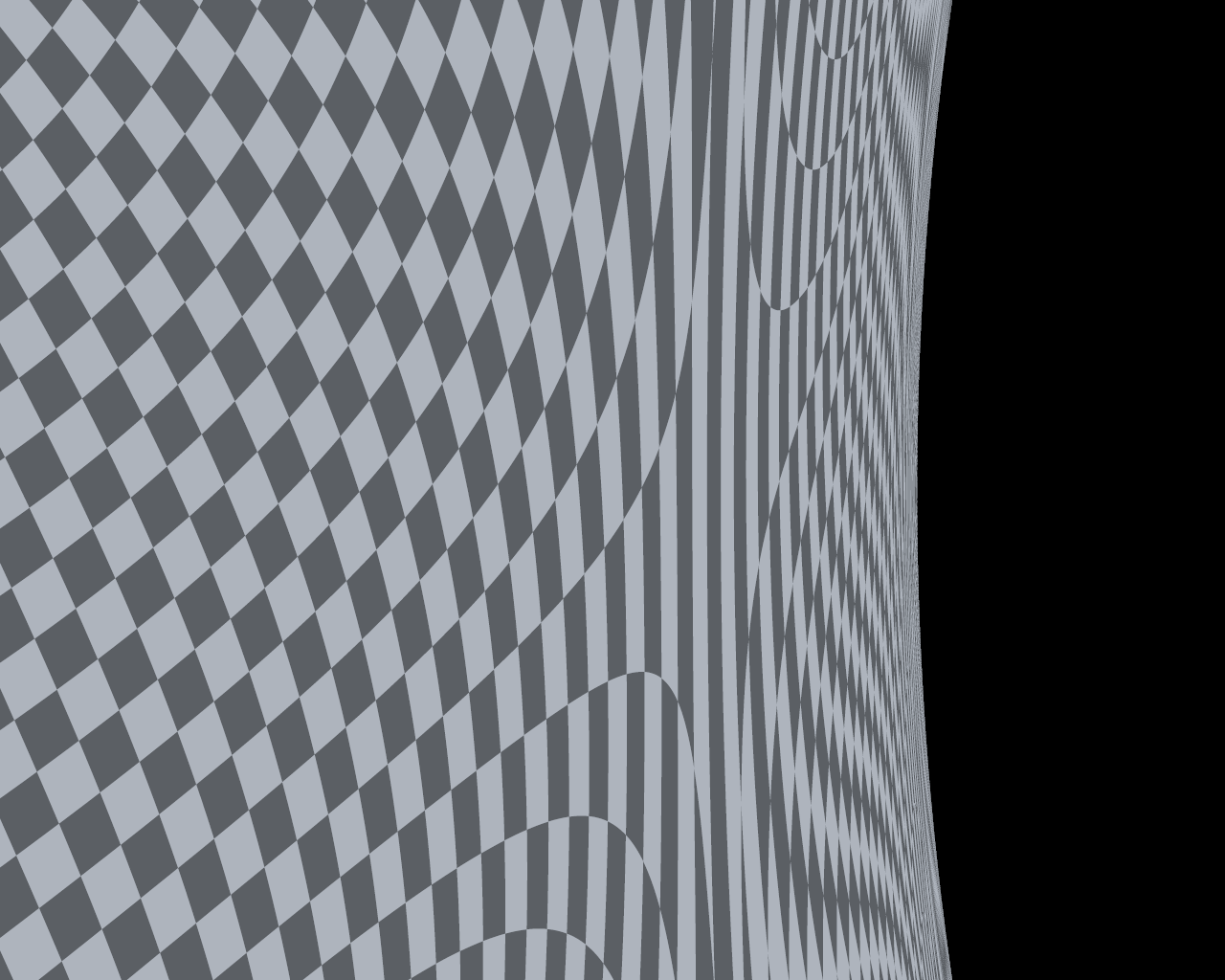}
\vskip 0.12cm
\includegraphics*[width=3.2in]{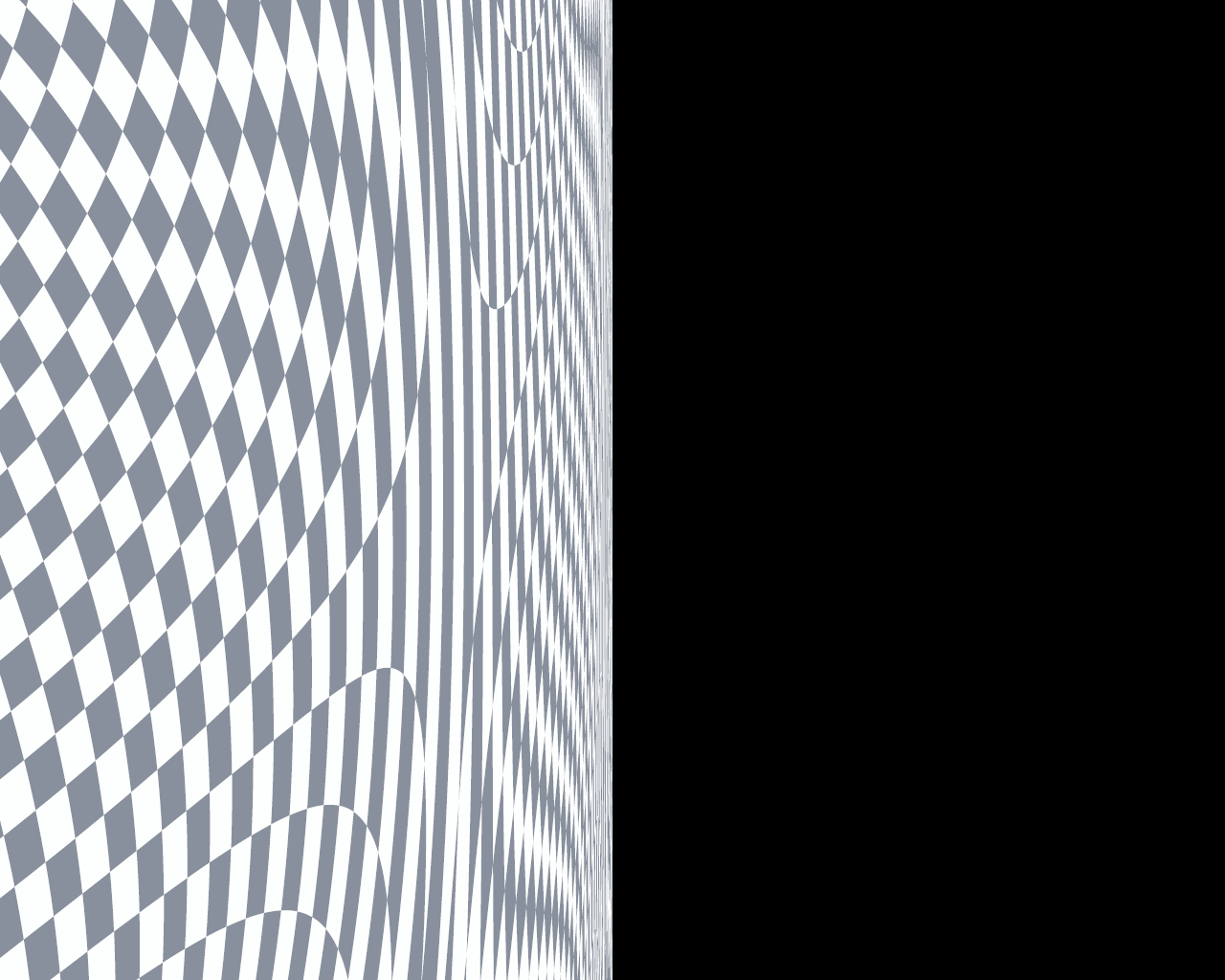}
\vskip 0.12cm
\includegraphics*[width=3.2in]{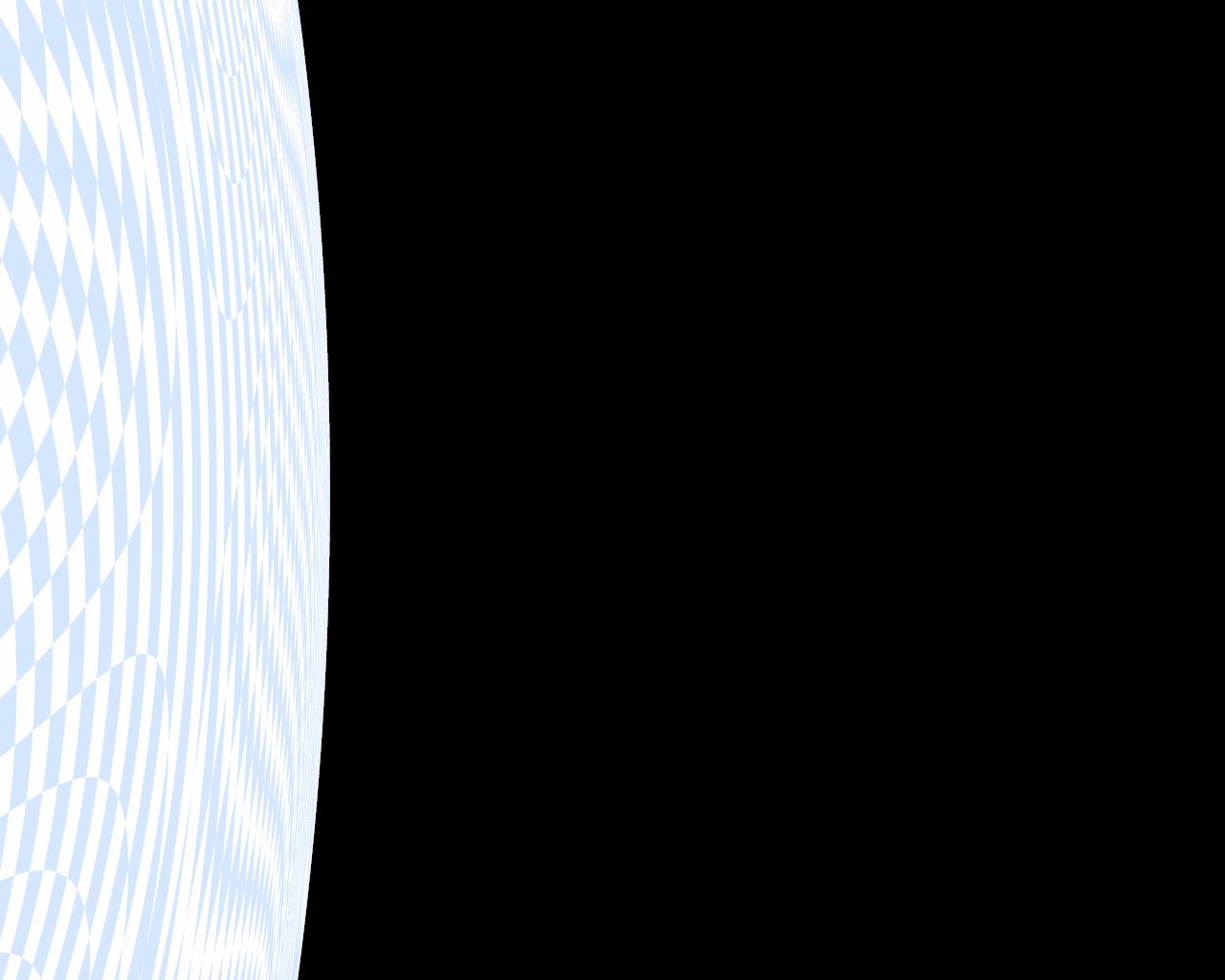}
\caption{Three views of the vicinity of the black hole as seen by a
  static observer that stands at $r = 4 M$, $r = 3 M$, and $r = 2.5 M$
  (from top to bottom). As is well-known, the $r = 3 M$ radial
  coordinate corresponds to that where photons can have (unstable)
  circular orbits around the black hole, so that in practice the black
  hole silhouette spreads over a half sphere for a static
  observer. Below this value of $r$, the silhouette no longer looks
  convex, but concave instead. The increasing brightness as $r$
  decreases is due to the increasing blueshift $z_\GRAV$, which takes
  values $\simeq -0.293$, $\simeq -0.423$ and $\simeq -0.553$,
  respectively. Because of our choice of rendering, highly blueshifted
  stars would have smeared the view of last image, and have therefore
  been removed, and region~I celestial sphere was replaced by a
  coordinate grid which makes more explicit the deformation at the
  edge of the black hole silhouette.}
\label{fig_blue}
\end{center}
\end{figure}

For any geodesic, the quantity $E \equiv \dot t (1 - 2 M / r)$ is
conserved. If we denote by $\omega$ the $\dot t_\gamma$ component of
some null geodesic, which means that the frequency of such photon
measured by an observer at rest and at infinity will be $\omega$, then
the $\dot t_\gamma$ component will be everywhere given by
\begin{equation}
\dot t_\gamma = \frac{\omega}{1 - \frac{2 M}{r}} .
\end{equation}
Now, if we consider a static observer outside the black hole, the only
component of his/her four-velocity will be
$\dot t_\OBS = (1 - 2 M / r)^{-\frac{1}{2}}$. Therefore the frequency
that such observer crossing a null geodesic coming from infinity will
be
\begin{equation}
\omega' = g_{tt} \dot t_\gamma \dot t_\OBS
        = \omega \dot t_\OBS
        = \omega \left(1 - \frac{2 M}{r} \right)^{- \frac{1}{2}} .
\end{equation}
The corresponding blueshift that will be measured is therefore
\begin{equation}
\label{zgrav}
z_\GRAV = - 1 + \frac{\omega}{\omega'} = - 1 + \sqrt{1 - \frac{2 M}{r}} .
\end{equation}
A few pictures on the increasing blueshift (and hence, the increasing
brightness) as a static observer approaches to horizon is shown on
Fig.~\ref{fig_blue}

\subsubsection{Lensing}

\begin{figure}[htbp]
\begin{center}
\includegraphics*[width=3.2in]{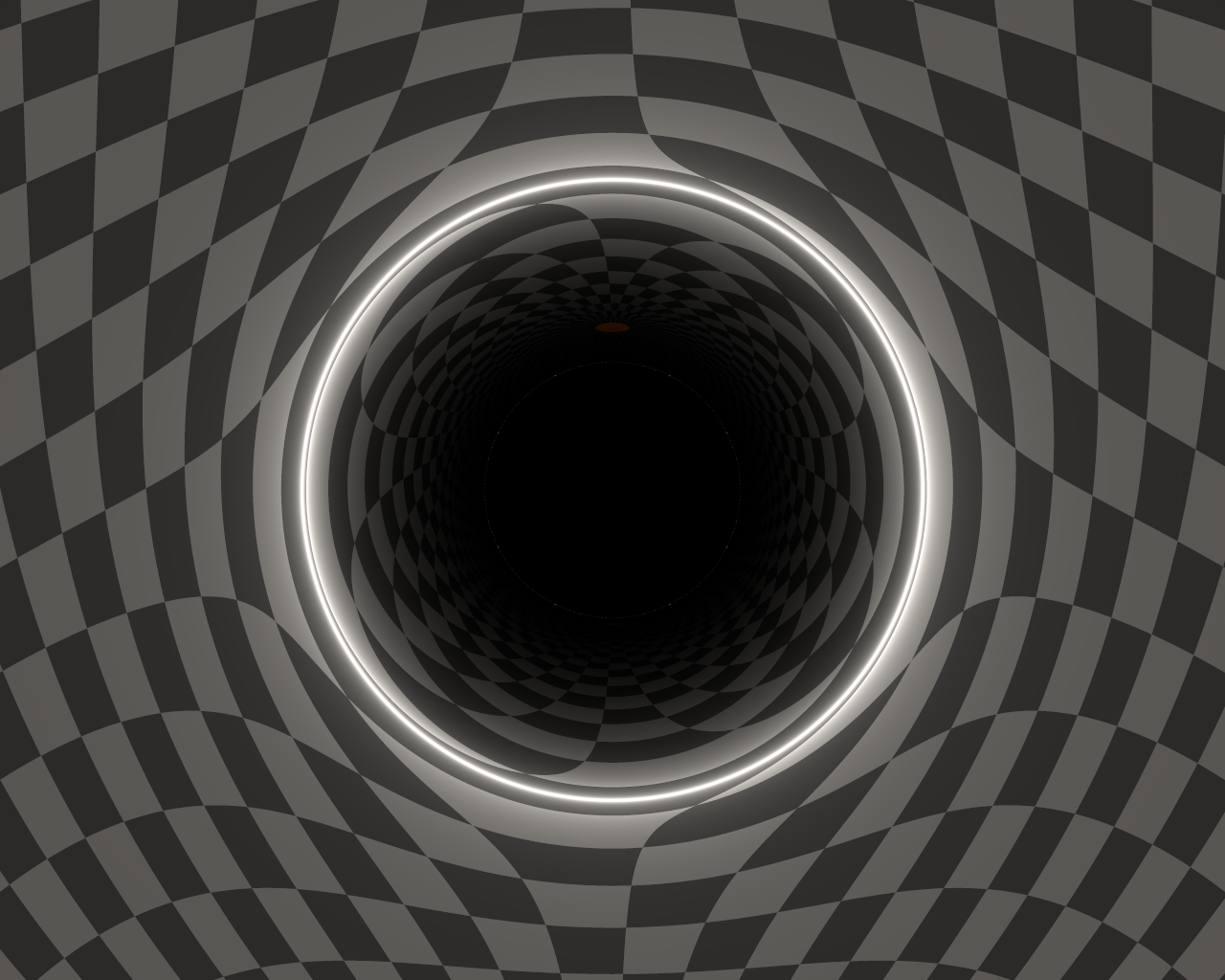}
\vskip 0.12cm
\includegraphics*[width=3.2in]{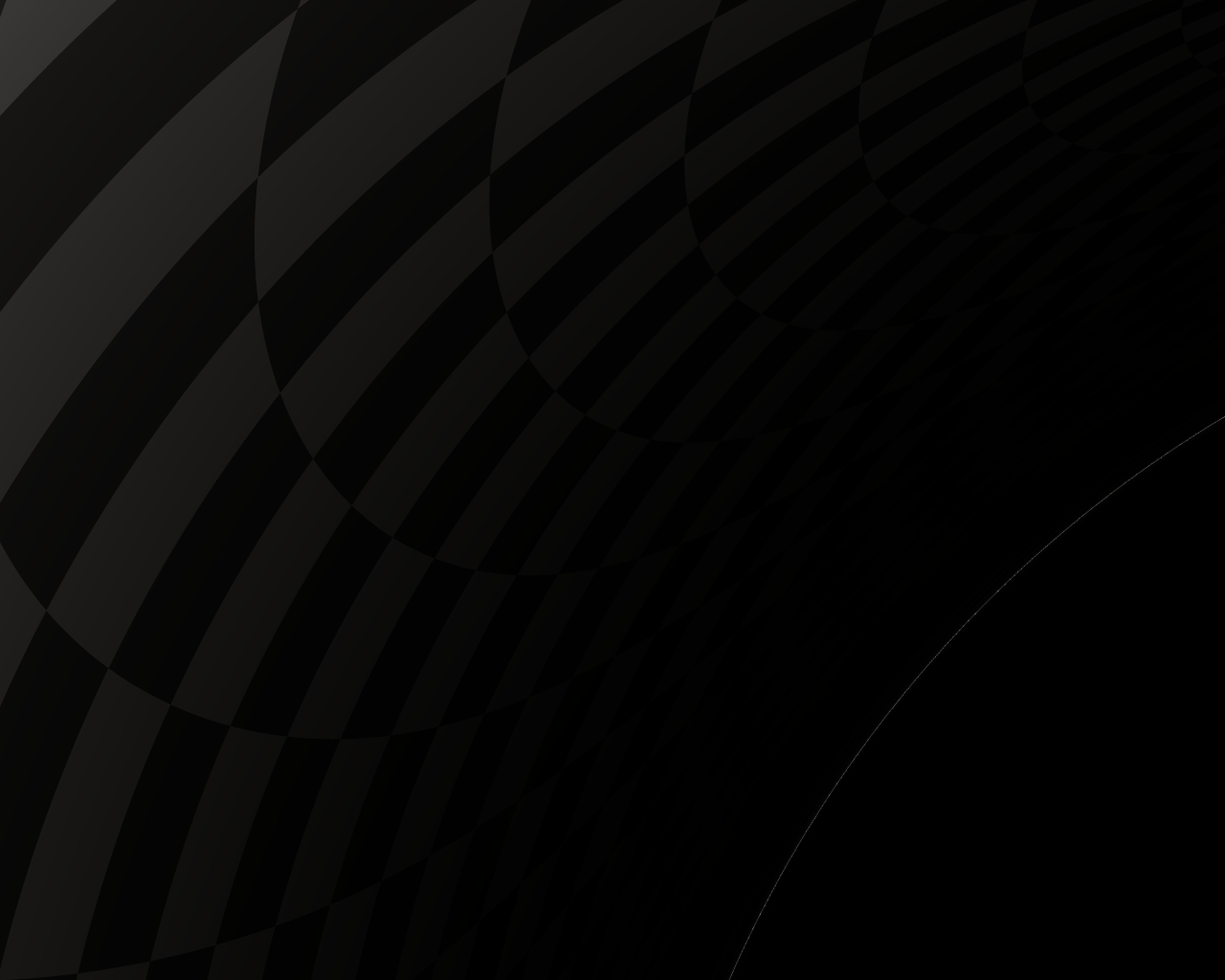}
\caption{An illustration of the amplification and de-amplification
  phenomena induced by deflection of light. Standard calculation
  indicate that the region opposite to the black hole is slightly
  de-amplified, whereas amplification increases as one approaches the
  Einstein ring. Within the Einstein ring, amplification decreases
  drastically and de-amplification kicks in till one goes very close
  to the black hole silhouette (top picture). Then, a tiny region of
  amplification appears, which corresponds to the Einstein ring of the
  black hole anticenter. Because the width on the second amplification
  region is extremely small, one needs a fairly large zooming factor in
  order to see it (bottom picture, showing the upper left quadrant of
  the black hole silhouette). }
\label{fig_dev3}
\end{center}
\end{figure}

As it is obvious from the distortion of central dark square of
Fig.~\ref{fig_dev2}, the distortion of image induced by the black hole
also modifies the angular size of background objects. This is the
so-called gravitational lensing effect. In order to illustrate it, we
reproduce the second image of Fig.~\ref{fig_dev2} by modulating its
intensity by the factor $f$ defined in Eq.~(\ref{eq_ampl}). The result
is shown in Fig.~\ref{fig_dev3}. As is well known, the point of the
celestial sphere which happens to be exactly behind the black hole
experiences an infinite amplification, at least a long as one
considers geometric optics, and appears as an infinitely bright circle
around the black hole, i.e., the so-called Einstein ring. The opposite
point on the celestial sphere experiences the same behaviour and can
be seen much closer to the black hole silhouette as shown in the
accompanying Figure.

\subsubsection{Multiple images}

\begin{figure}[htbp]
\begin{center}
\includegraphics*[width=3.2in]{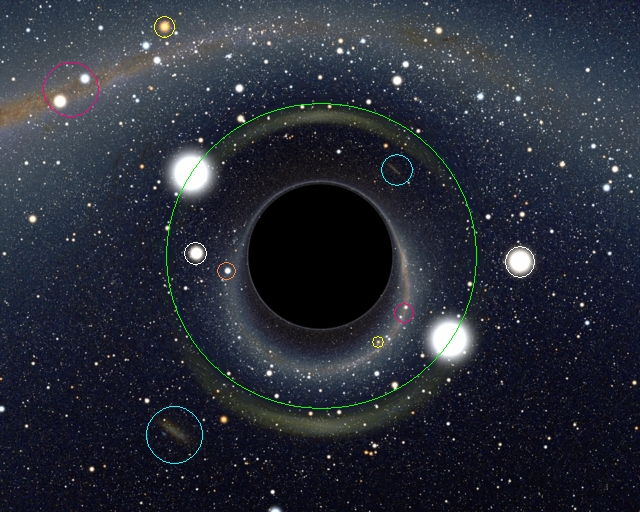}
\caption{A few examples of multiple images in an astronomically
  realistic simulated view. In the upper left quadrant, the circled
  pair of stars correspond to $\alpha$ and $\beta$~Centauri, and the
  single circled star is $\gamma$~Crucis. In the lower left quadrant,
  the the circled structure is the Small Magellanic cloud, the Large
  Magellanic cloud being the very large U-shaped structure above and
  below the black hole silhouette. The bright, circled star to in the
  right is $\alpha$~Carinae. Each pair of these stars are, for one
  outside the Einstein ring (large circle around the black hole
  silhouette), and for the other, inside. The only exception is for a
  star that is almost exactly behind the black hole, whose two image
  lie (almost) exactly on the Einstein ring. This is the case for the
  highly lensed, otherwise anonymous mag 7 star HD49359. The circled
  star inside the Einstein ring whose primary image is not seen is
  Sirius, whose primary image is off-screen (too much to the right),
  but whose secondary image is still easily visible despise
  de-amplification thanks to the very low magnitude of the unlensed
  star. }
\label{mult}
\end{center}
\end{figure}

\begin{figure}[htbp]
\begin{center}
\includegraphics*[width=3.2in]{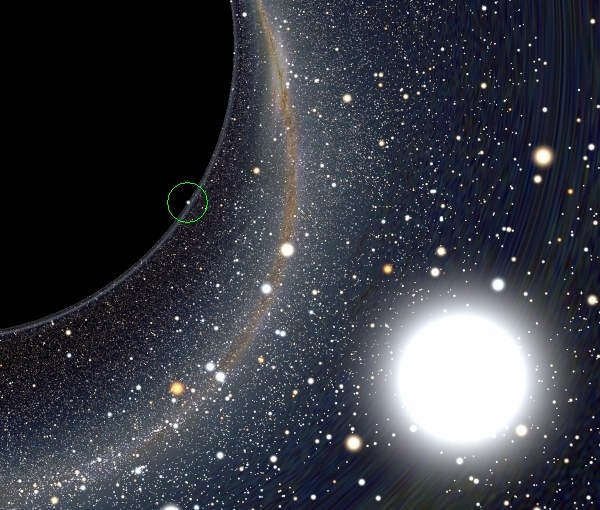}
\vskip 0.12cm
\includegraphics*[width=3.2in]{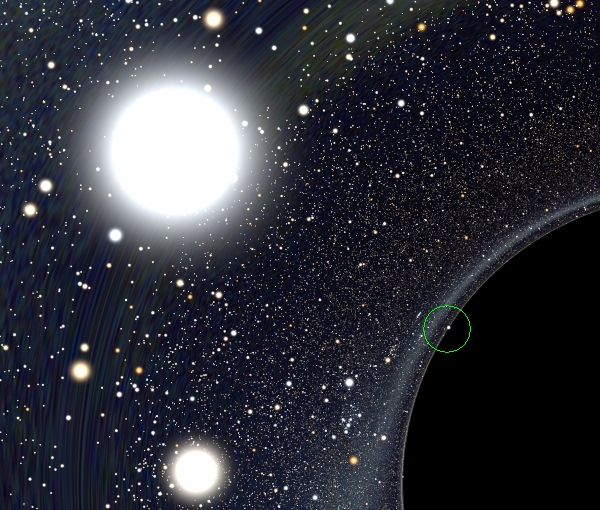}
\caption{When zooming in between HD49359 primary or secondary image
  and the silhouette of the black hole, one sees two ghost images of
  this star, all of which lie within the segment joining the previous
  two images. }
\label{HD49359}
\end{center}
\end{figure}

\begin{figure}[htbp]
\begin{center}
\includegraphics*[width=1.6in]{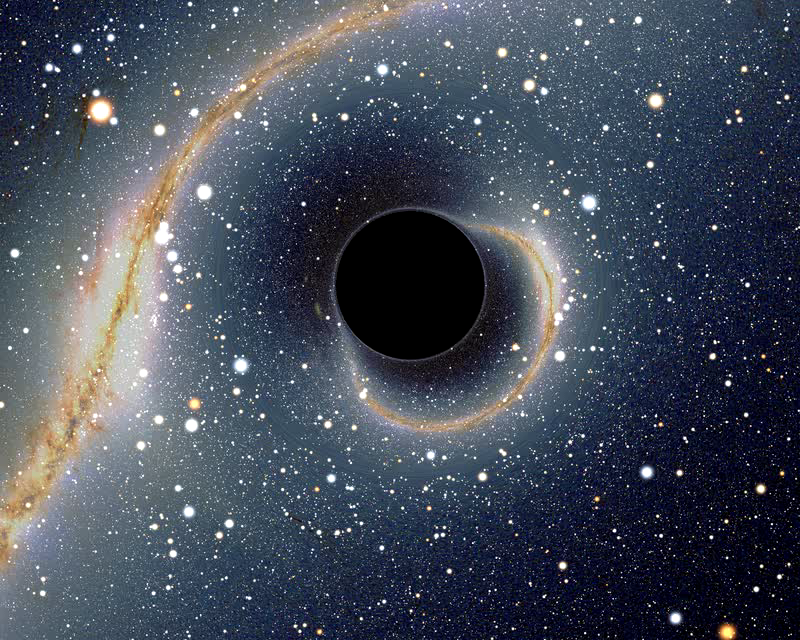}
\includegraphics*[width=1.6in]{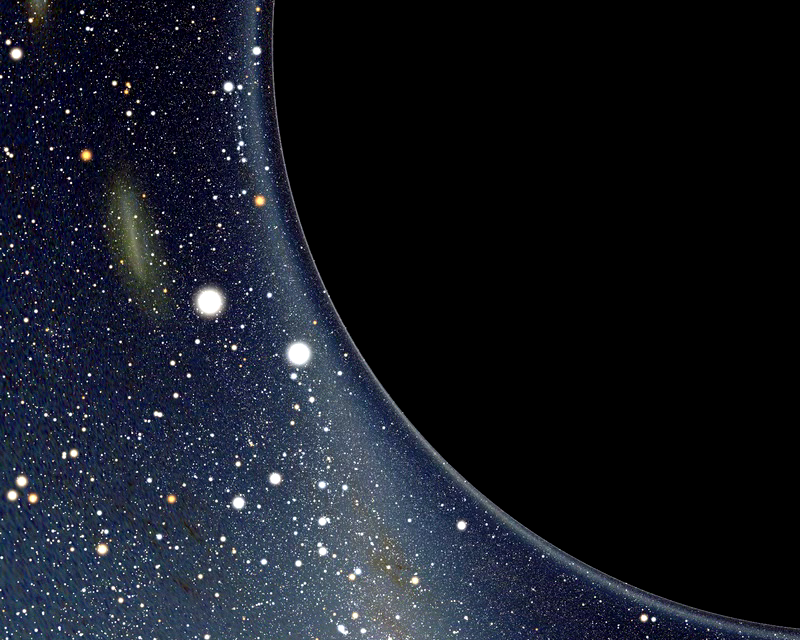}
\vskip 0.12cm
\includegraphics*[width=1.6in]{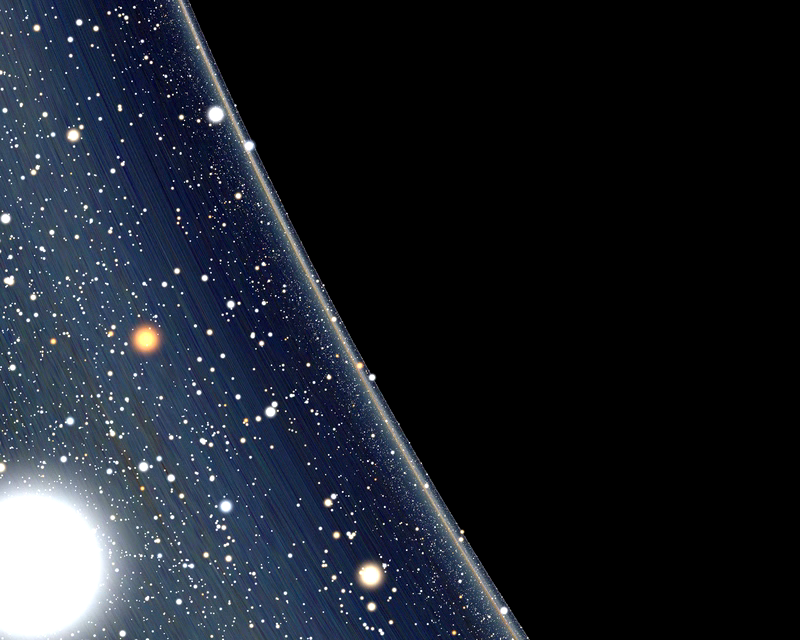}
\includegraphics*[width=1.6in]{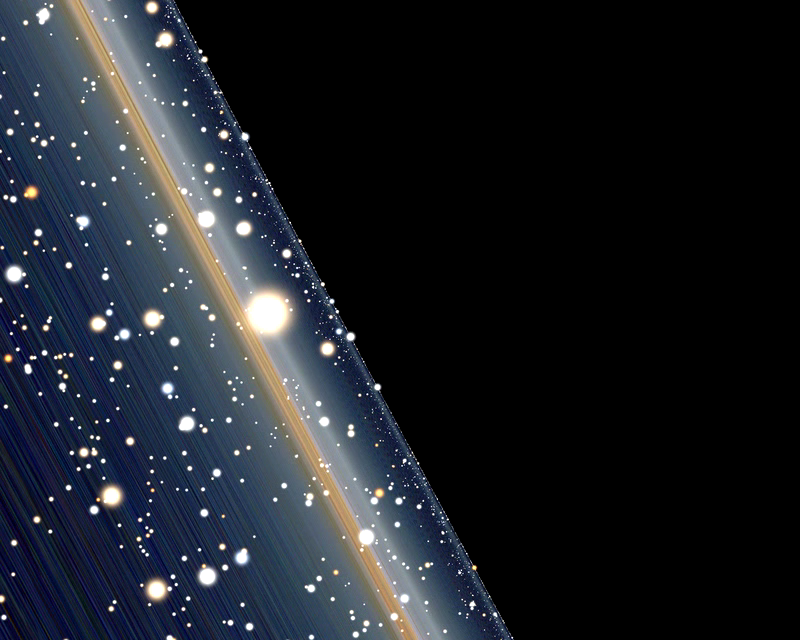}
\vskip 0.12cm
\includegraphics*[width=1.6in]{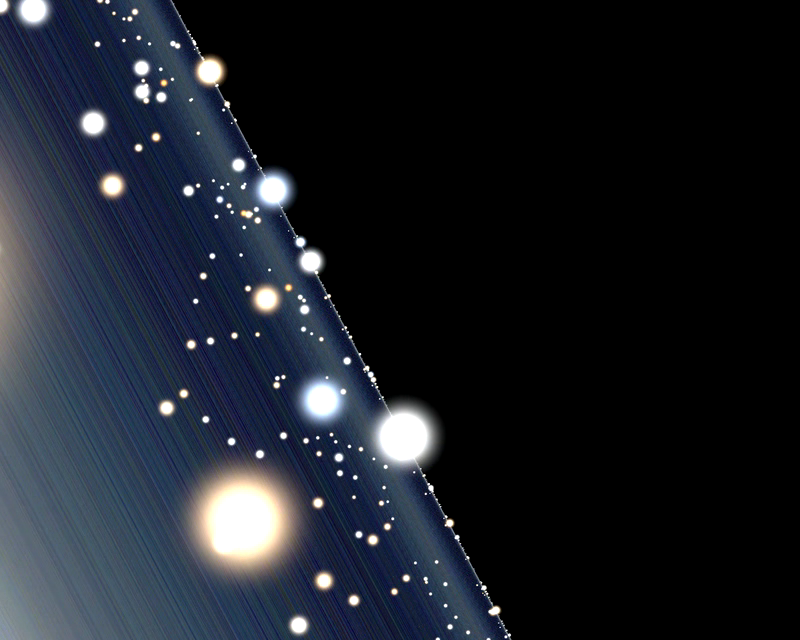}
\includegraphics*[width=1.6in]{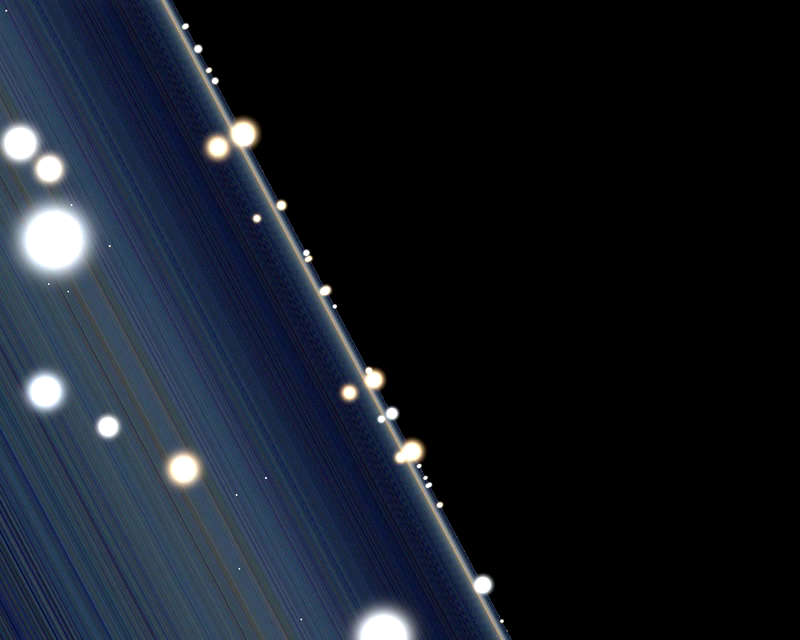}
\vskip 0.12cm
\includegraphics*[width=1.6in]{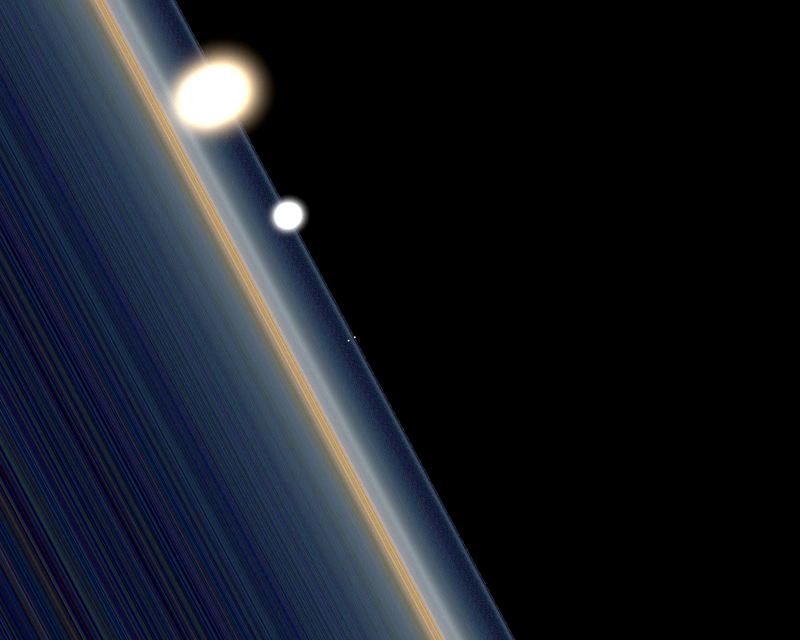}
\includegraphics*[width=1.6in]{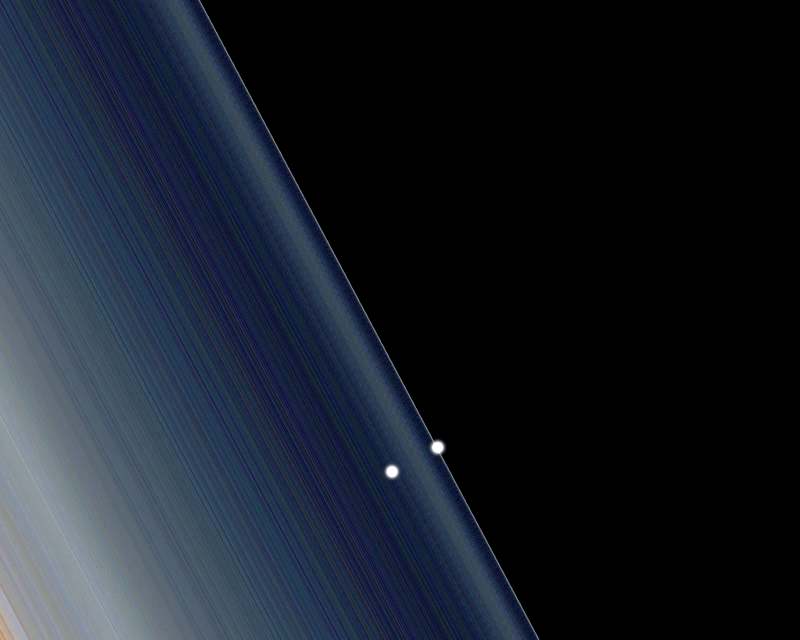}
\caption{An example of the multiple image phenomenon in the
  Schwarzschild metric. One starts with an image with a 90~degrees
  field of view, and each subsequent view (from left to right, then
  top to bottom) has a zoom factor of around 4 with respect to the
  previous one, so that the last one has a field of view around
  $160\,{\rm mas}$. The first view shows the primary image of the
  Galactic disk, as well as its first ``C''-shaped ghost image. The
  thin halo around the black hole corresponds to the second ghost
  image. The 2\NND{} view shows part of the first ghost image of the
  disk, as well as the ghost image of the Large Magellanic cloud. The
  second ghost image is now seen to be slightly away from the edge of
  the black hole silhouette (especially in the bottom part of the
  view). In the 3\RRD{} view, one sees the first ghost image as a
  large band in the left of the image, and the second ghost image is
  now clearly visible. It is still visible in the 4\TTH{} view, which
  also shows the third ghost image. Note that each successive ghost
  image is alternatively bright then dim because they correspond to
  opposite parts of the Galactic disk, whose luminosity is not
  uniform. In the 5\TTH{} view, the third ghost image has been
  broadened on the left, whereas the fourth ghost image appears
  detached from the edge of the silhouette. This fourth image is now
  quite broadened on the 6\TTH{} view, which show the fifth ghost
  image of the disk, close to the edge of the silhouette. This fifth
  ghost image is then seen on the 7\TTH{} and 8\TTH{} views, together
  with the sixth ghost image, and, in the last view, the seventh ghost
  image. Note that the number of stars decreases as one zooms in,
  because of the finiteness of our star catalog. A deeper star catalog
  would be obviously necessary for the last views. Also, assuming that
  stars are pointlike sources may be questionable here since at that
  resolution stars angular size might possibly become visible, and
  make them appear as very elongated thin segments. }
\label{fig_z}
\end{center}
\end{figure}

A star that lies close to the direction that is exactly behind the
black hole from the observer point of view will show two distinct
images, corresponding to light rays that pass on each side of the
black hole, while belonging to the plane containing the black hole,
the observer and the star. These sets of double images can be easily
spotted on any image, especially when the star direction is
sufficiently close to the observer-black hole axis, so that each star
image experiences lensing. However, even when this is no longer the
case, double images can be spotted, although with more difficulty. It
is to be noted that because geodesics in the Schwarzschild metric are
planar, each star image lies on a great circle on which also lies the
null radial geodesic going from the observer to the black hole. If one
uses stereographic projection, these three points lies along a
straight line along the screen if seen by a static
observer. Figure~\ref{mult} shows an example of multiple image in a
simulated, visually realistic astronomical background.
In some rare occasion, on can see more than two images of a given
star. If the observer does not stand too close to the black hole, this
occurs mostly for stars that lie along the observer-black hole
axis, as it is the case for star HD49359 of Fig.~\ref{mult}. Zoomed-in
versions of this view showing two extra ghost images of this star are
shown in Fig.~\ref{HD49359}.
The amount of deflection a light ray experience when approaching the
black hole increases indefinitely as its impact parameters approaches
the critical value $3\sqrt{3} M$ (see
Fig.~\ref{fig_dev}). Consequently, there exists an infinite number of
multiple images, all of which appear increasingly close to the edge of
the lack hole silhouette. Such images can be seen only if one zooms in
by a fairly large factor, as exemplified in Fig.~\ref{fig_z}.

\subsection{Combined effects}

\subsubsection{Circular orbits}

\begin{figure}[htbp]
\begin{center}
\includegraphics*[width=3.2in]{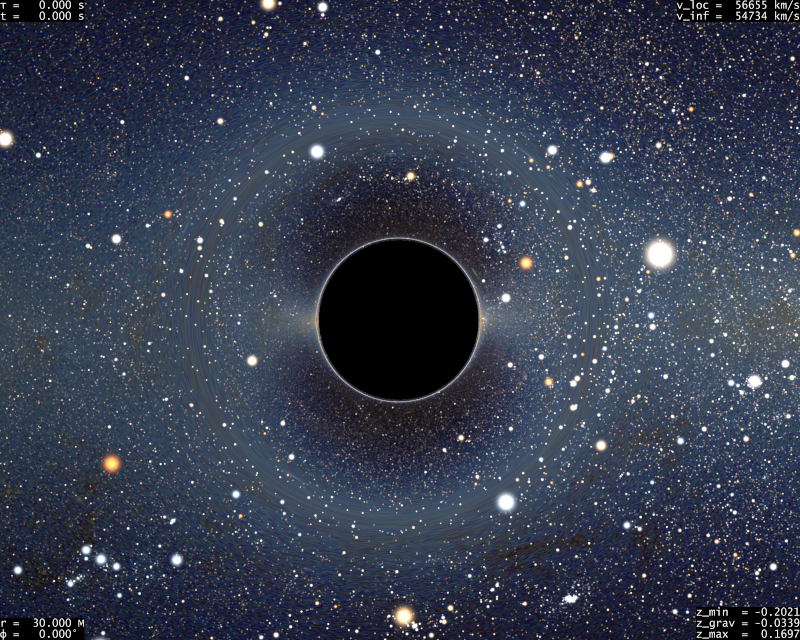}
\vskip 0.12cm
\includegraphics*[width=3.2in]{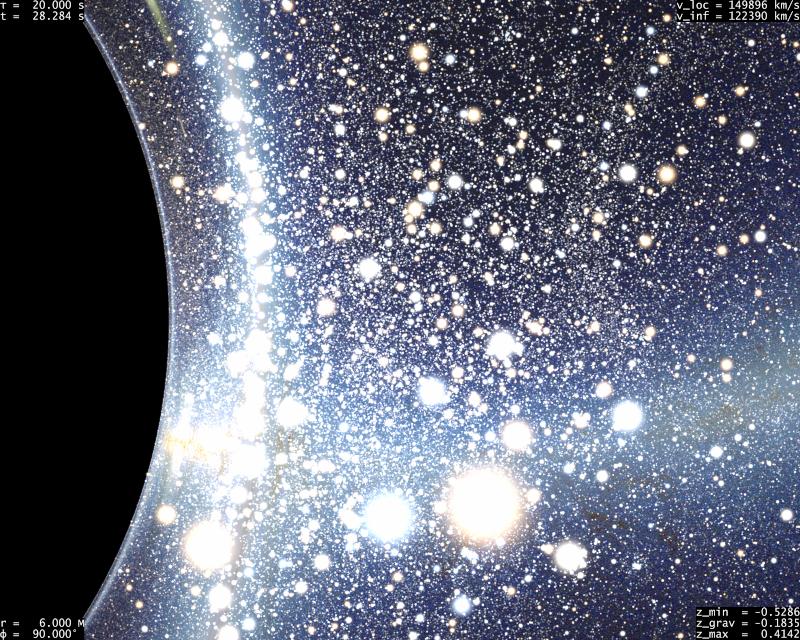}
\caption{Comparison of two circular orbits, with radii $r = 30 M$
  (top) and $r = 6 M$ (bottom). In the first trajectory, the velocity
  is weakly relativistic, so that special relativistic effects are
  moderate. The direction of motion, which is on the right of the
  image, is only thus only moderately brighter than the opposite
  one. On the contrary, there is a very strong brightening in the
  second image, since the observer is now subject to a large kinetic
  blueshift (since orbital velocity is half of the speed of light
  here) and a significant gravitational blueshift (since the observer
  is close to the black hole). Similarly, the star background is
  fairly recognizable in the first picture. One can for example spot a
  somewhat flattened version of Orion constellation in the lower left
  quadrant of the image as well as a flattened Taurus in the bottom
  centre. This is no longer the case in the second one, where even
  familiar constellations are difficult to spot because the star
  background is now so crowded. Large blueshift allows some highly
  blueshifted cold stars to overcome the usually brighter stars we are
  used to see.  There are, for example, many more bright, orange,
  stars (i.e., usually cold and faint stars) around Ursa Major
  constellation in the upper right quadrant of the image. }
\label{figcirc}
\end{center}
\end{figure}

One can combine special and general relativistic effects by
considering an observer who is moving around a black hole. The
simplest example corresponds to that of a circular orbit. As is well
known, a timelike observer around a Schwarzschild black hole
experiences a radial potential of the form
\begin{equation}
V (r) \propto
   \left(1 - \frac{2 M}{r} \right) \left(1 + \frac{L^2}{r^2} \right) ,
\end{equation}
where $L$ is the observer angular momentum per unit of mass. A stable
circular orbit corresponds to the local minima of $V$, which
correspond of the largest of the roots of equation
\begin{equation}
M r^2 - L^2 r + 3 L^3 M = 0 ,
\end{equation}
whose smallest value is $r = 6 M$. For a fixed orbit radius, the
observer's four-velocity $u_\CIRC^\mu$ is then given by its constants
of motion, with
\begin{equation}
L^2 = \frac{M r}{1 - \frac{3 M}{r}} ,
\end{equation}
and
\begin{equation}
E \equiv g_{tt} u_\CIRC^t 
  = \frac{1 - \frac{2 M }{r}}{\sqrt{1 - \frac{3 M}{r}}} ,
\end{equation}
so that one has, assuming that the orbital plane lies within the
equatorial $\theta = \pi / 2$ plane,
\begin{eqnarray}
u_\CIRC^t & = & \frac{1}{\sqrt{1 - \frac{3 M}{r}}} , \\
u_\CIRC^\varphi & = & \frac{M r^{-\frac{3}{2}}}{\sqrt{1 - \frac{3 M}{r}}} .
\end{eqnarray}
The orbital velocity $v_\CIRC$ with respect to an observer situated at
the same radial coordinate is then given by $u_{\STAT , \mu}
u_\CIRC^\mu = \gamma = (1 - v_\CIRC^2)^{-\frac{1}{2}}$, where
$u_\STAT^\mu$ is the static observer four-velocity, whose only non
zero component is $u_\CIRC^t = (1 - 2 M / r)^{-\frac{1}{2}}$. The
velocity is then
\begin{equation}
v_\CIRC = \sqrt{\frac{\frac{M}{r}}{1 - \frac{2 M}{r}}} .
\end{equation}
Neglecting the denominator gives the well-known formula of the non
relativistic third Kepler law. For orbital radii of $r = 30 M$ and $r
= 6 M$, the orbital velocity is therefore $v_\CIRC \simeq 0.189 c$ and
$v_\CIRC = 0.5 c$, respectively. In order to determine the maximum
redshift and blueshift an observer experiences, one has to combine the
gravitational redshift formula~(\ref{zgrav}) with the kinetic
one~(\ref{zkin}), the combined redshift $z_\TOT$ being given by
\begin{equation}
1 + z_\TOT = (1 + z_\GRAV) (1 + z_\KIN) .
\end{equation}
For an observer at $r = 6 M$, the maximum blueshift (in the front
direction) is thus $z_\MAX = \sqrt{2} / 3 - 1 \simeq - 0.529$, and the
maximum redshift (in the rear direction) is $z_\MIN = \sqrt{2} - 1
\simeq 0.414$. Fig.~\ref{figcirc} shows a comparison between the side
view seen by a $r = 30 M$ observer looking toward the black hole, and
the front view seen by a $r = 6 M$ observer, both in circular orbit.

\subsubsection{Non circular, non radial trajectories}

\begin{figure}[htbp]
\begin{center}
\includegraphics*[width=3.2in]{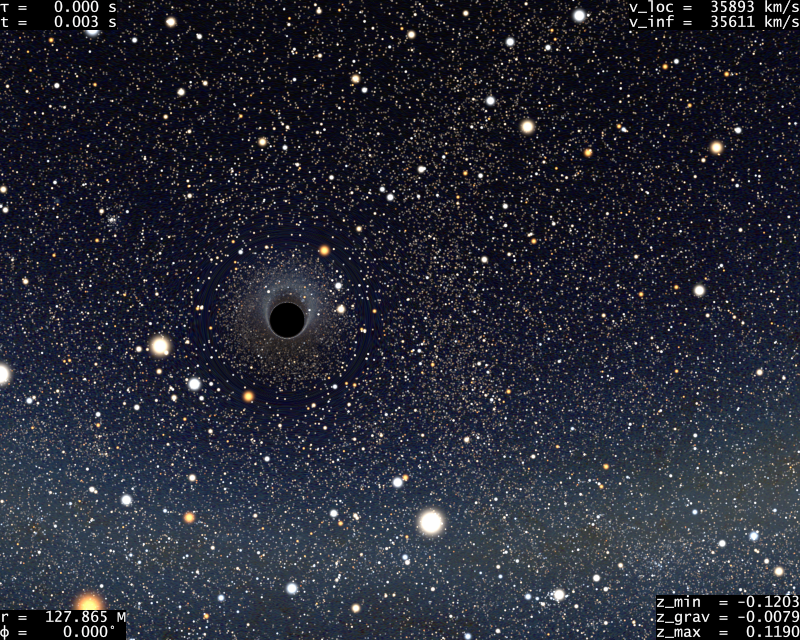}
\vskip 0.12cm
\includegraphics*[width=3.2in]{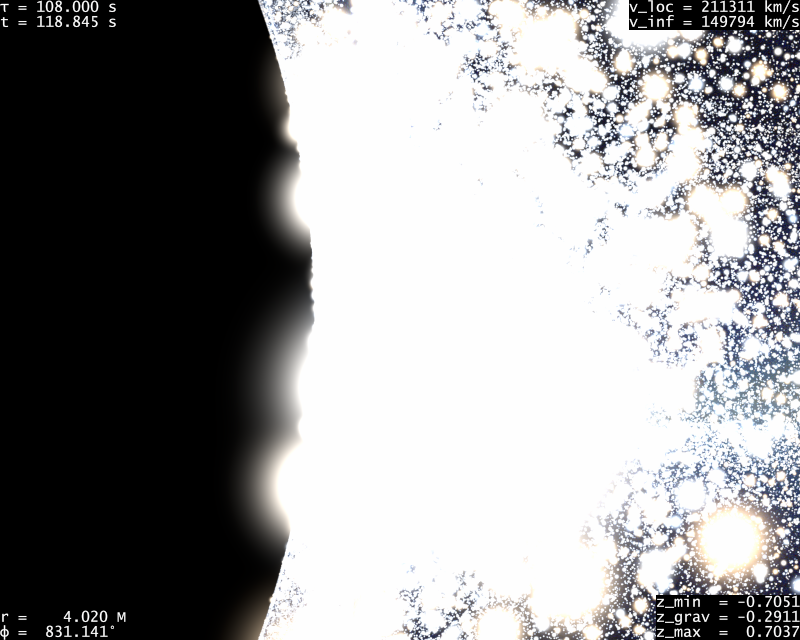}
\caption{Two views of an almost extremal pseudo parabolic orbit,
  starting from infinity a zero velocity and angular momentum close to
  the minimum allowed value of $4 M$ (see Eq.~(\ref{r_par})). Top
  image corresponds to a view far from the black hole, where the
  trajectory is close to radial. Bottom image corresponds to a view
  earlier than, but very close to periastron, at $r = 4.02 M$. Local
  velocity, given by Eq.~(\ref{vit_par}) is very close to $c /
  \sqrt{2}$, and velocity measured by a distant observer, given by
  $v_\INF = v_\PER \sqrt{1 - 2 M / r}$ is very close to $c / 2$. Note
  that the azimuthal angle shift between the two view is $\sim
  831$~deg., much larger than the Newtonian analog which should be
  slightly smaller than $180$~degrees.}
\label{figpar}
\end{center}
\end{figure}

Let us switch to geodesic non circular trajectories. Relativistic
equivalent of Newtonian parabolic trajectories correspond to
trajectories with zero velocity at infinity but non zero angular
momentum $L$ per unit of mass. Simple algebra then show that
periastron radial coordinate is given by
\begin{equation}
\label{r_par}
\frac{r_\PER}{M}
 = \frac{1}{\frac{1}{4} - \sqrt{\frac{1}{16} - \frac{M^2}{L^2}}} , 
\end{equation}
and velocity at periastron is 
\begin{equation}
\label{vit_par}
v_\PER = \sqrt{\frac{2 M}{r}} .
\end{equation}
Obviously, such a relativistic equivalent of a parabolic trajectory
has a periastron bounded by $r = 4 M$, and the maximal velocity at
periastron is then $v = c / \sqrt{2}$. Now, the trajectory in itself
turns much more around the black hole than a parabola because of the
extreme relativistic shift of periastron it experiences (the so-called
zoom-and-whirl effect), and the azimuthal angle shift between far from
the black hole and periastron is much bigger than $\pi /
2$. Fig.~\ref{figpar} gives an example of an almost extremal
pseudo-parabolic trajectory.

\begin{figure}[htbp]
\begin{center}
\includegraphics*[width=3.2in]{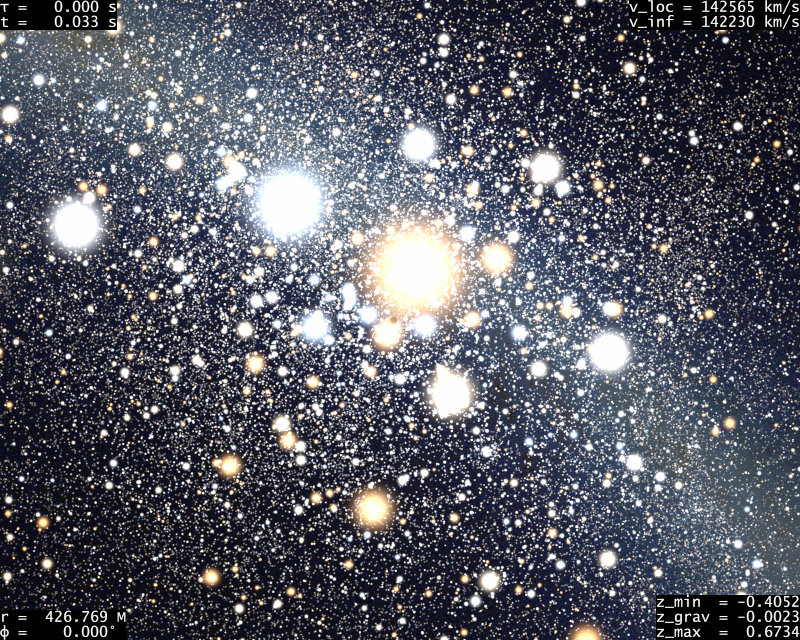}
\vskip 0.12cm
\includegraphics*[width=3.2in]{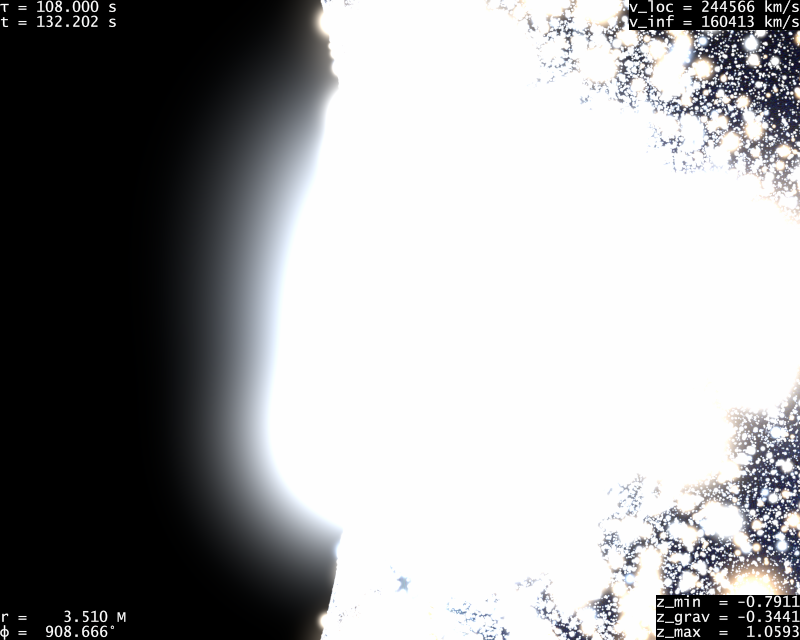}
\caption{Pseudo hyperbolic trajectory with a periastron radial
  coordinate of $3.5 M$, which necessitates a velocity at infinity at
  least larger than $\sqrt{2} c / 3 \sim 141323 \,{\rm km}/{\rm s}$
  (actual value chosen here is very slightly higher). A velocity close
  to that value is reached far from the black hole in the top
  view. Bottom view is computed close to periastron. The black hole
  mass is the same here as in the previous Figure, and the proper time
  interval between first and second view is also the
  same. Consequently, since bottom view in both figures are close to
  periastron, top view radial coordinate is large in this Figure than
  in the previous one because velocity far from the black hole is
  larger. Black hole angular size is in addition further reduced by
  the stronger aberration of this Figure, so that it is barely visible
  (below the brightest central, orange star). Note also that the
  background sky is much brighter, as expected.}
\label{fighy1}
\end{center}
\end{figure}

\begin{figure}[htbp]
\begin{center}
\includegraphics*[width=3.2in]{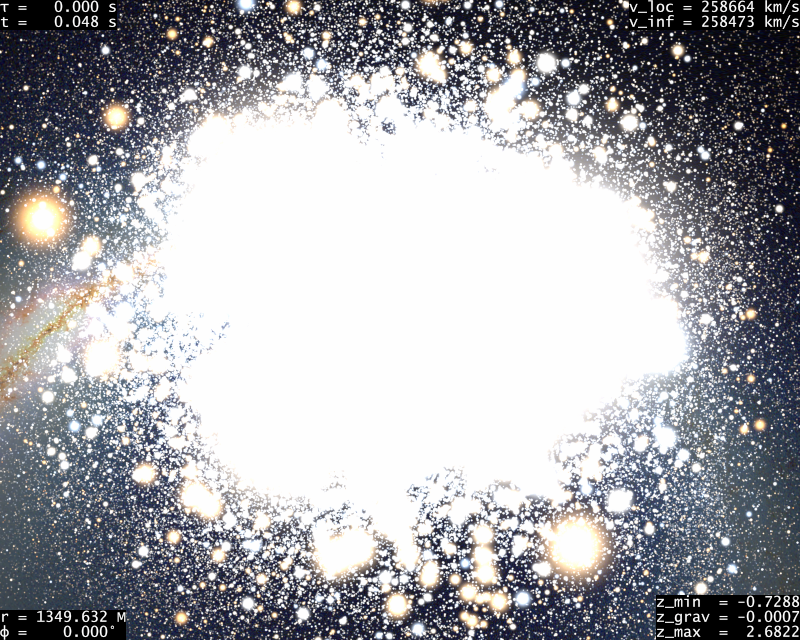}
\vskip 0.12cm
\includegraphics*[width=3.2in]{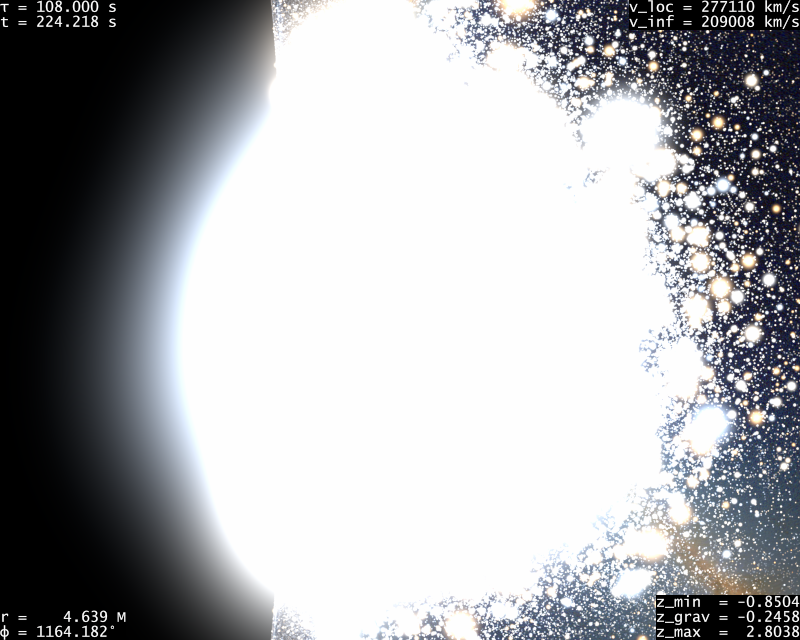}
\caption{Same as Fig.~\ref{fighy1}, except that periastron now occurs
  at $r = 3.1 M$, which necessitates an initial velocity greater than
  $v \sim 0.862 c \sim 258552\,{\rm km}/{\rm s}$. On the top picture,
  the black hole angular size is so small and the star background so
  bright that the black hole silhouette is invisible with our
  rendering choices. }
\label{fighy2}
\end{center}
\end{figure}

One can also have trajectories which are the relativistic equivalent
of Newtonian hyperbolic trajectories. In this case, a convenient set
of parameters describing the trajectory are the velocity at infinity
and the impact parameter. Conversely, one can determine which is the
minimum velocity at infinity, $v_\infty$, that allows to reach a given
value of the radial $r_\PER$ coordinate at periastron. After some
algebra, it appears that\footnote{See Ref.~\cite{chandrasekhar83} for
  more details.} this is conveniently done if we parametrize the
periastron radial coordinate by the quantity $e$, $1 < e < 3$, such
that
\begin{equation}
r_\PER = \frac{2 (3 + e)}{1 + e} M ,
\end{equation}
then the velocity at infinity must be greater than
\begin{equation}
v_\infty > \sqrt{\frac{e^2 - 1}{8}} .
\end{equation}
Equivalently, this minimum velocity can be expressed as a function of
$r_\PER / M$~:
\begin{equation}
v_\infty > \frac{\sqrt{4 - \frac{r_\PER}{M}}}{\frac{r_\PER}{M} - 2} .
\end{equation}
Figures~\ref{fighy1} and~\ref{fighy2} show two example of such near
extremal pseudo-hyperbolic trajectories.

\end{document}